\documentclass[a4paper,11pt]{article}
\usepackage{jheppub,esint,shuffle,psfrag}
 \usepackage[utf8]{inputenc}
\usepackage{xcolor}
\usepackage{varioref}
\usepackage{amsmath,amsfonts,amsthm,mathrsfs}
\usepackage{amssymb,amsthm}
\usepackage{enumerate}
\usepackage{float}
\usepackage{fancyvrb}
\usepackage{verbatim}
\usepackage{wrapfig}
\usepackage{appendix}
\usepackage{amstext}
\usepackage{amssymb}
\usepackage{braket}
\usepackage{graphicx}
\usepackage{color}
\usepackage{varioref}
\usepackage{multirow,graphics}
\usepackage{epstopdf}

\usepackage{bbm}
\usepackage{slashed}
\usepackage{relsize}

\usepackage{hyperref} 
\usepackage[english]{babel}
\usepackage{mathtools, amssymb} 
\usepackage{mathtext} 
\usepackage{bm} 
\usepackage{amsthm} 
\usepackage{dsfont}
\usepackage{tikz}
\usetikzlibrary{patterns, calc, decorations.markings, arrows.meta}

\tikzset{->-/.style={line width=0.25mm,decoration={
			markings,
			mark=at position 0.53 with {\arrow{Stealth}}},postaction={decorate}}}
\tikzset{-<-/.style={line width=0.25mm,decoration={
			markings,
			mark=at position 0.47 with {\arrowreversed{Stealth}}},postaction={decorate}}}

\hypersetup{colorlinks=true, linktocpage=false, linkcolor=blue!80!black, citecolor=red!85!black, urlcolor=blue} 

\def\blue{\color{blue}}
\def\red{\color{red}}

\renewcommand{\imath}{\mathrm{i}}
\renewcommand{\Re}{\operatorname{Re}}
\renewcommand{\Im}{\operatorname{Im}}
\renewcommand{\phi}{\varphi}

\def\II{\hbox{{1}\kern-.25em\hbox{l}}}

\usepackage{tocloft}

\numberwithin{equation}{section}

\begin{document}
\thispagestyle{empty}

\begin{center}
{\bf \large BC-type open $SL(2,\mathbb{C})$ spin chain}
\vspace{0.7cm}

{P. Antonenko$^{\dagger\ast}$, S. Derkachov$^{\dagger\diamond}$, P. Valinevich$^{\dagger}$}
\vspace{0.7cm}

{\small \it
	$^\dagger$Steklov Mathematical Institute, Fontanka 27, \\St.~Petersburg, 191023, Russia\vspace{0.3cm}\\
	$^\ast$Leonhard Euler International Mathematical Institute,\\ Pesochnaya nab. 10, St.~Petersburg, 197022, Russia\vspace{0.3cm}\\ 
$^\diamond$Saint Petersburg State University, Saint Petersburg, Russia
}
	
\end{center}

\vspace{0.1cm}

\begin{abstract} \noindent
We diagonalize the $B$-element of monodromy matrix for noncompact open $SL(2,\mathbb{C})$ spin chain with boundary interaction.
The monodromy matrix is defined in terms of $SL(2,\mathbb{C})$ $L$-operator and boundary $K$-matrix. 
The eigenfunctions of $B$-operator are constructed iteratively using 
raising $\Lambda$-operators. The key role in the calculations plays the Baxter $Q$-operator commuting with the $B$-operator. 
The main building blocks for $\Lambda$- and $Q$-operators are $\mathcal{K}$-operator  -- the general solution of reflection equation and 
$\mathcal{R}$-operator -- the reduction of the general solution 
of the Yang-Baxter equation.

Two types of the symmetry of eigenfunctions are established. 
The first kind is the invariance under permutations and reflections 
of spectral variables, or in other words, under the action of Weyl 
group of B and C root systems.
The second kind is the symmetry with respect to transformation $(s,g) \to (1-s,1-g)$, where $s$ is the spin variable and $g$ is the parameter of $K$-matrix.

We prove that obtained system of eigenfunctions is orthogonal and complete.
The calculation of the scalar product of eigenfunction is given in initial coordinate representation. 
We derive the Mellin-Barnes integral representation for eigenfunctions 
and use it to prove the comleteness.
\end{abstract}

\newpage

 \tableofcontents

 \newpage


\section{Introduction}

In the present paper we continue the study of noncompact $SL(2,\mathbb{C})$ spin magnet of BC-type, which was started in \cite{ABDV}.
This quantum-mechanical model describes the two-dimensional system of $n$ interacting particles including the boundary interaction \cite{Skl:Boundary, Ch}.
The Hilbert space $H$ of the model is the direct product of $n$ copies of infinite-dimensional functional space $\mathrm{L}^2(\mathbb{C})$, that is $H=V_1\otimes\ldots\otimes V_n$, where $V_k=\mathrm{L}^2(\mathbb{C})$. 
This model can be treated in the framework of the quantum inverse scattering method \cite{Skl91,Fad:BetheAns,FST,KulSk,TaFa}. For the spin chain with $n$ sites we assign to the each site the quantum $L$-operator
\begin{align*}
L(u) =
\left (\begin{array}{cc}
u + S & S_{-} \\
S_{+} & u - S \end{array} \right ) ,
\end{align*}
where $S_{\pm}\,,S$ are generators of principal series representation of $SL(2,\mathbb{C})$ group \cite{GGV, Gelfand-Naimark}. The representation is defined in the space $\mathrm{L}^2(\mathbb{C})$, it is parametrized by (in general complex) spin~$s$.
Following Sklyanin's approach \cite{Skl:Boundary} we consider the matrix
\begin{equation} \label{Kint}
	K(u) = \begin{pmatrix}
		\gamma\bigl(g - \frac{1}{2} \bigr) & u - \frac{1}{2} \\[6pt]
		\gamma^2 \bigl(u - \frac{1}{2} \bigr) & \gamma\bigl(g - \frac{1}{2} \bigr)
	\end{pmatrix}
\end{equation}
that contains information about boundary interaction.
Momodromy matrix is constructed as follows \cite{Skl:Boundary, KS, PP}
\begin{equation} \nonumber
	T_n(u) = L_n(u) \cdots L_1(u) K(u) L_1(u) \cdots L_n(u) = \begin{pmatrix}
		A_n(u) & B_n(u) \\
		C_n(u) & D_n(u)
	\end{pmatrix} ,
\end{equation}
where $L_k$ denotes the $L$-operator acting nontrivially in the space $V_k$.
The key property of $K$-matrix, which provides the integrability of the model, is that it solves the reflection equation with Yang's $R$-matrix acting in the tensor product $\mathbb{C}^2\otimes\mathbb{C}^2$, which is a solution of Yang-Baxter equation.
From this property and the $RLL$-relation between the $R$-matrix and $L$-operator follows the similar reflection relation for the monodromy matrix.
It implies the commutativity
\begin{equation} \label{Bcomm} 
	[B_n(u), B_n(v)] = 0 \,.
\end{equation}
The operator $B_n(u)$ is a polynomial in the spectral parameter $u$, and by construction this polynomial has $n$ independent nontrivial coefficients.
From \eqref{Bcomm} it follows that these coefficients form the set of commuting operators. 

In this paper we construct mutual eigenfunctions of the mentioned commuting operators, or, in other words, the eigenfunctions of the operator $B_n(u)$.
Since $B_n(u)$ is a polynomial in $u$, its eigenvalues are also polynomials in the spectral parameter. The eigenfunctions are parametrized by quantum numbers $x_1,\ldots,x_n$, such that $\pm x_k$ are roots of these polynomials.

We should note that we call the model \textit{the spin chain of BC-type} in the full 
analogy with the BC-type Toda chain \cite{KuzSk,IS}, because the eigenfunctions and corresponding eigenvalues are invariant under the action of Weyl group of B and C root systems. 
That is, they are symmetric with respect to permutations of parameters $x_1,\ldots,x_n$ and reflections $x_k \to -x_k$.
In addition, for proving the completeness of obtained set of eigenfunctions we use the BC-type Gustafson integral generalized to the complex field~\cite[eq.~(4.2)]{DMV:Gust3}, \cite[eq.~(2.3b)]{DM:Gust4}, which possesses the same symmetry.

The motivation for constructing the eigenfunctions of $B$-operator is that these functions are used for diagonalizing the transfer matrix $t(u)=A(u)+D(u)$, which is the trace of monodromy matrix.
By analogy with the operator $B(u)$, the transfer matrix is a polynomial in $u$, and its diagonalizability follows from the commutation $[t(u),t(v)]=0$, which in turn follows from the reflection equation for the monodromy matrix.
In the case of infinite-dimensional principal series representations the well-known technique of the Algebraic Bethe Ansatz \cite{Fad:BetheAns,FST,KS}
does not work, and to obtain the eigenfunctions of $t(u)$ one needs to use the Sklyanin’s method of Separation of variables \cite{Skl:Toda,Skl:VarSep,KK,KL2,KL1}.
Eigenfunction of $B(u)$ plays the role of the integral kernel of the transition operator to the representation of separated variables.
The orthogonality and completeness of the set of eigenfunctions, which we prove in the present paper, is equivalent to the unitarity of the transition operator.
In \cite{DKM2} the SoV method was applied to a resembling model -- BC-type open $SL(2,\mathbb{R})$ spin chain with trivial boundary matrix and 
in \cite{IS} to the BC-type Toda chain.

We construct the set of eigenfunctions iteratively.
The function corresponding to the spin chain consisting of $n$ sites is obtained by application of raising integral operator
to the function for $n-1$ sites.
This operator is expressed in terms of simple building blocks of two types:
the $\mathcal{R}$-operator, which is defined by the $\mathcal{R}LL$-relation of Yang-Baxter type with the $L$-operator of the model, 
and the $\mathcal{K}$-operator (reflection operator) satisfying the reflection equation with $K$-matrix and $L$-operator.
We prove the orthogonality and completeness of the obtained set of eigenfunctions and investigate their symmetries.

The similar program was realized for other spin chains with $SL(2,\mathbb{C})$ symmetry group as well. In \cite{DMV:Gust3} all the above mentioned steps,
except the proof of completeness, were done for the eigenfunctions of $B(u)$ in the case of open BC-type chain with $K$-matrix equal to identity. 
The diagonalization of all elements of the monodromy matrix for the $A$-type spin chain as well as proofs of symmetries and orthogonality of their eigenfunctions can be found in \cite{DKM,DM14}, the completeness was proven in \cite{M}.
In these papers the eigenfunctions are constructed inductively with the help of integral operators of the same type.

The calculations in all mentioned works were performed using the Feynman diagram technique. The kernels of integral operators, and, consequently, 
the eigenfunctions, were expressed in terms of Feynman diagrams, and all computations were reduced to simple graphical transformations. 
However, in our case the diagrams become too cumbersome, so that we apply 
the algebraic approach of our previous work \cite{ADV1} used for A-type 
$SL(2,\mathbb{C})$ spin chain. 
By means of this approach the diagrammatic transformations are reformulated in the language of local operator relations like Yang-Baxter equation for
$\mathcal{R}$-operator.

The paper is organized as follows. In Section~\ref{sect:Rep} we give all the necessary information about principal series representations of the group $SL(2,\mathbb{C})$.

In Section~\ref{sect:ReflEq} we consider the reflection equation for $\mathcal{K}$- and $\mathbb{R}$-operators. The last one is the general
solution of Yang-Baxter equation, corresponding to the symmetry group $SL(2,\mathbb{C})$, and $\mathcal{R}$-operator is its special case.
This general reflection equation was formulated in~\cite{Oliv}
on the basis of the particular case from~\cite[Section~7]{ABDV}. Its proof is given in Appendix~\ref{Reflection}. We also show the whole hierarchy
of relations of Yang-Baxter and reflection (boundary Yang-Baxter) type, starting from the mentioned general relations and ending with their simplest versions for $R$- and $K$-matrices.
This hierarchy includes the identities which are used in the construction of eigenfunctions: the $\mathbb{R}LL$-relation between $\mathbb{R}$-operator and $L$-operators and the reflection relation between $\mathcal{K}$-, $L$-operators and $R$-matrix.
It seems that our consideration is far from the full generality 
\cite{Weston,VlaarWest,CoopVlaarWest} but it is enougt for our goals.

The definition of the model in the language of quantum inverse scattering method is given in Section~\ref{sect:Monodromy}. We also specify the form of the operator $B(u)$ and state its eigenvalue problem.
Section~\ref{sect:EigenfuncConstruct} is devoted to the inductive construction of eigenfunctions of $B(u)$.

In Section~\ref{qop} we introduce one of the key objects -- Baxter's $Q$-operator. Using the mentioned algebraic approach we prove
its characteristic properties: commutativity of $Q$-operators, commutation with $B$-operator and Baxter equation.
We should note that using similar construction it is possible 
to obtain $Q$-operator commuting with transfer matrix $t(u)=A(u)+D(u)$. 
It will be very instructive to compare constructions of 
$\mathcal{K}$- and $Q$-operators with \cite{Weston,VlaarWest,CoopVlaarWest}.

In Section~\ref{sect:Eigenunc} we show that the constructed functions are common eigenfunctions of $Q$- and $B$-operators. We also study different symmetries of eigenfunctions.
As it was noted, they are invariant under permutations and reflections of spectral variables $x_1,\ldots,x_n$. Besides, there is a simple relation between
the eigenfunctions corresponding to parameters $s, g$ and $1-s, 1-g$, where $g$ is the parameter of boundary matrix~\eqref{Kint}. They differ by a product of power functions.
Derivation of above-listed properties relies on the commutation relations between $Q$-operators and raising operators. The raising operator and
$Q$-operator are closed relatives, their formulas almost coincide, and they can be easily expressed in terms of each other.
Therefore, the mentioned relations can be deduced from one fundamental identity -- the commutativity of $Q$-operators.

Section~\ref{ort} is devoted to the proof of orthogonality relation for eigenfunctions and calculation of the Sklyanin measure.
The scalar product of eigenfunctions is calculated recursively, an important role is played by the $Q$-operator. We use three its properties:
it is closely related to the raising operator, which provides the recursive construction of eigenfunctions, it is diagonalized by eigenfunctions, and its hermitian conjugate is expressed in terms of its inverse.

In Section~\ref{sect:MB} we derive the alternative integral representation for eigenfunctions -- the Mellin-Barnes representation. We expand the constructed
functions of BC-type spin chain over the set of eigenfunctions of the operator $a(u)+\gamma\,b(u)$ for the chain of A-type \cite{ADV1}. 
Here $a(u)$ and $b(u)$ are the elements of monodromy matrix of the A-type spin chain and $\gamma$ is the parameter of $K$-matrix \eqref{Kint}.
The kernel of the corresponding integral expansion is expressed in the closed form in terms of gamma functions associated with the complex field.
It is given by scalar product of eigenfunctions of A-type and BC-type spin chains, which is again calculated recursively. The use of $Q$-operators drastically simplifies the computation.

The section~\ref{sect:compl} is devoted to the proof of the completeness of constructed eigenfunctions for BC-type spin chain. The proof is based on the use of the 
Mellin-Barnes representation obtained in the previous section and the generalization of BC-type Gustafson integral~\cite[eq.~(2.3b)]{DM:Gust4}. 
This allows to reduce everything to the proven~\cite{M} completeness of eigenfunctions of A-type spin chain.

The main results of the paper are listed in Section~\ref{sect:Conclus} and 
Appendices contain some useful formulae.

We should note that this work is the second one from the series of two papers. 
In~\cite{ADV1} the simpler case of A-type $SL(2,\mathbb{C})$ spin chain 
is considered in details. In fact the present paper is organized in exactly the same way as \cite{ADV1} in order to see evident parallels. 
Moreover, we extensively use formulae from the first paper.

\section{Principal series representations of $SL(2,\mathbb{C})$} \label{sect:Rep}

In the definition of the model the central role is played by \textit{principal series representations} of the group $SL(2,\mathbb{C})$~\cite[Ch.~III]{GGV}.
These representations are defined on the space $\mathrm{L}^2(\mathbb{C})$ of square integrable functions on $\mathbb{C}$ with the scalar product
\begin{equation} \label{sp}
	\langle \Phi|\Psi\rangle = \int \mathrm{d}^2z \; \overline{\Phi(z,\bar{z})}\,\Psi(z, \bar{z}) \,,
\end{equation}
where $\mathrm{d}^2z$ is the Lebesque measure in $\mathbb{C}$: $\mathrm{d}^2 z = \mathrm{d}\Re z \,\mathrm{d}\Im z$. In what follows, the entry $\int \mathrm{d}^2z$ would mean the integration over the whole complex plane.

Let us introduce some notations which will be used throughout the text. The complex conjugation of a number $a \in \mathbb{C}$ would be denoted as $a^*$ so that in generic situation $\bar{a}$ does not mean complex conjugated to $a$.
We make only one traditional exception for the variable $z$, that is~\mbox{$\bar{z} \equiv z^*$}, and for the complex conjugation of functions $\overline{\Psi(z, \bar{z})}$. In order to define the representation, we introduce for a pair $(a,\bar{a}) \in \mathbb{C}$ the power function in complex plane -- the double power
\begin{align} \label{power}
	[z]^a \equiv z^a \bar{z}^{\bar{a}} = |z|^{a+\bar{a}} \, {e}^{\imath(a-\bar{a}) \arg z}.
\end{align}
From the condition of single-valuedness of this function follows the restriction $a-\bar{a} \in \mathbb{Z}$. The parameter $a$ we call ``holomorphic'' and parameter $\bar{a}$ -- ``antiholomorphic'', but the numbers $a$ and $\bar{a}$ are not complex conjugate in general. For brevity we display only the ``holomorphic'' exponent. In addition, for $\rho \in \mathbb{R}$ we denote
\begin{equation} \nonumber
	[z]^{\rho+a} \equiv z^{\rho+a} \bar{z}^{\rho+\bar{a}} .
\end{equation}

The representation is parametrized by the pair of ``spins'' $(s, \bar{s})$ such that $2(s - \bar{s}) \in \mathbb{Z}$.
The operator $T^{(s,\bar{s})}(g)$ corresponding to the group element
\begin{equation} \nonumber
	g =
	\begin{pmatrix}
		a & b \\
		c & d
	\end{pmatrix}, \qquad ad - bc = 1,
\end{equation}
acts on functions $\Psi(z\,,\bar{z})$ according to the formula
\begin{equation} \nonumber
	[T^{(s,\bar{s})}(g) \, \Psi] (z) =
	[d-bz]^{-2s} \, \Psi \biggl(\frac{-c+az}{d-bz}\biggr) \,,
\end{equation}
where $[d-bz]^{-2s}\equiv(d-bz)^{-2s}(d^\ast-b^\ast\bar{z})^{-2\bar{s}}$. Note that hereinafter for the sake of simplicity we display only the holomorphic argument $z$, so that ~\mbox{$\Psi(z) \equiv \Psi(z, \bar{z})$}.

The unitarity of principal series representation imposes the following condition on its parameters
\begin{equation} \label{scond}
	s^\ast + \bar{s} = 1 ,
\end{equation}
where $s^\ast$ denotes the complex conjugation of $s$. Thus, bearing in mind the restriction $2(s - \bar{s}) \in \mathbb{Z}$ one obtains the parametrization of spins
\begin{equation}\label{sparam}
	s = \frac{1+n_s}{2} + \imath\nu_s, \qquad
	\bar{s} = \frac{1-n_s}{2} + \imath\nu_s \,,
	\qquad
	n_s \in \mathbb{Z}+\sigma, \quad \nu_s \in \mathbb{R},
\end{equation}
where for the rest of the paper we fix the parameter $\sigma$
\begin{equation} \label{sigma}
	\sigma \in \left\{0, \tfrac{1}{2}\right\} 
\end{equation}
so that $n_s$ is integer if $\sigma=0$, and half-integer if $\sigma=\tfrac{1}{2}$.

To define the model we will need the generators of the representation
\begin{align} \label{gen}
	\begin{aligned}
		&S = z\partial_z + s\,, \qquad S_- = -\partial_z\,, \qquad
		S_+ = z^2\partial_z+2sz\,, \\[2pt]
		&
		\bar{S} = \bar{z}\partial_{\bar{z}} + \bar{s}\,, \qquad \bar{S}_- = -\partial_{\bar{z}}\,, \qquad
		\bar{S}_+ = \bar{z}^2\partial_{\bar{z}}+2\bar{s}\bar{z} \,.
	\end{aligned}
\end{align}
The generators from the set $S, S_{\pm}$, which we call holomorphic, commute with antiholomorphic generators $\bar{S}, \bar{S}_\pm$. And each of these sets satisfies the standard commutation relations of the Lie algebra $sl(2,\mathbb{C})$, that is,
\begin{align} \nonumber
	[S_+, S_-] = 2S, \qquad [S,S_\pm] = \pm S_\pm,
\end{align}
and similarly for the antiholomorphic generators. Two types of generators are hermitian conjugate to each other with respect to the scalar product~\eqref{sp}
\begin{equation} \nonumber
	S^\dagger = -\bar{S}, \qquad S_-^\dagger = -\bar{S}_-, \qquad
	S_+^\dagger = -\bar{S}_+
\end{equation}
due to relation~\eqref{scond}.

It is well known \cite{Gelfand-Naimark} that the representations characterized by the
parameters $(s, \bar s)$ and $(1-s, 1-\bar{s})$ are equivalent.
There exists an integral operator $W$ which intertwines 
equivalent principal series representations $T^{(s,\bar{s})}$
and $T^{(1-s,1-\bar{s})}$ for generic complex $s$ and~$\bar{s}$
\begin{equation} \nonumber
	W(s,\bar{s})\,T^{(s,\bar{s})}(g) =
	T^{(1-s,1-\bar{s})}(g)\,W(s,\bar{s})\,.
\end{equation}
It has the form \cite{GGV}
\begin{align} \label{W1}
	W(s,\bar{s}) := \left[\hat{p}\right]^{1-2s} \,,
\end{align}
where for a pair $(\alpha,\bar{\alpha})$, such that $\alpha-\bar{\alpha}\in\mathbb{Z}$, we define the operator $[\hat{p}]^{\alpha}$ as follows
\begin{align} \label{d} 
	\left[\hat{p}\right]^{\alpha}\Phi(z,\bar z) :=
	c(\alpha)\,
	\int \mathrm{d}^2 w\,
	\frac{\Phi(w,\bar w)}{[z-w]^{1+\alpha}} \,.
\end{align}
In \eqref{W1} we have $(\alpha,\bar{\alpha})=(1-2s,1-2\bar{s})$. Again, for brevity we display only the holomorphic parameter $\alpha$.
The normalization coefficient $c(\alpha)$ depends on $\alpha$ and $\bar{\alpha}$, it has the form
\begin{align}\label{c}
	c(\alpha) = \frac{[\imath]^{\alpha}\,\bm{\Gamma}(\alpha+1)}{\pi} \,,
\end{align}
where $[\imath]^{\alpha} = \imath^{\alpha-\bar{\alpha}}$ and $\bm{\Gamma}(\alpha)$ is the gamma-function of the complex field~\cite{GGR,N}
\begin{equation} \nonumber
	\bm{\Gamma}(\alpha):=\frac{\Gamma(\alpha)}{\Gamma(1-\bar\alpha)} \,.
\end{equation}
By means of Fourier transformation the operator \eqref{d} can be interpreted as the double power $\hat{p}^\alpha\,\hat{\bar{p}}^{\bar{\alpha}}$
of momentum operator $\hat{p}=-\imath\partial_z \,, \, \hat{\bar{p}}=-\imath\partial_{\bar{z}}$, see \cite[Section~2.1]{ADV1}, \cite[Section~2.3]{DM09}. For this reason we use the notation $[\hat{p}]^\alpha$. The motivation for normalization \eqref{c} is given in \cite[Section~2.1]{ADV1}.

Operators of the type \eqref{d} are widely used throughout the text. One can verify using the chain relation \eqref{Chain} and its limiting case \eqref{delta2} that in the taken normalization this operator obeys the relations
\begin{equation} \nonumber
	[\hat{p}]^\alpha\,[\hat{p}]^\beta=[\hat{p}]^{\alpha+\beta} \,, \qquad
	[\hat{p}]^\alpha\,[\hat{p}]^{-\alpha} = \II \,,
\end{equation}
where $\II$ is the identity operator.
From \eqref{W1} and relation $[p]^\alpha [p]^{-\alpha} = \II$ it follows that
\begin{equation} \nonumber
	W^{-1}(s,\bar{s}) = W(1-s,1-\bar{s}) \,.
\end{equation}
At the same time, using the definition of $[\hat{p}]^\alpha$ one can obtain the following conjugation rule
\begin{align} \label{conj} 
	\left(\left[\hat{p}\right]^{\alpha}\right)^{\dagger} =  
	\left[\hat{p}\right]^{\bar{\alpha}^*} \,.
\end{align}
In parametrization \eqref{sparam} one has $\bar{s}^* = 1-s$ and therefore 
\begin{equation} \nonumber
	W^{\dagger}(s,\bar{s}) = W(1-s,1-\bar{s}) = W^{-1}(s,\bar{s}) \,,
\end{equation}
so that $W(s,\bar{s})$ is unitary operator 
$W^{\dagger}(s,\bar{s})\,W(s,\bar{s}) = \II$.
In the following we will also need the conjugation rule for operator of multiplication by power function
\begin{equation} \label{zconj}
	([z]^\alpha)^\dagger=([z]^\alpha)^\ast=[z]^{\bar{\alpha}^\ast}=z^{\bar{\alpha}^\ast}\bar{z}^{\alpha^\ast} \,.
\end{equation}

\section{$SL(2,\mathbb{C})$ spin chain of BC type}

We are going to introduce the model in section~\ref{sect:Monodromy}. But first let us consider in section~\ref{sect:ReflEq} some universal objects which contain all the ingredients used to define our model, including $L$-operator, Yang's $R$-matrix and Sklyanin's $K$-matrix.

The first object is the general $\mathbb{R}$-operator acting in the tensor product of two principal series representations. It is the general solution of the Yang-Baxter equation corresponding to the group $SL(2,\mathbb{C})$. The $L$-operator and $R$-matrix can be obtained as certain reductions of this universal $\mathbb{R}$-operator. The second object is reflection operator which also acts in the space of principal series representation. Together with $\mathbb{R}$-operator it satisfies the reflection equation. Its finite-dimensional reduction is boundary $K$-matrix \cite{Ch,Skl:Boundary,KS}.

The reflection operator and a special case of $\mathbb{R}$-operator also play the role of elementary building blocks in the construction of eigenfunctions of the model described in section~\ref{sect:EigenfuncConstruct}.
The relations between $K$-matrix, $L$-operator and these building blocks, which are used in that construction, were derived in previous works with the help of representation theory of the group $SL(2,\mathbb{C})$. However, these relations are also particular cases of the general Yang-Baxter and reflection equations.
In even more degenerate case one obtains from the latter identities the $RLL$-relation between $L$-operator and $R$-matrix and reflection equation for $R$- and $K$-matrices, which yield the commutativity of model's quantum integrals of motion. In section~\ref{sect:ReflEq} we describe all special cases of general reflection and Yang-Baxter equations corresponding to the above-mentioned relations.

The general reflection equation has another important application. In section~\ref{sect:Qcomm} we show that it is equivalent to the commutativity of $Q$-operators for the spin chain consisting of $2$ sites. Thus it is used as a base for the inductive proof of this commutativity for arbitrary number of sites.

On top of everything, in section~\ref{sect:ReflEq} we complete the study which was started in section~7 of the paper~\cite{ABDV}. In that work a rather general version of reflection equation is formulated with all operators acting in infinite dimensional spaces of principal series representations. However, that result is a special case of the relation \eqref{Refl}, which is the most general form of the reflection equation in the case of symmetry group $SL(2,\mathbb{C})$.

\subsection{Reflection equation} \label{sect:ReflEq}

The general reflection equation has the following form \cite{Skl:Boundary,PP,Oliv}
\begin{align}\label{Refl}
	\mathbb{R}_{12}(u-v)\,K_1(u)\,\mathbb{R}_{12}(u+v)\,K_2(v) = 
	K_2(v)\,\mathbb{R}_{12}(u+v)\,K_1(u)\,\mathbb{R}_{12}(u-v) \,.
\end{align}
Operator $\mathbb{R}_{12}(u)$ acts in the tensor product $V_1\otimes V_2$ of two principal series representations of $SL(2,\mathbb{C})$ with spins $s_1$ and $s_2$.
It is the function of spectral parameter $u$ and the spins which determine the representations. It also depends on the antiholomorphic counterparts of these parameters which we do not show for brevity.
Reflection operator $K_1(u)$ is defined in the space $V_1$ of representation with spin $s_1$, and the similar operator $K_2(v)$ acts in the space $V_2$ corresponding to spin $s_2$.
The $\mathbb{R}$-operator is the general solution of the Yang-Baxter equation with symmetry group $SL(2,\mathbb{C})$~\cite{DM09,Derkachov:2005hw}
\begin{align} \label{YBgen}
	\mathbb{R}_{12}(u-v)\,\mathbb{R}_{13}(u)\,\mathbb{R}_{23}(v) = 
	\mathbb{R}_{23}(v)\,\mathbb{R}_{13}(u)\,\mathbb{R}_{12}(u-v) \,.
\end{align}
In \eqref{YBgen} we have the operator relation in the tensor product of three representations $V_1\otimes V_2\otimes V_3$, each operator $\mathbb{R}_{ij}$ acts nontrivially in spaces $V_i$ and $V_j$ and as the identity operator in the third space.

In the full analogy with Yang-Baxter equation there are different variants 
of the general reflection equation depending on the choice of 
representations $V_1$ and $V_2$.

In the simplest case of two-dimensional representations $V_1=V_2=\mathbb{C}^2$ the operator $\mathbb{R}_{12}(u)$ degenerates \cite[Section~2.3]{CDS} into the finite-dimensional solution of Yang-Baxter equation -- the Yang's $R$-matrix acting 
in the tensor product $\mathbb{C}^2\otimes \mathbb{C}^2$
\begin{equation} \nonumber
	R_{12}(u) = R(u) = u  + P \,, \qquad P\,a \otimes b = b \otimes a \,.
\end{equation}
The reflection relation \eqref{Refl} is reduced in this case to the form
\begin{align} \label{ReflFundam}
	R(u-v)\, \bigl(K(u)\otimes \bm{1}\bigr)\,R(u+v)\, 
	\bigl(\bm{1} \otimes K(v)\bigr) 
	= \bigl(\bm{1} \otimes K(v)\bigr)\,R(u+v)\,\bigl(K(u)\otimes \bm{1}\bigr)\, R(u-v) \,,
\end{align}
where $\bm{1}$ is a $2\times 2$ identity matrix, and its general solution is \cite{Ch, Skl:Boundary, KS} 
\begin{align} \label{K0}
	K(u) = \begin{pmatrix}
		\gamma\bigl(g - \frac{1}{2} \bigr) & u  \\
		\gamma^2\, u  & \gamma\bigl(g - \frac{1}{2} \bigr)
	\end{pmatrix} .
\end{align}

The more complicated solution of Yang-Baxter 
relation -- the $L$-operator, appears in the case when $V_1 = V_2 = \mathbb{C}^2$ and $V_3$ 
is the space of arbitrary representation  
$T^{(s,\bar{s})}$ with generators~\eqref{gen}
\begin{align*}
	S = z\partial_z + s \,, \qquad S_- = -\partial_z \,, \qquad 
	S_+ = z^2\partial_z+2sz \,.
\end{align*}
In this case $\mathbb{R}_{12}(u-v)$ is given by the Yang's $R$-matrix 
and the general $\mathbb{R}$-operators $\mathbb{R}_{13}(u)$ and $\mathbb{R}_{23}(v)$ are reduced up to the shift of spectral parameters to the $L$-operators~\cite[Section~2.3]{CDS}
\begin{align*}
	\textstyle R(u-v)\, \bigl(L(u+\frac{1}{2})\otimes \bm{1}\bigr)\, 
	\bigl(\bm{1} \otimes L(v+\frac{1}{2})\bigr)
	= \bigl(\bm{1} \otimes L(v+\frac{1}{2})\bigr)\,
	\bigl(L(u+\frac{1}{2})\otimes \bm{1}\bigr)\, R(u-v) \,,
\end{align*}
where the explicit expression for the $L$-operator reads
\begin{align} \label{L}
	L(u) =
	\left( \begin{array}{cc}
		u + S & S_{-} \\
		S_{+} & u - S \end{array} \right) .
\end{align}
Note that after the shifts $u\to u-\frac{1}{2}$ and $v\to v-\frac{1}{2}$
one obtains the previous relation in a more common form 
\begin{align*}
	\textstyle R(u-v)\, \bigl(L(u)\otimes \bm{1}\bigr)\, 
	\bigl(\bm{1} \otimes L(v)\bigr)
	= \bigl(\bm{1} \otimes L(v)\bigr)\,
	\bigl(L(u)\otimes \bm{1}\bigr)\, R(u-v) \,.
\end{align*}
As for the reflection relation, if $V_1 = \mathbb{C}^2$ and $V_2$ 
is the space of arbitrary representation $T^{(s,\bar{s})}$,
one obtains the defining relation for the general 
reflection operator $K_2(v) = K(v\,,s)$ 
\begin{align} \nonumber
	\textstyle L(u-v+\frac{1}{2})\,K(u)\,L(u+v+\frac{1}{2})\,K(v\,,s) = 
	K(v\,,s)\,L(u+v+\frac{1}{2})\,K(u)\,L(u-v+\frac{1}{2}) \,.
\end{align}
After the shift of spectral parameter $u \to u - \frac{1}{2}$ the defining relation takes a more convenient form
\begin{align} \label{ReflK}
	\textstyle L(u-v)\,K(u)\,L(u+v)\,K(v\,,s) = 
	K(v\,,s)\,L(u+v)\,K(u)\,L(u-v) \,,
\end{align}
here and in what follows we use the expression for the $K$-matrix
\begin{align*}
	K(u) = \begin{pmatrix}
		\gamma\bigl(g - \frac{1}{2} \bigr) & u - \frac{1}{2} \\
		\gamma^2 \bigl(u - \frac{1}{2} \bigr) & \gamma\bigl(g - \frac{1}{2} \bigr)
	\end{pmatrix} \,,
\end{align*}
which differs from the canonical one \eqref{K0} by the shift of argument $u\to u-\frac{1}{2}$.
The operator $K(v\,,s)$ is constructed explicitly in \cite{ABDV} 
and is given by the following formula 
\begin{align}\label{Kvs}
	K(v\,,s) =
	[z+\gamma]^{v-s+g} \,
	[z-\gamma]^{v-s+1-g} \,
	[\hat{p}]^{2v} \,
	[z+\gamma]^{v+s-g} \,
	[z-\gamma]^{v+s+g-1} \,.
\end{align}

In the case when $V_1$ and $V_2$ are spaces of two 
arbitrary representations  
$T^{(s_1,\bar{s}_1)}$ and $T^{(s_2,\bar{s}_2)}$ and $V_3 = \mathbb{C}^2$ 
the Yang-Baxter relation is reduced to the defining relation for 
the general $\mathbb{R}$-operator
\begin{align} \nonumber
	\mathbb{R}_{12}(u-v)\,L_1(u)\,L_2(v) = 
	L_2(v)\,L_1(u)\,\mathbb{R}_{12}(u-v) \,,
\end{align}
where by $L_1(u)$ and $L_2(v)$ we denote the $L$-operators corresponding to representations $T^{(s_1,\bar{s}_1)}$ and $T^{(s_2,\bar{s}_2)}$
\begin{align*}
	L_1(u) = 
	\left(\begin{array}{cc} u+s_1+z_1 \partial_1 & -\partial_1\\
		z_1^2 \partial_1 + 2s_1 z_1 & u-s_1 -z_1
		\partial_1
	\end{array}\right) , \quad
	L_2(v) = 
	\left(\begin{array}{cc} v+s_2+z_2 \partial_2 & -\partial_2\\
		z_2^2 \partial_2 + 2s_2 z_2 & v-s_2 -z_2
		\partial_1
	\end{array}\right) .
\end{align*}
The general solution of this relation was constructed in \cite[\S3]{DM09}, it has the form
\begin{align}\label{RYB}
	\mathbb{R}_{12}(u) = \mathbb{P}_{12}\,\check{R}_{12}(u) = \mathbb{P}_{12}\,
	[z_{12}]^{u+1-s_1-s_2}\,
	[\hat{p}_1]^{u+s_2-s_1}\,[\hat{p}_2]^{u+s_1-s_2}\,
	[z_{12}]^{u+s_1+s_2-1} \,,
\end{align}
where $\mathbb{P}_{12}$ is the operator of permutation 
\begin{align*}
	\mathbb{P}_{12}\,\Psi(z_1\,,z_2) = \Psi(z_2\,,z_1) \,.
\end{align*}
The proof of the Yang-Baxter equation \eqref{YBgen} for $\mathbb{R}$-operator \eqref{RYB} can also be found in~\cite[\S3]{DM09}.

It remains to prove that the reflection operator \eqref{Kvs} obeys the general reflection relation \eqref{Refl}.
Substituting into this equation the formula \eqref{RYB} for $\mathbb{R}$-operator and removing all permutation operators one reduces it to the identity
\begin{multline}\label{Refl1}
	\check{R}_{12}(u-v)\,K_1(u\,,s_1)\,\check{R}_{21}(u+v)\,K_1(v\,,s_2) \\
	= K_1(v\,,s_2)\,\check{R}_{12}(u+v)\,K_1(u\,,s_1)\,\check{R}_{21}(u-v) \,,
\end{multline}
which we rewrite in the following explicit form for clarity
\begin{multline*}
	[z_{12}]^{u-v+1-s_1-s_2}\,
	[\hat{p}_1]^{u-v+s_2-s_1}\,[\hat{p}_2]^{u-v+s_1-s_2}\,
	[z_{12}]^{u-v+s_1+s_2-1}\,\\
	K_1(u\,,s_1)\,[z_{21}]^{u+v+1-s_1-s_2}\,
	[\hat{p}_2]^{u+v+s_2-s_1}\,[\hat{p}_1]^{u+v+s_1-s_2}\,
	[z_{21}]^{u+v+s_1+s_2-1}\, K_1(v\,,s_2) = \\ 
	K_1(v\,,s_2)\,[z_{12}]^{u+v+1-s_1-s_2}\,
	[\hat{p}_1]^{u+v+s_2-s_1}\,[\hat{p}_2]^{u+v+s_1-s_2}\,
	[z_{12}]^{u+v+s_1+s_2-1}\, K_1(u\,,s_1)\,\\
	[z_{21}]^{u-v+1-s_1-s_2}\,
	[\hat{p}_2]^{u-v+s_2-s_1}\,[\hat{p}_1]^{u-v+s_1-s_2}\,
	[z_{21}]^{u-v+s_1+s_2-1} \,.
\end{multline*}

Now let us rewrite the equation \eqref{Refl1} using the parametrization of the reflection operator from the article \cite{ABDV}, in which it was constructed. This parametrization will be applied in the rest of the paper.
We have two free parameters in the $L$-operator: 
spin $s$ and spectral parameter $u$, but equivalently 
one can use another pair of parameters
\begin{align*}
	u_1 = u+s-1 \,, \quad  u_2 = u-s \,,
\end{align*}
which appear in a very natural way.
The $L$-operator \eqref{L} can be 
expressed in a factorized form~\cite{Derkachov:2005hw}
\begin{align} \label{Lfact}
	L(u_1, u_2) = 
	\left(\begin{array}{cc} u_1+1+z \partial_z & -\partial_z\\
		z^2 \partial_z + (u_1-u_2+1) z & u_2-z\partial_z
	\end{array}\right) 
	= 
	\left(%
	\begin{array}{cc}
		1 & 0 \\
		z & 1 \\
	\end{array}%
	\right)\,\left(%
	\begin{array}{cc}
		u_1 & -\partial_{z} \\
		0 & u_2 \\
	\end{array}%
	\right)\, \left(%
	\begin{array}{cc}
		1 & 0 \\
		-z & 1 \\
	\end{array}%
	\right) .
\end{align}
In \cite{ABDV} we used the following defining relation 
for the reflection operator 
\begin{multline} \label{Kdef}
	\mathcal{K}(s,x) \, L(u +x - 1, u - s) \, K(u) \, L(u  + s - 1, u - x) \\
	= L(u + s - 1, u - x) \, K(u) \, L(u + x - 1, u - s) \, 
	\mathcal{K}(s,x) \,,
\end{multline}
where the $K$-matrix has the same form as in \eqref{ReflK}
\begin{align*}
	K(u) = \begin{pmatrix}
		\gamma\bigl(g - \frac{1}{2} \bigr) & u - \frac{1}{2} \\
		\gamma^2 \bigl(u - \frac{1}{2} \bigr) & \gamma\bigl(g - \frac{1}{2} \bigr)
	\end{pmatrix} .
\end{align*}
It was shown that the solution of \eqref{Kdef} has the form 
\begin{align}\label{Kop-d}
	\mathcal{K}(s,x) =
	[z+\gamma]^{g-s} \,
	[z-\gamma]^{1-s-g} \,
	[\hat{p}]^{x-s} \,
	[z+\gamma]^{x-g} \,
	[z-\gamma]^{x+g-1}.
\end{align}
As we already mentioned, the reflection operator $\mathcal{K}(s,x)$ is an elementary building block in the construstion of eigenfunctions of the model in section~\ref{sect:EigenfuncConstruct}. The identity \eqref{Kdef} plays an important role in this construction.

The defining relation \eqref{ReflK} for the reflection operator $K(v\,,s)$ is equivalent to defining relation \eqref{Kdef} for $\mathcal{K}(s,x)$ and 
is obtained from \eqref{Kdef} after the substitution $x\to s+v \,,\, s\to s-v$.
Using the same substitution in the formula \eqref{Kop-d} for operator $\mathcal{K}(s,x)$ one obtains the expression \eqref{Kvs} for $K(v\,,s)$. 

After the corresponding reparametrization in the general reflection equation \eqref{Refl1}
\begin{align} \nonumber
	u+s_1 = x \,, \ \ u-s_1 = -s \,, \qquad v+s_2 = y \,, \ \ v-s_2 = -s^{\prime}
\end{align}
we obtain the following equivalent identity
\begin{align*}
	&
	[z_{12}]^{1-s-y}\,
	[\hat{p}_1]^{s^{\prime}-s}\,[\hat{p}_2]^{x-y}\,
	[z_{12}]^{x+s^{\prime}-1}\,
	\mathcal{K}_1(s,x)\,[z_{12}]^{1-s-s^{\prime}}\,
	[\hat{p}_1]^{x-s^{\prime}}\,[\hat{p}_2]^{y-s}\,
	[z_{12}]^{x+y-1}\, \mathcal{K}_1(s^{\prime},y) \\
	&
	= \mathcal{K}_1(s^{\prime},y)\,[z_{12}]^{1-s-s^{\prime}}
	[\hat{p}_1]^{y-s} [\hat{p}_2]^{x-s^{\prime}}
	[z_{12}]^{x+y-1} \mathcal{K}_1(s,x)\,
	[z_{12}]^{1-s-y} [\hat{p}_1]^{x-y} [\hat{p}_2]^{s^{\prime}-s}
	[z_{12}]^{x+s^{\prime}-1} .
\end{align*}
It is proven in Appendix~\ref{Reflection}.
In the paper \cite{ABDV} we have stated 
the particular case of this relation corresponding to $s^{\prime} = s$
\begin{multline}\label{Q2}
	[z_{12}]^{1-s-y}\,[\hat{p}_2]^{x-y}\,
	[z_{12}]^{x+s-1}\,
	\mathcal{K}_1(s,x)\,[z_{12}]^{1-2s}\,
	[\hat{p}_1]^{x-s}\,[\hat{p}_2]^{y-s}\,
	[z_{12}]^{x+y-1}\, \mathcal{K}_1(s,y) = \\ 
	\mathcal{K}_1(s,y)\,[z_{12}]^{1-2s}\,
	[\hat{p}_1]^{y-s}\,[\hat{p}_2]^{x-s}\,
	[z_{12}]^{x+y-1}\, \mathcal{K}_1(s,x)\,
	[z_{12}]^{1-s-y}\,
	[\hat{p}_1]^{x-y}\,
	[z_{12}]^{x+s-1} \,.
\end{multline}

\subsection{Monodromy matrix} \label{sect:Monodromy}

The Hilbert space of the model is given by the direct product of the $\mathrm{L}^2(\mathbb{C})$ spaces,
\begin{equation}\label{HN}
	H=V_1\otimes V_2\otimes\cdots\otimes V_n,  \qquad V_k=\mathrm{L}^2(\mathbb{C})\,,\qquad
	k=1,\ldots,n.
\end{equation}
To each site $k$ we associate the quantum $L$-operators \eqref{L}
with subscript $k$, acting nontrivially in the $k$-th component of the tensor product~(\ref{HN})
\begin{equation} \nonumber
	L_k(u) =
	\left (\begin{array}{cc}
		u + S^{(k)} & S_{-}^{(k)} \\
		S_{+}^{(k)} & u - S^{(k)} \end{array} \right )\ \,, \ 
	\bar L(\bar u) = \left(\begin{array}{cc}
		\bar u + \bar S^{(k)} & \bar S_{-}^{(k)}\\
		\bar S_{+}^{(k)}& \bar u - \bar S^{(k)}
	\end{array}\right)
\end{equation}
Functions in the space $V_k$ depend on variable $z_k$, and copies of generators \eqref{gen} acting in this space have the form
\begin{equation} \nonumber
	S^{(k)} = z_k\partial_k + s, \qquad  S^{(k)}_- = -\partial_k, \qquad
	S^{(k)}_+ = z_k^2\partial_k+2sz_k \,,
\end{equation}
expressions for the antiholomorphic generators are similar. That is, we consider only homogeneous chains with spin parameters $(s,\bar{s})$ being equal in all sites.

The monodromy matrix is defined as follows \cite{Skl:Boundary,KS,PP}
\begin{equation} \label{Top}
	T_n(u) = L_n(u) \cdots L_1(u) K(u) L_1(u) \cdots L_n(u) = \begin{pmatrix}
		A_n(u) & B_n(u) \\
		C_n(u) & D_n(u)
	\end{pmatrix} .
\end{equation}
The antiholomorphic counterpart $\bar{T}_n(\bar{u})$ is defined by 
means of $\bar{L}_k(\bar{u})$ and $\bar{K}(\bar{u})$ in the same way.
The $K$-matrix is a solution of the reflection equation~\eqref{ReflFundam} with Yang's $R$-matrix. 
The simplest possible choice is $K(u) = \bm{1}$ but we will consider 
the general case \cite{Ch,Skl:Boundary,KS,PP} where 
\begin{align}\label{K}
	K(u) = \begin{pmatrix}
		\gamma\bigl(g - \frac{1}{2} \bigr) & u - \frac{1}{2} \\[6pt]
		\gamma^2 \bigl(u - \frac{1}{2} \bigr) & \gamma\bigl(g - \frac{1}{2} \bigr)
	\end{pmatrix},
	\qquad
	\bar{K}(\bar{u}) = \begin{pmatrix}
		\bar{\gamma}\bigl(\bar{g} - \frac{1}{2} \bigr) & \bar{u} - \frac{1}{2} \\[6pt]
		\bar{\gamma}^2 \bigl(\bar{u} - \frac{1}{2} \bigr) & \bar{\gamma}\bigl(\bar{g} - \frac{1}{2} \bigr)
	\end{pmatrix},
\end{align}

The matrix $T_n(u)$ satisfies the reflection equation \cite{Ch,Skl:Boundary,KS,PP}
\begin{multline}\label{refl_eq}
	R(u-v)\, \bigl(T_n(u)\otimes \bm{1}\bigr)\,R(u+v-1)\, \bigl(\bm{1} \otimes T_n(v)\bigr) \\[6pt]
	= \bigl(\bm{1} \otimes T_n(v)\bigr)\,R(u+v-1)\,\bigl(T_n(u)\otimes \bm{1}\bigr)\, R(u-v) \,,
\end{multline}
where the Yang's $R$-matrix $R(u)$ acts in the tensor product $\mathbb{C}^2\otimes \mathbb{C}^2$
\begin{equation} \label{Rmat}
	R(u) = u  + P, \qquad P \, a \otimes b = b \otimes a.
\end{equation}
From \eqref{refl_eq} and its antiholomorphic analogue follows the commutativity
\begin{equation} \nonumber
	[B_n(u), B_n(v)] = 0, \qquad [\bar{B}_n(\bar{u}), \bar{B}_n(\bar{v})] = 0 .
\end{equation}
The holomorphic and antiholomorphic operators also commute with each other
\begin{equation} \nonumber
	[B_n(u), \bar{B}_n(\bar{v})] = 0 \,.
\end{equation}

Now, denote by $u$ and $\bar{u}$ the pair of complex conjugate variables
\begin{equation} \label{uuConj}
	\bar{u} = u^\ast \,.
\end{equation}
By analogy with the first work devoted to BC-type open spin chain \cite{ABDV} we choose the same parametrization for constants $g,\bar{g}$ as for spins
\begin{equation} \label{gparam}
	g = \frac{1+n_g}{2} + \imath\nu_g \,, \qquad
	\bar{g} = \frac{1-n_g}{2} + \imath\nu_g \,,
	\qquad
	n_g \in \mathbb{Z}+\sigma \,, \quad \nu_g \in \mathbb{R}
\end{equation}
where the parameter $\sigma$ was introduced in \eqref{sigma}, and assume that the constants $\gamma,\bar{\gamma}$ are complex conjugate to each other
\begin{equation} \nonumber
	\gamma^\ast = \bar{\gamma} \,.
\end{equation}
First, the pair \eqref{gparam} satisfies the same requirement as spins $g-\bar{g} \in \mathbb{Z}+\sigma$ so that the double powers in formulas like \eqref{Kop-d} are well defined.
Second, for such values of $g,\bar{g}$ and $\gamma,\bar{\gamma}$ the operators $B_n(u)$ and $\bar{B}_n(\bar{u})$ are hermitian conjugate to each other
\begin{equation} \label{Bconj}
	B_n^\dagger(u) = \bar{B}_n(\bar{u}) \,,
\end{equation}
provided that the condition \eqref{uuConj} holds. We will prove the relation \eqref{Bconj} at the end of the section.
Hence, the linear combinations $B_n(u)+\bar{B}_n(\bar{u})$ and $\imath(B_n(u)-\bar{B}_n(\bar{u}))$ form the set of commuting self-adjoint operators, and thus can be diagonalized simultaneously.
This way, $B_n(u)$ and $\bar{B}_n(\bar{u})$ have the common system of eigenfunctions, which are the main objects of our research.

By construction, the operator $B_n(u)$ is a polynomial of degree $2n+1$ in the spectral parameter $u$,
and further in this section we will show that $B_n(u)/(u-\frac{1}{2})$ is an even polynomial in $u$. Of course, the same holds for $\bar{B}_n(\bar{u})$.
Consequently, the eigenvalues of $B_n(u)$ and $\bar{B}_n(\bar{u})$ are also polynomials of degree $2n+1$ in $u$ and $\bar{u}$, which have the mentioned property.
The corresponding common eigenfunctions are parametrized by zeros of these polynomials
\begin{align} \label{PsiDef}
	\begin{aligned}
		& \textstyle
		B_{n}(u)\,\Psi_{\bm x_n}(\bm z_n) = 
		\left(u - \frac{1}{2} \right)\left(u^2-x^2_1\right)\cdots
		\left(u^2-x^2_{n}\right)\,
		\Psi_{\bm x_n}(\bm z_n) \,, \\
		& \textstyle
		\bar{B}_{n}(\bar{u})\,\Psi_{\bm x_n}(\bm z_n) = 
		\left(\bar{u} - \frac{1}{2} \right)\left(\bar{u}^2-\bar{x}^2_1\right)\cdots
		\left(\bar{u}^2-\bar{x}^2_{n}\right)\,
		\Psi_{\bm x_n}(\bm z_n) \,,
	\end{aligned}
\end{align}
where we use the following compact notations for tuples of variables
\begin{align}\label{xn}
	\bm z_n=(z_1,\bar{z}_1, \ldots, z_n, \bar{z}_n) \ \,, \  
	\bm x_n=(x_1,\bar{x}_1, \ldots, x_n, \bar{x}_n)  \,.
\end{align}
On parameters $x_k,\bar{x}_k$ we impose the similar restriction $x_k-\bar{x}_k\in\mathbb{Z}+\sigma$ as on $s,\bar{s}$ and $g,\bar{g}$. In addition, due to \eqref{Bconj} one has $(x_k^\ast)^2=\bar{x}_k^2$, and we choose $x_k^\ast=-x_k$.
Thus we obtain the following parametrization for spectral variables
\begin{equation} \label{x}
	x_k = \tfrac{n_k}{2}+\imath\nu_k \,, \quad
	\bar{x}_k = -\tfrac{n_k}{2}+\imath\nu_k \,,
\end{equation}
where $n_k\in\mathbb{Z}+\sigma$ and $\nu_k$ is real.

It is worth making some remarks:
\begin{itemize}
	\item Since $\Psi_{\bm x_n}(\bm z_n)$ form the set of eigenfunctions of (at least formally) self-adjoint operators $B_n(u)+\bar{B}_n(\bar{u})$ and $\imath(B_n(u)-\bar{B}_n(\bar{u}))$, we expect this set to be orthogonal if the spectrum is simple. This property actually takes place, and in Section~\ref{ort} we derive the corresponding orthogonality relation.
	Moreover, in Section~\ref{sect:compl} it is proven that the eigenfunctions determined by parameters of the form \eqref{x} form a complete set in the Hilbert space \eqref{HN}.
	
	\item As a function of spectral variables, the eigenvalue of $B$-operator is invariant under permutations of $x_1,\ldots,x_n$. It is also invariant under reflections $x_i\to -x_i$.
	Thus the complete group of eigenvalue's symmetries is $\mathbb{Z}_2\ltimes\mathfrak{S}_n$ -- the Weyl group of B and C root systems.
	Owing to the simplicity of the spectrum, the eigenfunctions are also symmetric with respect to such transformations of parameters $\bm{x}_n$.
	This property is proven in Section~\ref{sect:Symmetry}.
	
	\item By diagonalizing the operators $B(u)$ and $\bar{B}(\bar{u})$, which are polynomials in spectral parameters $u$ and $\bar{u}$, we diagonalize simultaneously the operator-valued coefficients of these polynomials. That is, $\Psi_{\bm x_n}(\bm z_n)$ are common eigenfunctions for the following set of operators
	\begin{equation} \label{Bcoeff}
		\frac{1}{(2j)!}\left.\frac{\partial^{2j}}{\partial^{2j}u}\right|_{u=0}\frac{B_n(u)}{u-\frac{1}{2}} \,, \quad
		\frac{1}{(2j)!}\left.\frac{\partial^{2j}}{\partial^{2j}\bar{u}}\right|_{\bar{u}=0}\frac{\bar{B}_n(\bar{u})}{\bar{u}-\frac{1}{2}} \,, \qquad
		j=0,\dots,n \,.
	\end{equation}
	Therefore, the eigenvalue equations \eqref{PsiDef} hold for arbitrary $u$ and $\bar{u}$, not necessary complex conjugate to each other.
	
	\item By construction, the operators in \eqref{Bcoeff} are differential operators of order $2j$ in the variables $z_1,\dots,z_n$ and $\bar{z}_1,\dots,\bar{z}_n$. The leading coefficients ($j=0$) are identities. Besides, from \eqref{Bconj} it follows that the holomorphic and antiholomorphic operators in \eqref{Bcoeff} are hermitian conjugate to each other. Thus the eigenfunctions $\Psi_{\bm x_n}(\bm z_n)$ diagonalize $n$ self-adjoint differential operators
	\begin{equation} \nonumber
		\frac{1}{(2j)!}\left(
		\left.\frac{\partial^{2j}}{\partial^{2j}u}\right|_{u=0}\frac{B_n(u)}{u-\frac{1}{2}} +
		\left.\frac{\partial^{2j}}{\partial^{2j}\bar{u}}\right|_{\bar{u}=0}\frac{\bar{B}_n(\bar{u})}{\bar{u}-\frac{1}{2}}\right) , \qquad
		j=1,\dots,n \,.
	\end{equation}
\end{itemize}

Now we are going to show that $B_n(u)/(u-\frac{1}{2})$ is an even polynomial function of $u$
\begin{align}\label{-u}
	\frac{B_n(u)}{u-\frac{1}{2}} = \frac{B_n(-u)}{-u-\frac{1}{2}} \,.
\end{align}
The relation \eqref{-u} implies that $B_n(u)/(u-\frac{1}{2})$ is a polynomial in $u$, because it yields $B_n(\frac{1}{2})=0$, and thus $B_n(u)$ must be proportional to $(u-\frac{1}{2})$.
To derive the identity \eqref{-u} we express the monodromy matrix
\begin{equation} \nonumber
	T_n(u) = L_n(u) \cdots L_1(u) K(u) L_1(u) \cdots L_n(u)
\end{equation}
in terms of the monodromy matrix $t_n(u)$ of A-type open spin chain
\begin{equation} \label{t}
	t_n(u) = L_1(u) \cdots L_n(u) = \begin{pmatrix}
		a_n(u) & b_n(u) \\
		c_n(u) & d_n(u)
	\end{pmatrix} \,.
\end{equation}
We have
\begin{align}
	\nonumber
	T_n(u) & = (-1)^n\,\sigma_2\,t^{\prime}_n(-u)\,\sigma_2\,K(u)\,t_n(u)
	\\ 
	\nonumber
	& =
	(-1)^n \begin{pmatrix}
		d_n(-u) & -b_n(-u) \\
		-c_n(-u) & a_n(-u)
	\end{pmatrix} \begin{pmatrix}
		\gamma\bigl(g - \frac{1}{2} \bigr) & u - \frac{1}{2} \\[6pt]
		\gamma^2 \bigl(u - \frac{1}{2} \bigr) & \gamma\bigl(g - \frac{1}{2} \bigr)
	\end{pmatrix} \begin{pmatrix}
		a_n(u) & b_n(u) \\
		c_n(u) & d_n(u)
	\end{pmatrix} \,,
\end{align}
where the prime denotes the matrix transposition $(t^{\prime}_n)_{ij}=(t_n)_{ji}$, and $\sigma_2$ is the Pauli matrix.
This expression is a simple consequence of the relation for the $L$-operator
\begin{align} \nonumber
	L_k(u) = -\sigma_2\,L^{\prime}_k(-u)\,\sigma_2 \,.
\end{align}
We use the explicit formula for the operator $B_n(u)$ in terms of elements of monodromy matrix $t_n(u)$
\begin{multline} \label{B}
	\frac{B_n(u)}{u-\frac{1}{2}}
	= (-1)^n\,\gamma{\textstyle\left(g - \frac{1}{2} \right)}\,
	\frac{d_n(-u)\,b_n(u) - b_n(-u)\,d_n(u)}{u-\frac{1}{2}} \\
	+ (-1)^n\,d_n(-u)\,d_n(u) + (-1)^{n+1}\gamma^2\,b_n(-u)\,b_n(u).
\end{multline}
and the Yang-Baxter equation for $t_n(u)$ \cite{Fad:BetheAns,TaFa,Skl:VarSep,Skl91,KulSk}
\begin{align} \nonumber
	R(u-v)\, \bigl(t_n(u)\otimes \bm{1}\bigr)\,\bigl(\bm{1} \otimes t_n(v)\bigr) = 
	\bigl(\bm{1} \otimes t_n(v)\bigr)\,\bigl(t_n(u)\otimes \bm{1}\bigr)\, R(u-v) \,,
\end{align}
where $R(u-v)$ is the Yang's $R$-matrix \eqref{Rmat}.
In the explicit form the relations of Yang-Baxter algebra read
\begin{align}\label{tt}
	(u-v)\,t(u)_{ij}\,t(v)_{kl} + t(u)_{kj}\,t(v)_{il} = 
	(u-v)\,t(v)_{kl}\,t(u)_{ij} + t(v)_{kj}\,t(u)_{il} \,.
\end{align}
First of all, it can be seen from \eqref{tt} that operators $b(u)$ and $d(u)$ form commutative families $\left[b(u)\,,b(v)\right] = \left[d(u)\,,d(v)\right] = 0$, so that the 
last two terms in \eqref{B} are invariant under the substitution 
$u\to -u$.  
Next, using \eqref{tt} one can derive the following relation
\begin{align} \nonumber
	(u-v-1)\,\left[\sigma_2\,t^{\prime}(u)\,\sigma_2\,t(v)\right]_{ij} = 
	(u-v)\,\left[t^{\prime}(v)\,\sigma_2\,t(u)\,\sigma_2\right]_{ji} - \left[\sigma_2\,t^{\prime}(v)\,\sigma_2\,t(u)\right]_{ij} \,.
\end{align}
In the case $v=-u$ and $i=1\,,j=2$ the last formula reduces to
\begin{align} \nonumber
	(2u-1)\,\bigl[d_n(u)\,b_n(-u) - b_n(u)\,d_n(-u)\bigr] = 
	-(2u+1)\,\bigl[d_n(-u)\,b_n(u) - b_n(-u)\,d_n(u)\bigr] \,.
\end{align}
This relation shows the invariance of the remaining part 
of expression \eqref{B} under reflection 
$u\to -u$. 

It remains to derive the hermitian conjugation formula \eqref{Bconj} for $B$-operators
\begin{equation} \label{Bconj1}
	B_n^\dagger(u) = \bar{B}_n(\bar{u}) \,,
\end{equation}
where we recall that $\bar{u} \equiv u^\ast$.
We use the explicit expression \eqref{B} for $B_n(u)$ once more
\begin{multline} \label{B1}
	B_n(u)
	= (-1)^n\,\gamma{\textstyle\left(g - \frac{1}{2} \right)}\left(d_n(-u)\,b_n(u) - b_n(-u)\,d_n(u)\right) \\
	+ (-1)^n{\textstyle\left(u - \frac{1}{2} \right)}\left(d_n(-u)\,d_n(u) -\gamma^2\,b_n(-u)\,b_n(u)\right).
\end{multline}
The formula for $\bar{B}_n(\bar{u})$ is similar, one just needs to write down the parameters $\bar{g}, \bar{\gamma}, \bar{u}$ instead of $g, \gamma, u$, and replace $b_n, d_n$ by the same elements $\bar{b}_n, \bar{d}_n$ 
of the antiholomorphic monodromy matrix $\bar{t}_n(\bar{u})$, which is defined in terms of $L$-operators $\bar{L}_k(\bar{u})$ analogously to~\eqref{t}.

According to \eqref{B1}, for proving \eqref{Bconj1} one needs to know the conjugation rules for operators $b_n(u)$ and $d_n(u)$.
Recall that the holomorphic and antiholomorphic generators of principal series representations are related via conjugation
\begin{equation} \nonumber
	S^\dagger = -\bar{S}, \qquad S_-^\dagger = -\bar{S}_-, \qquad
	S_+^\dagger = -\bar{S}_+ \,,
\end{equation}
so we obtain the following property for matrix elements of every $L$-operator
\begin{equation} \nonumber
	(L_k)_{ij}^\dagger(u) = -(\bar{L}_k)_{ij}(-\bar{u}) \,.
\end{equation}
By virtue of the last formula and commutativity of matrix entries of $L$-operators corresponding to different $L^2(\mathbb{C})$ spaces one gets the conjugation rule for the needed elements of monodromy matrix $t_n(u)$
\begin{equation} \label{bdDagger}
	b_n^\dagger(u) = (-1)^n\,\bar{b}_n(-\bar{u}) \,, \qquad
	d_n^\dagger(u) = (-1)^n\,\bar{d}_n(-\bar{u}) \,.
\end{equation}
Using \eqref{bdDagger} and the relations $\gamma^\ast=\bar{\gamma}$, $u^\ast=\bar{u}$, $g^\ast=1-\bar{g}$, where the last one follows from the parametrization of $g$ and $\bar{g}$ given by \eqref{gparam}, one can easily show that the hermitian conjugate of the expression \eqref{B1} for $B_n(u)$ is equal to $\bar{B}_n(\bar{u})$, just as expected.

\section{Iterative construction of eigenfunctions of $B$-operator} \label{sect:EigenfuncConstruct}

In this section we present the construction 
of eigenfunctions \eqref{PsiDef} of the operator $B_n(u)$
\begin{align}\label{BPsi}
	B_{n}(u)\,\Psi_{\bm x_n}(\bm z_n) = 
	\left(u - \frac{1}{2} \right)\left(u^2-x^2_n\right)\cdots
	\left(u^2-x^2_{1}\right)\,
	\Psi_{\bm x_n}(\bm z_n) \,.
\end{align}
We obtain iterative expression for the eigenfunction   
\begin{equation*}
	\Psi_{\bm x_n}(\bm z_n) = \Psi_{x_1,  \ldots,  \, x_n}(z_1, \dots, z_n) = 
	\Lambda_n(x_n) \, \Lambda_{n-1}(x_{n-1}) \cdots \Lambda_{1}(x_1) \cdot 1 \,,
\end{equation*} 
where the raising operator $\Lambda_k(x)$ for $k=2,3\ldots$ has the following 
expression in terms of $\mathcal{R}$-operators
\begin{align}\label{L_k}
\Lambda_k(x)=  \mathcal{R}_{k\,k-1}(x)\,
\mathcal{R}_{k-1\, k-2}(x)\ldots \mathcal{R}_{2 1}(x)\,
\mathcal{K}_1(s,x)\,\mathcal{R}_{1 2}(x) \, \mathcal{R}_{2 3}(x) 
\ldots \mathcal{R}_{k-1\, k}(x)
\end{align} 
and for $k=1$ coincides with reflection operator 
$\Lambda_1(x) = \mathcal{K}_1(s,x)$.
The product of $\Lambda$-operators is applied to the function identically equal to $1$.

As in the case of A-type spin chain \cite{ADV1} the main building block is 
the local $\mathcal{R}$-operator.
It is the integral operator acting on functions $\Psi(z_k, z_j)$
\begin{align} \label{R}
	\left[\mathcal{R}_{k j}(x,\bar{x}) \, \Psi\right](z_k, z_j) = 
	c(x-s)\,
	\int \mathrm{d}^2 w \,
	\frac{[z_k-z_j]^{1-2s}}{[z_k-w]^{1-s+x} \, [w-z_j]^{1-s-x}} \,
	\Psi(w,z_j) \,,
\end{align}
where $x = \frac{n}{2}+\imath\nu\,,\bar{x} = -\frac{n}{2}+\imath\nu$ and 
$c(x-s)$ is given by \eqref{c} for $\alpha=x-s$.
In the simpler form \eqref{d} this operator 
can be represented in a two equivalent ways ($z_{kj} = z_k-z_j$)
\begin{align}\label{R1}
	\mathcal{R}_{k j}(x,\bar{x}) = 
	[z_{kj}]^{1-2s}\,[\hat{p}_k]^{x-s}\,[z_{kj}]^{s+x-1} = 
	[\hat{p}_k]^{x+s-1}\,[z_{kj}]^{x-s}
	\,[\hat{p}_k]^{1-2s} \,.
\end{align}
The equivalence of these two representations 
is essentially the statement of the star-triangle 
relation \eqref{star-tr}.
Note that initial variables $z_k,\bar{z}_k$ and spectral variables 
$x_k,\bar{x}_k$ enter almost everywhere in pairs. 
For the sake of simplicity we will omit dependence on $\bar{z}_k$ 
and $\bar{x}_k$ in explicit notations. For example we will 
use simpler notation $\mathcal{R}_{k j}(x)$ instead of 
$\mathcal{R}_{k j}(x,\bar{x})$.

The $\mathcal{R}$-operator is defined as solution of the following key relation \cite{DM09}
\begin{align}\label{defR}
	\mathcal{R}_{1 2}(x) \, L_1(u_1, u_2) \, L_2(u_1, u-x) = 
	L_1(u_1, u-x) \, L_2(u_1, u_2) \, \mathcal{R}_{1 2}(x) \,,
\end{align}
so that it interchanges the arguments in the product of 
two holomorphic $L$-operators in a very specific way. 
The same relation holds for the antiholomorphic $L$-operators, 
but we skip it for simplicity. 
Operator $\mathcal{R}_{i j}(x)$ is related to the general $SL(2,\mathbb{C})$-invariant solution~\eqref{RYB} of the Yang-Baxter equation \cite{DM09,Derkachov:2005hw}.

The defining relation \eqref{defR}
can be rewritten in an equivalent form
\begin{align}\label{equiv}
	\mathcal{R}_{1 2}(x) \, L_2(u+x-1, u_2) \, 
	L_1(u_1, u_2) =  
	L_2(u_1, u_2) \, L_1(u+x-1, u_2) \, 
	\mathcal{R}_{1 2}(x) \,.
\end{align}
The relation \eqref{equiv} can be derived from the relation \eqref{defR} as follows.
The first step is overall inversion
\begin{align*}
	L^{-1}_2(u_1,u-x)\, L^{-1}_1(u_1,u_2)\,
	\mathcal{R}^{-1}_{1 2}(x) =
	\mathcal{R}^{-1}_{1 2}(x)\,  
	L^{-1}_2(u_1,u_2)\, L^{-1}_1(u_1,u-x) \,,
\end{align*}
then using formula for the inverse $L$-operator
\begin{align*}
	L^{-1}(u_1, u_2) = -(u_1 u_2)^{-1}\,L(-u_2,-u_1)
\end{align*}
and returning $R$-operator back on its place we obtain
\begin{align*}
	\mathcal{R}_{1 2}(x) \, L_2(x-u,-u_1) \, 
	L_1(-u_2,-u_1) = 
	L_2(-u_2,-u_1) \, L_1(x-u,-u_1) \, 
	\mathcal{R}_{1 2}(x)
\end{align*}
The obtained relation differs from \eqref{equiv} by the 
change of the spectral parameter $u \to 1-u$ because under this change 
$u_2 = u-s \to  -u_1 = 1-u-s$ and $u_1 = u-1+s \to -u_2 = -u+s$. 
The $R$-operator does not depend on the spectral parameter 
so that it is possible to perform this change and finally 
obtain \eqref{equiv}.

Now we are going to derive the commutation relation 
of the raising operator \eqref{L_k} with the product of $L$-operators 
which is very similar to the monodromy matrix \eqref{Top}. 
Note that raising operator is constructed from three different blocks 
and in fact its construction mimics the construction of the monodromy matrix.
The first block is the product of $R$-operators $\mathcal{R}_{12}(x)\cdots \mathcal{R}_{k-1\,k}(x)$, 
the middle block is reflection operator $\mathcal{K}_1(s,x)$ and 
the last block is the product of $R$-operators in opposite order 
$\mathcal{R}_{k\,k-1}(x)\,\ldots \mathcal{R}_{2 1}(x)$.
We will derive commutation relation with the product of $L$-operators  
step by step according to this decomposition of the raising operator.

The operator
$\mathcal{R}_{12}(x)\mathcal{R}_{23}(x)\cdots \mathcal{R}_{n-1\,n}(x)$
transfers the parameter $u-x$ from the right hand side of the product of 
$L$-operators to the left hand side (here we need the relation \eqref{defR})
\begin{eqnarray}\nonumber
	\mathcal{R}_{12}(x)\mathcal{R}_{23}(x)\cdots \mathcal{R}_{n-1\,n}(x)\,
	L_{1}\left(u_{1},u_{2}\right)\cdots
	L_{n-1}\left(u_{1},u_{2}\right)
	\,L_{n}\left(u_{1},u-x\right) = \\
	\nonumber
	= L_{1}\left(u_{1},u-x\right)
	L_{2}\left(u_{1},u_{2}\right)\cdots
	L_{n}\left(u_{1},u_{2}\right)\,
	\mathcal{R}_{12}(x)\mathcal{R}_{23}(x)\cdots \mathcal{R}_{n-1\,n}(x)
\end{eqnarray}
and transfers the parameter $u+x-1$ from the left hand side of the 
opposite product of $L$-operators to the right hand side 
(here we need the relation \eqref{equiv})
\begin{eqnarray}\nonumber
	\mathcal{R}_{12}(x)\mathcal{R}_{23}(x)\cdots \mathcal{R}_{n-1\,n}(x)\,
	L_{n}\left(u+x-1,u_{2}\right)
	L_{n-1}\left(u_{1},u_{2}\right)\cdots
	\,L_{1}\left(u_{1},u_2\right) = \\
	\nonumber
	= L_{n}\left(u_{1},u_2\right)
	L_{n-1}\left(u_{1},u_{2}\right)\cdots
	L_{1}\left(u+x-1,u_{2}\right)\,
	\mathcal{R}_{12}(x)\mathcal{R}_{23}(x)\cdots \mathcal{R}_{n-1\,n}(x)
\end{eqnarray}
If we consider the whole monodromy matrix we observe that this operator transfers the parameter $u+x-1$ from the left hand side to the center and at the same time 
transfers the parameter $u-x$ from the right hand side to the center
\begin{multline*}
	\mathcal{R}_{12}(x)\cdots\mathcal{R}_{n-1\,n}(x)\,
	L_{n}\left(u+x-1,u_{2}\right)\cdots
	\,L_{1}\left(u_{1},u_2\right)\,K(u)\,
	L_{1}\left(u_{1},u_{2}\right)\cdots L_{n}\left(u_{1},u-x\right) = \\
	L_{n}\left(u_{1},u_2\right)\cdots
	L_{1}\left(u+x-1,u_{2}\right)\,K(u)\,
	L_{1}\left(u_{1},u-x\right)\cdots
	L_{n}\left(u_{1},u_{2}\right)\,
	\mathcal{R}_{12}(x)\cdots\mathcal{R}_{n-1\,n}(x)
\end{multline*}
Next step the reflection operator performs the needed 
interchange in the centrum
\begin{multline*}
	\mathcal{K}_1(s,x)\,L_1(u + x - 1,u_2)\,K(u)\,L_1(u_1,u - x) = \\ 
	= L_1(u_1,u - x) \, K(u) \, L_1(u + x - 1, u_2) \, 
	\mathcal{K}_1(s,x)
\end{multline*}
and then the second operator 
$\mathcal{R}_{n\,n-1}(x)\mathcal{R}_{n-1\,n-2}(x)\cdots 
\mathcal{R}_{21}(x)$ transfers the parameter $u+x-1$ from
the centrum to the right hand side and at the same time 
transfers the parameter $u-x$ from the centrum to the left hand side  
\begin{multline*}
	\mathcal{R}_{n\,n-1}(x)\cdots 
	\mathcal{R}_{21}(x)\,
	L_{n}\left(u_1,u_{2}\right)\cdots
	\,L_{1}\left(u_{1},u-x\right)\,K(u)\,
	L_{1}\left(u + x - 1,u_{2}\right)\cdots 
	L_{n}\left(u_{1},u_2\right) = \\
	= L_{n}\left(u_1,u - x\right)\cdots
	L_{1}\left(u_1,u_{2}\right)\,K(u)\,
	L_{1}\left(u_{1},u_2\right)\cdots
	L_{n}\left(u+x-1,u_{2}\right)\,
	\mathcal{R}_{n\,n-1}(x)\cdots \mathcal{R}_{21}(x)
\end{multline*}
Finally we obtain the commutation relation for 
the raising operator  
$$
\Lambda_n(x) = \mathcal{R}_{n\,n-1}(x)\mathcal{R}_{n-1\,n-2}(x)\cdots 
\mathcal{R}_{21}(x)
\mathcal{K}_1(s,x)\,
\mathcal{R}_{12}(x)\mathcal{R}_{23}(\bm x)\cdots \mathcal{R}_{n-1\,n}(x)
$$
and the needed product of the $L$-operators 
\begin{multline*}
	\Lambda_n(x)\,
	L_{n}\left(u+x-1,u_{2}\right)\cdots
	\,L_{1}\left(u_{1},u_2\right)\,K(u)\,
	L_{1}\left(u_{1},u_{2}\right)\cdots L_{n}\left(u_{1},u-x\right) = \\
	= L_{n}\left(u_1,u - x\right)\cdots
	L_{1}\left(u_1,u_{2}\right)\,K(u)\,
	L_{1}\left(u_{1},u_2\right)\cdots
	L_{n}\left(u+x-1,u_{2}\right)\,
	\Lambda_n(x)
\end{multline*}
In terms of the monodromy matrix 
\begin{align*}
	T_n(u) = L_n(u) \cdots L_1(u)\, K(u)\, L_1(u) \cdots L_n(u) =
	\begin{pmatrix}
		A_n(u) & B_n(u) \\
		C_n(u) & D_n(u)
	\end{pmatrix}
\end{align*}
we have 
\begin{multline*}
	\Lambda_n (x)\,
	L_{n}\left(u+x-1,u_{2}\right)\,T_{n-1}(u)\,
	L_{n}\left(u_{1},u-x\right) = \\
	= L_{n}\left(u_1,u-x\right)\,T_{n-1}(u)\, 
	L_{n}\left(u+x-1,u_{2}\right)\,
	\Lambda_n(x)
\end{multline*}
The matrix element $(1,2)$ of previous matrix relation 
results in the following relation
\begin{align*}
	\Lambda_n(x)\,\left(\begin{array}{cc} u+x+z_n \partial_n & -\partial_n
	\end{array}\right)\,T_{n-1}(u)\,
	\left(\begin{array}{c} -\partial_n\\
		u-x-z_n\partial_n
	\end{array}\right) = B_n(u)\,
	\Lambda_n(x) 
\end{align*}
Next step is the action on the function
$\Psi(z_1\ldots z_{n-1})$ which does not depend on $z_n$
\begin{align*}
	\Lambda_n(x)\,\left(\begin{array}{cc} u+x & 0
	\end{array}\right)T_{n-1}(u)
	\left(\begin{array}{c} 0\\
		u-x
	\end{array}\right)\Psi(z_1\ldots z_{n-1}) = B_n(u)
	\Lambda_n(x)\Psi(z_1\ldots z_{n-1})
\end{align*}
and one obtains 
$$
B_{n}(u)\,\Lambda_{n}\left(x\right)\,\Psi(z_1\ldots z_{n-1}) =
\left(u^2-x^2\right)\,\Lambda_{n}\left(x\right)\,B_{n-1}(u)\,
\Psi(z_1\ldots z_{n-1})
$$
This iteration can be continued up to the last step
\begin{eqnarray}
	\nonumber
	B_{n}(u)\,
	\Lambda_{n}\left(x_n\right)\cdots
	\Lambda_{k}(x_{k})\cdots
	\Lambda_{2}(x_{2})\,\Psi(z_1) =\\
	\nonumber
	= \left(u^2-x^2_n\right)\cdots\left(u^2-x^2_{2}\right)\,
	\Lambda_{n}\left(x_n\right)\cdots
	\Lambda_{k}(x_{k})\cdots
	\Lambda_{2}(x_{2})\,B_{1}(u)\,\Psi(z_1)
\end{eqnarray}
where 
\begin{align*}
	&\Lambda_k(x)=  \mathcal{R}_{k\,k-1}(x)\,
	\mathcal{R}_{k-1\, k-2}(x)
	\ldots \mathcal{R}_{2 1}(x)\,
	\mathcal{K}_1(s,x)\,
	\mathcal{R}_{1 2}(x) \, \mathcal{R}_{2 3}(x) \ldots 
	\mathcal{R}_{k-1\, k}(x)\,, \\ 
	&\Lambda_1(x) = \mathcal{K}_1(s,x)
\end{align*} 
The eigenfunction of the last operator
$B_{1}(u)$ is $\Lambda_1(x)\cdot 1$
\begin{align*}
	B_{1}(u) \Lambda_1(x)\cdot 1 = 
	\left(u - \frac{1}{2} \right)\left(u^2 - x^2\right)\,
	\Lambda_1(x)\cdot 1
\end{align*}
To avoid misunderstanding we should to note that 
$\Lambda_1(x)\cdot 1$ means the result of application of the operator 
$\Lambda_1(x)$ to the constant function $\Psi(z_1) = 1$.

Finally we obtain iterative expression for the eigenfunction   
\begin{equation*}
	\Psi_{\bm x_n}(\bm z_n) = \Psi_{x_1,  \ldots,  \, x_n}(z_1, \dots, z_n) = 
	\Lambda_n(x_n) \, \Lambda_{n-1}(x_{n-1}) \cdots \Lambda_{1}(x_1) \cdot 1
\end{equation*} 
of the operator $B_n(u)$
\begin{align*}
	B_{n}(u)\,\Psi_{\bm x_n}(\bm z_n) = 
	\left(u - \frac{1}{2} \right)\left(u^2-x^2_n\right)\cdots
	\left(u^2-x^2_{1}\right)\,
	\Psi_{\bm x_n}(\bm z_n)
\end{align*}
We have constructed the reach set of eigenfunctions. 
It remains to prove that it is orthogonal and complete set.

\section{$Q$-operator}
\label{qop}

In this section we construct the $Q$-operator and 
prove all its characteristic properties.
The construction of the $Q$-operator is parallel to the similar 
construction in the case of A-type spin chain \cite{ADV1}.
The composite operator
\begin{align}\label{Qop}
	Q_n(x) = \Lambda_n(x)\,[\hat{p}_n]^{x-s} = 
	\mathcal{R}_{n\,n-1}(x)\cdots 
	\mathcal{R}_{21}(x)
	\mathcal{K}_1(s,x)\,
	\mathcal{R}_{12}(x)
	\cdots \mathcal{R}_{n-1\,n}(x)\,[\hat{p}_n]^{x-s}
\end{align}
has all characteristic properties of the Baxter $Q$-operator:
\begin{itemize}
	\item
	it commutes with operators $B_n(u)$ and 
	$\bar{B}_n(\bar{u})$
	\begin{eqnarray}\label{comQB}
		Q_n(x)\, B_n(u) = B_n(u) \, Q_n(x) \,, \\  
		Q_n(x)\, \bar{B}_n(\bar{u}) = 
		\bar{B}_n(\bar{u}) \, Q_n(x) \,,
	\end{eqnarray}
	\item these operators commute at different 
	values of the spectral parameters
	\begin{eqnarray}\label{comQQ}
		Q_n(x)\,Q_n(y) = Q_n(y)\,Q_n(x) \,,
	\end{eqnarray}
	\item it obeys the Baxter relations
	\begin{eqnarray}\label{BaxB}
		&B_n(u) \, Q_n(u,\bar{u}) = \imath^{2(n-1)}\,{\textstyle \left(u-\frac{1}{2}\right)}\,
		Q_n(u+1\,,\bar{u}) \,,
		\\
		\nonumber
		&\bar{B}_n(\bar{u})\,Q_n(u,\bar{u}) = 
		\imath^{2(n-1)}\,\,{\textstyle \left(\bar{u}-\frac{1}{2}\right)}
		Q_n(u\,,\bar{u}+1) \,.
	\end{eqnarray}
\end{itemize}
In fact all these relations are consequence of the defining 
relation~(\ref{defR}) for the $R$-operator. 

To prove the first relation \eqref{comQB} we start from the formula  
\begin{align*}
	\Lambda_n(x)\,\left(\begin{array}{cc} u+x+z_n \partial_n & -\partial_n
	\end{array}\right)\,T_{n-1}(u)\,
	\left(\begin{array}{c} -\partial_n\\
		u-x-z_n\partial_n
	\end{array}\right) = B_n(u)\,
	\Lambda_n(x)
\end{align*}
obtained in the previous section in the process of deriving the 
recurrence relations for the eigenfunctions of $B$-operator. 
Next we use the similarity transformation 
\begin{align*}
	[\hat{p}]^{-\alpha}\,z \partial\,[\hat{p}]^{\alpha} = 
	z \partial -\alpha
\end{align*}
to rewrite the operator $x+z_n \partial_n$ in the form
\begin{align}\label{sim}
	[\hat{p}_n]^{s-x}\,\left(x+z_n \partial_n\right)\,[\hat{p}_n]^{x-s} &= 
	s+z_n \partial_n \,.
\end{align} 
Finally we obtain the following relation 
\begin{align*}
	\Lambda_n(x)\,[\hat{p}_n]^{x-s}\,
	\left(\begin{array}{cc} u+s+z_n \partial_n & -\partial_n
	\end{array}\right)\,T_{n-1}(u)\,
	\left(\begin{array}{c} -\partial_n\\
		u-s-z_n\partial_n
	\end{array}\right) =  
	B_n(u)\,
	\Lambda_n(x)\,[\hat{p}_n]^{x-s} \,.
\end{align*}
The matrix element in the left hand side is $B_n(u)$ and this relation 
is exactly the needed commutation relation 
\begin{align*}
	Q_n(x)\, B_n(u) = B_n(u) \, Q_n(x) \,.
\end{align*}

Further in the text we will also use the formula for hermitian conjugation of $Q$-operator with respect to the scalar product in the Hilbert space \eqref{HN} of the model
\begin{equation*}
	\langle\Phi|\Psi\rangle = 
	\int \mathrm{d}^2\bm{z}_n \, 
	\overline{\Phi(\bm{z}_n)} \, \Psi(\bm{z}_n) \,,
\end{equation*}
where $\mathrm{d}^2\bm{z}_n = \prod_{k=1}^n \mathrm{d}^2 z_k$.
First, it is easy to obtain the conjugation rules for $\mathcal{R}$- and $\mathcal{K}$-operators. One needs to use the explicit expressions \eqref{R1} and \eqref{Kop-d} for these operators and apply the conjugation rules \eqref{conj}, \eqref{zconj} for $[\hat{p}]^\alpha$ and $[z]^{\alpha}$ together with the properties $\bar{s}^\ast=1-s, \, \bar{g}^\ast=1-g$
\begin{align}
	\label{Rconj}
	& \mathcal{R}_{k j}^\dagger(x) = 
	[z_{kj}]^{\bar{x}^\ast-s}\,[\hat{p}_k]^{\bar{x}^\ast+s-1}\,[z_{kj}]^{2s-1} =
	 \mathcal{R}_{k j}^{-1}(1-\bar{x}^\ast) \,, \\
	 \label{Kconj}
	& \mathcal{K}_1^\dagger(s,x) =
	[z_1+\gamma]^{\bar{x}^\ast+g-1}\,[z_1-\gamma]^{\bar{x}^\ast-g}\,[\hat{p}_1]^{\bar{x}^\ast+s-1} \,[z_1+\gamma]^{s-g} \, [z_1-\gamma]^{s+g-1}
	=\mathcal{K}_1^{-1}(s,1-\bar{x}^\ast).
\end{align}
Using these relations one obtains from the definition \eqref{Qop} of $Q$-operator the following rule
\begin{equation} \label{Qconj}
	Q_n^\dagger(x) = Q(1-\bar{x}^\ast) \,.
\end{equation}

\subsection{Commutativity} \label{sect:Qcomm}

In this section we present the iterative proof of the commutation 
relation for $Q$-operators 
\begin{align*}
	Q_n( x)\,Q_n( y) = Q_n(y)\,Q_n(x)\,.
\end{align*}
In some sense it is the central commutation relation because all needed commutation relations between raising operators itself and 
raising operators and $Q$-operators will be obtained by its reductions. 
Due to importance of \eqref{comQQ} we will present its detailed proof but 
restrict ourselves by considering representative examples to avoid lengthy formulae. 

First of all, the operator $Q_1(x)$ has the form 
\begin{align*}
	Q_1(x) = \mathcal{K}(x)\,[\hat{p}]^{x-s}\,,
\end{align*}
so that the commutation relation in the case $n=1$
\begin{align}\label{QQ1}
	\mathcal{K}(x)\,[\hat{p}]^{x-s}\,
	\mathcal{K}(y)\,[\hat{p}]^{y-s} = 
	\mathcal{K}(y)\,[\hat{p}]^{y-s}\,
	\mathcal{K}(x)\,[\hat{p}]^{x-s}
\end{align}
is in fact the relation for reflection operators itself.
It can be derived by some reduction from the reflection 
relation \eqref{Q2}, but in Appendix~\ref{Reflection} we present the direct proof.

The operator $Q_2(x)$ is of the form 
\begin{align*}
	Q_2(x) = \mathcal{R}_{21}(x)\,\mathcal{K}_1(x)\,
	\mathcal{R}_{12}(x)\,[\hat{p}_2]^{x-s} \,,
\end{align*}
and it is remarkable that the commutation relation in the case $n=2$ 
\begin{multline}\label{n=2}
	\mathcal{R}_{21}(x)
	\mathcal{K}_1(x)\,
	\mathcal{R}_{12}(x)\,[\hat{p}_2]^{x-s}\,
	\mathcal{R}_{21}(y)
	\mathcal{K}_1(y)\,
	\mathcal{R}_{12}(y)\,[\hat{p}_2]^{y-s} \\ 
	= \mathcal{R}_{21}(y)
	\mathcal{K}_1(y)\,
	\mathcal{R}_{12}(y)\,[\hat{p}_2]^{y-s}\,
	\mathcal{R}_{21}(x)
	\mathcal{K}_1(x)\,
	\mathcal{R}_{12}(x)\,[\hat{p}_2]^{x-s}\,
\end{multline}
is equivalent to the relation \eqref{Q2}.
Indeed we have 
\begin{align} \nonumber
	\mathcal{R}_{12}(x) = 
	[z_{12}]^{1-2s}\,[\hat{p}_1]^{x-s}\,[z_{12}]^{x+s-1} \,, \quad
	\mathcal{R}_{21}(x) = 
	[z_{21}]^{1-2s}\,[\hat{p}_2]^{x-s}\,[z_{21}]^{x+s-1} \,,
\end{align}
and using the star-triangle relation \eqref{star-tr}
it is possible to derive the identities
\begin{align}
	& \label{RinvR} \mathcal{R}^{-1}_{21}(y)\,\mathcal{R}_{21}(x) = [z_{21}]^{1-s-y}\,[\hat{p}_2]^{x-y}\,
	[z_{21}]^{x+s-1} \\
	& \label{RpR1} \mathcal{R}_{12}(x)\,[\hat{p}_2]^{x-s}\,
	\mathcal{R}_{21}(y) = [-1]^{y-s}\,[z_{12}]^{1-2s}\,
	[\hat{p}_1]^{x-s}\,[\hat{p}_2]^{y-s}\,
	[z_{12}]^{x+y-1}\,[\hat{p}_2]^{x-s} 
\end{align}
and the simplest version of commutativity relation \eqref{comm}
\begin{equation} \label{comm2}
	\mathcal{R}_{12}(x)\,[\hat{p}_2]^{x-s}\,
	\mathcal{R}_{12}(y)\,[\hat{p}_2]^{y-s} = 
	\mathcal{R}_{12}(y)\,[\hat{p}_2]^{y-s}\,
	\mathcal{R}_{12}(x)\,[\hat{p}_2]^{x-s} \,.
\end{equation}
We use these formulae to show the equivalence of \eqref{n=2} and \eqref{Q2}. The first step is to multiply both sides of \eqref{n=2} by $\mathcal{R}_{21}^{-1}(y)$ from the left and by $[\hat{p}_2]^{s-y}\mathcal{R}_{12}^{-1}(y)$ from the right
\begin{multline*}
	{\red \mathcal{R}_{21}^{-1}(y)\,\mathcal{R}_{21}(x)}\,
	\mathcal{K}_1(x)\,
	{\blue \mathcal{R}_{12}(x)\,[\hat{p}_2]^{x-s}\,
		\mathcal{R}_{21}(y)}\,
	\mathcal{K}_1(y) \\ 
	= 
	\mathcal{K}_1(y)\,
	{\blue \mathcal{R}_{12}(y)\,[\hat{p}_2]^{y-s}\,
		\mathcal{R}_{21}(x)}\,
	\mathcal{K}_1(x)\,
	\mathcal{R}_{12}(x)\,[\hat{p}_2]^{x-s}\,[\hat{p}_2]^{s-y}\,\mathcal{R}_{12}^{-1}(y)
\end{multline*}
Second, we use \eqref{RinvR} and \eqref{RpR1} to rewrite the corresponding coloured products and apply \eqref{comm2} to commute $\mathcal{R}_{12}(x)\,[\hat{p}_2]^{x-s}$ with $[\hat{p}_2]^{s-y}\mathcal{R}_{12}^{-1}(y)$ in the right hand side
\begin{multline*}
	[z_{12}]^{1-s-y}\,[\hat{p}_2]^{x-y}\,
	[z_{12}]^{x+s-1}\,
	\mathcal{K}_1(x)\,
	[z_{12}]^{1-2s}\,[\hat{p}_1]^{x-s}\,[\hat{p}_2]^{y-s}\,
	[z_{12}]^{x+y-1}\,{\blue [\hat{p}_2]^{x-s}}
	\mathcal{K}_1(y) \\ 
	= 
	\mathcal{K}_1(y)\,
	[z_{12}]^{1-2s}\,[\hat{p}_1]^{y-s}\,[\hat{p}_2]^{x-s}\,
	[z_{12}]^{x+y-1}\,{\red [\hat{p}_2]^{y-s}}\,
	\mathcal{K}_1(x)\,{\red [\hat{p}_2]^{s-y}}
	\mathcal{R}_{12}^{-1}(y)\,\mathcal{R}_{12}(x)\,{\blue [\hat{p}_2]^{x-s}}
\end{multline*}
Canceling the coloured factors and rewriting the product of $\mathcal{R}$-operators in the right hand side by means of \eqref{RinvR} we finally reduce \eqref{n=2} to \eqref{Q2}.

Let us prove the commutation relation for $n=3$ using \eqref{n=2}.
We have to prove that the product of operators   
\begin{multline*}
	Q_3(x)\,Q_3(y) = \mathcal{R}_{32}(x)\,\mathcal{R}_{21}(x)
	\mathcal{K}_1(x)\,
	\mathcal{R}_{12}(x)\,\mathcal{R}_{23}(x)\,[\hat{p}_3]^{x-s}\,\\
	\times\mathcal{R}_{32}(y)\,\mathcal{R}_{21}(y)\,
	\mathcal{K}_1(y)\,
	\mathcal{R}_{12}(y)\,\mathcal{R}_{23}(y)\,[\hat{p}_3]^{y-s} 
\end{multline*}
can be transformed to the same product but with the change $x \rightleftarrows y$.
The starting point is the transformation of the product of $\mathcal{R}$-operators 
between two $\mathcal{K}$-operators using \eqref{I}
\begin{align} \nonumber
	\mathcal{R}_{23}(x)\,[\hat{p}_3]^{x-s}\,
	\mathcal{R}_{32}(y)\,\mathcal{R}_{21}(y) = 
	\mathcal{R}_{32}(y)\,[\hat{p}_2]^{x-s}\,
	\mathcal{R}_{21}(y)\,[\hat{p}_2]^{s-x}\,
	\mathcal{R}_{23}(x)\,[\hat{p}_3]^{x-s} \,,
\end{align}
Thus, we obtain 
\begin{multline*}
	Q_3(x)\,Q_3(y) =  
	\mathcal{R}_{32}(x)\,\mathcal{R}_{21}(x)\,\mathcal{R}_{32}(y)\,
	{\blue \mathcal{K}_1(x)\,
		\mathcal{R}_{12}(x)\,[\hat{p}_2]^{x-s}\,
		\mathcal{R}_{21}(y)\,
		\mathcal{K}_1(y)}\,\\ 
	\times [\hat{p}_2]^{s-x}\,
	\mathcal{R}_{23}(x)\,[\hat{p}_3]^{x-s}\,
	\mathcal{R}_{12}(y)\,\mathcal{R}_{23}(y)\,[\hat{p}_3]^{y-s} \,.
\end{multline*}
Note that the product of operators in the middle also enters into $Q_2(x)Q_2(y)$ 
\begin{align*}
	Q_2(x)\,Q_2(y) =
	\mathcal{R}_{21}(x)\,\mathcal{K}_1(x)\,\mathcal{R}_{12}(x)\,[\hat{p}_2]^{x-s}\,
	\mathcal{R}_{21}(y)\,\mathcal{K}_1(y)\,\mathcal{R}_{12}(y)\,[\hat{p}_2]^{y-s} \,.
\end{align*}
Therefore, it remains to transform the product of $\mathcal{R}$-operators
from the left by means of~\eqref{YB}
\begin{align*}
	\mathcal{R}_{32}(x)\,\mathcal{R}_{21}(x)\,\mathcal{R}_{32}(y) = 
	\mathcal{R}_{32}(y)\,\mathcal{R}_{21}(y)\,\mathcal{R}_{32}(x)\,
	\mathcal{R}^{-1}_{21}(y)\,\mathcal{R}_{21}(x)
\end{align*}
and the product of $\mathcal{R}$-operators from the right using the commutation relation \eqref{comm}
\begin{align}
	& \nonumber
	[\hat{p}_2]^{s-x}\,
	\mathcal{R}_{23}(x)\,[\hat{p}_3]^{x-s}\,
	\mathcal{R}_{12}(y)\,\mathcal{R}_{23}(y)\,[\hat{p}_3]^{y-s} \\
	& \nonumber
	= [\hat{p}_2]^{s-x}\,\mathcal{R}_{12}^{-1}(x)\,
	{\blue \mathcal{R}_{12}(x)\mathcal{R}_{23}(x)\,[\hat{p}_3]^{x-s}}\,\,
	{\blue \mathcal{R}_{12}(y)\,\mathcal{R}_{23}(y)\,[\hat{p}_3]^{y-s}} \\
	& \nonumber
	= {\red [\hat{p}_2]^{s-x}\,\mathcal{R}_{12}^{-1}(x)}\,\,
	{\red \mathcal{R}_{12}(y)\,[\hat{p}_2]^{y-s}}\,[\hat{p}_2]^{s-y}\,\mathcal{R}_{23}(y)\,[\hat{p}_3]^{y-s}
	\mathcal{R}_{12}(x)\mathcal{R}_{23}(x)\,[\hat{p}_3]^{x-s}
	\\
	& \label{Rtrans}
	= \mathcal{R}_{12}(y)\,[\hat{p}_2]^{y-s}\,
	[\hat{p}_2]^{s-x}\,\mathcal{R}^{-1}_{12}(x)\,
	[\hat{p}_2]^{s-y}\, \mathcal{R}_{23}(y)\,[\hat{p}_3]^{y-s}\,\mathcal{R}_{12}(x)\,\mathcal{R}_{23}(x)\,[\hat{p}_3]^{x-s} \,.
\end{align}
As the result we arrive at
\begin{align*}
	Q_3(x)\,Q_3(y) & =   
	{\blue \mathcal{R}_{32}(y)\,\mathcal{R}_{21}(y)\,\mathcal{R}_{32}(x)\,
		\mathcal{R}^{-1}_{21}(y)}\, \\ 
	& \times \mathcal{R}_{21}(x)\mathcal{K}_1(x)\,
	\mathcal{R}_{12}(x)\,[\hat{p}_2]^{x-s}\,
	\mathcal{R}_{21}(y)\,
	\mathcal{K}_1(y)\,\mathcal{R}_{12}(y)\,[\hat{p}_2]^{y-s}\,\\
	& \times {\blue [\hat{p}_2]^{s-x}\,\mathcal{R}^{-1}_{12}(x)\,
		[\hat{p}_2]^{s-y}\,
		\mathcal{R}_{23}(y)\,[\hat{p}_3]^{y-s}\,
		\mathcal{R}_{12}(x)\,\mathcal{R}_{23}(x)\,[\hat{p}_3]^{x-s}}  \,.
\end{align*}
In the middle line one has $Q_2(x)Q_2(y)$, so that it is possible to change 
$x \rightleftarrows y$ over there
\begin{align*}
	Q_3(x)\,Q_3(y) & =   
	{\blue \mathcal{R}_{32}(y)\,\mathcal{R}_{21}(y)\,\mathcal{R}_{32}(x)\,
		\mathcal{R}^{-1}_{21}(y)}\, \\ 
	& \times \mathcal{R}_{21}(y)\mathcal{K}_1(y)\,
	\mathcal{R}_{12}(y)\,[\hat{p}_2]^{y-s}\,
	\mathcal{R}_{21}(x)\,
	\mathcal{K}_1(x)\,\mathcal{R}_{12}(x)\,[\hat{p}_2]^{x-s}\,\\
	& \times {\blue [\hat{p}_2]^{s-x}\,\mathcal{R}^{-1}_{12}(x)\,
		[\hat{p}_2]^{s-y}\,
		\mathcal{R}_{23}(y)\,[\hat{p}_3]^{y-s}\,
		\mathcal{R}_{12}(x)\,\mathcal{R}_{23}(x)\,[\hat{p}_3]^{x-s}}  \,.
\end{align*}
Then, first, we cancel $\mathcal{R}_{21}(y)$ from the left and $\mathcal{R}_{12}(x)\,[\hat{p}_2]^{x-s}$ from the right. Second, we commute $\mathcal{R}_{32}(x)$ with $\mathcal{K}_1(y)\,
\mathcal{R}_{12}(y)$ and $\mathcal{K}_1(x)$ with $[\hat{p}_2]^{s-y}\,
\mathcal{R}_{23}(y)\,[\hat{p}_3]^{y-s}$.
The first commutation follows directly from the definition of $\mathcal{R}$-operator, since $\mathcal{R}_{32}$ and $\mathcal{R}_{12}$ act with respect to $z_2$ as operators of multiplication by a function, and the second commutation is obvious, because the operators act with respect to different variables.
This way, we finally obtain
\begin{align*}
	& Q_3(x)\,Q_3(y) =  
	{\blue \mathcal{R}_{32}(y)\,\mathcal{R}_{21}(y)}\,
	\mathcal{K}_1(y)\,
	\mathcal{R}_{12}(y)\,\\
	& \times\mathcal{R}_{32}(x)\,[\hat{p}_2]^{y-s}\,
	\mathcal{R}_{21}(x)\,[\hat{p}_2]^{s-y}\,
	\mathcal{R}_{23}(y)\,[\hat{p}_3]^{y-s}\,
	\mathcal{K}_1(x)\,
	{\blue \mathcal{R}_{12}(x)\,\mathcal{R}_{23}(x)\,[\hat{p}_3]^{x-s}} \\
	& =\mathcal{R}_{32}(y)\,\mathcal{R}_{21}(y)
	\mathcal{K}_1(y)\,
	\mathcal{R}_{12}(y)\,\mathcal{R}_{23}(y)\,[\hat{p}_3]^{y-s}\,
	\mathcal{R}_{32}(x)\,\mathcal{R}_{21}(x)\,
	\mathcal{K}_1(x)\,
	\mathcal{R}_{12}(x)\,\mathcal{R}_{23}(x)\,[\hat{p}_3]^{x-s} \\
	& = Q_3(y)\,Q_3(x) \,,
\end{align*}
where in the last step we applied \eqref{I} with $x\rightleftarrows y$ 
in the opposite way
\begin{align*} 
	\mathcal{R}_{32}(x)\,[\hat{p}_2]^{y-s}\,
	\mathcal{R}_{21}(x)\,[\hat{p}_2]^{s-y}\,
	\mathcal{R}_{23}(y)\,[\hat{p}_3]^{y-s} = 
	\mathcal{R}_{23}(y)\,[\hat{p}_3]^{y-s}\,
	\mathcal{R}_{32}(x)\,\mathcal{R}_{21}(x) \,.
\end{align*}

In general situation the recurrent procedure contains the same steps and uses the same identities. To avoid the general cumbersome formulae we will illustrate everything using next example $n=4$. We start from 
\begin{multline*}
	Q_4(x)\,Q_4(y) = \mathcal{R}_{43}(x)\,\mathcal{R}_{32}(x)\,\mathcal{R}_{21}(x)
	\mathcal{K}_1(x)\,
	\mathcal{R}_{12}(x)\,\mathcal{R}_{23}(x)\,{\blue \mathcal{R}_{34}(x)\,[\hat{p}_4]^{x-s}}\,\\
	\times
	{\blue \mathcal{R}_{43}(y)\,\mathcal{R}_{32}(y)}\,\mathcal{R}_{21}(y)\,
	\mathcal{K}_1(y)\,
	\mathcal{R}_{12}(y)\,\mathcal{R}_{23}(y)\,\mathcal{R}_{34}(y)\,[\hat{p}_4]^{y-s}\,, 
\end{multline*}
and then rewrite the product of $\mathcal{R}$-operators between two $\mathcal{K}$-operators with the help of the similar identity \eqref{I}
\begin{align*}
	\mathcal{R}_{34}(x)\,[\hat{p}_4]^{x-s}\,
	\mathcal{R}_{43}(y)\,\mathcal{R}_{32}(y) = 
	\mathcal{R}_{43}(y)\,[\hat{p}_3]^{x-s}\,
	\mathcal{R}_{32}(y)\,[\hat{p}_3]^{s-x}\,
	\mathcal{R}_{34}(x)\,[\hat{p}_4]^{x-s} \,.
\end{align*}
Consequently,
\begin{multline*}
	Q_4(x)\,Q_4(y) = \mathcal{R}_{43}(x)\,\mathcal{R}_{32}(x)\,\mathcal{R}_{43}(y)\,
	\mathcal{R}_{21}(x)
	\mathcal{K}_1(x)\,
	\mathcal{R}_{12}(x)\,\mathcal{R}_{23}(x)\,[\hat{p}_3]^{x-s}\,\\
	\times \mathcal{R}_{32}(y)\,\mathcal{R}_{21}(y)\,
	\mathcal{K}_1(y)\,
	\mathcal{R}_{12}(y)\,
	[\hat{p}_3]^{s-x}\,\mathcal{R}_{34}(x)\,[\hat{p}_4]^{x-s}\,
	\mathcal{R}_{23}(y)\,\mathcal{R}_{34}(y)\,[\hat{p}_4]^{y-s} \,.
\end{multline*}
Next step we transform the product of $\mathcal{R}$-operators from the left using \eqref{YB}
\begin{align*}
	\mathcal{R}_{43}(x)\,\mathcal{R}_{32}(x)\,\mathcal{R}_{43}(y) = 
	\mathcal{R}_{43}(y)\,\mathcal{R}_{32}(y)\,\mathcal{R}_{43}(x)\,
	\mathcal{R}^{-1}_{32}(y)\,\mathcal{R}_{32}(x)
\end{align*}
and the product of $\mathcal{R}$-operators from the right analogously to \eqref{Rtrans}
\begin{multline*}
	[\hat{p}_3]^{s-x}\,
	\mathcal{R}_{34}(x)\,[\hat{p}_4]^{x-s}\,
	\mathcal{R}_{23}(y)\,\mathcal{R}_{34}(y)\,[\hat{p}_4]^{y-s} \\
	=\mathcal{R}_{23}(y)\,[\hat{p}_3]^{y-s}\,
	[\hat{p}_3]^{s-x}\,\mathcal{R}^{-1}_{23}(x)\,
	[\hat{p}_3]^{s-y}\,
	\mathcal{R}_{34}(y)\,
	[\hat{p}_4]^{y-s}\,\mathcal{R}_{23}(x)\,\mathcal{R}_{34}(x)\,[\hat{p}_4]^{x-s}
\end{multline*}
getting the following
\begin{multline*}
	Q_4(x)\,Q_4(y) =  
	\mathcal{R}_{43}(y)\,\mathcal{R}_{32}(y)\,\mathcal{R}_{43}(x)\,
	\mathcal{R}^{-1}_{32}(y)\,\\
	\times \mathcal{R}_{32}(x)\mathcal{R}_{21}(x)
	\mathcal{K}_1(x)\,
	\mathcal{R}_{12}(x)\,\mathcal{R}_{23}(x)\,[\hat{p}_3]^{x-s}\,
	\mathcal{R}_{32}(y)\,\mathcal{R}_{21}(y)\,
	\mathcal{K}_1(y)\,
	\mathcal{R}_{12}(y)\,\mathcal{R}_{23}(y)\,[\hat{p}_3]^{y-s}\,\\
	\times [\hat{p}_3]^{s-x}\,\mathcal{R}^{-1}_{23}(x)\,
	[\hat{p}_3]^{s-y}\,
	\mathcal{R}_{34}(y)\,
	[\hat{p}_4]^{y-s}\,\mathcal{R}_{23}(x)\,\mathcal{R}_{34}(x)\,[\hat{p}_4]^{x-s}\,.
\end{multline*}
In the middle line we recognize the product $Q_3(x)Q_3(y)$, so that it is possible to use $n=3$ commutativity and change $x\rightleftarrows y$
\begin{multline*}
	Q_4(x)\,Q_4(y) =  
	\mathcal{R}_{43}(y)\,\mathcal{R}_{32}(y)\,\mathcal{R}_{43}(x)\,
	\mathcal{R}^{-1}_{32}(y)\,\\
	\times \mathcal{R}_{32}(y)\mathcal{R}_{21}(y)
	\mathcal{K}_1(y)\,
	\mathcal{R}_{12}(y)\,\mathcal{R}_{23}(y)\,[\hat{p}_3]^{y-s}\,
	\mathcal{R}_{32}(x)\,\mathcal{R}_{21}(x)\,
	\mathcal{K}_1(x)\,
	\mathcal{R}_{12}(x)\,\mathcal{R}_{23}(x)\,[\hat{p}_3]^{x-s}\,\\
	\times [\hat{p}_3]^{s-x}\,\mathcal{R}^{-1}_{23}(x)\,
	[\hat{p}_3]^{s-y}\,
	\mathcal{R}_{34}(y)\,
	[\hat{p}_4]^{y-s}\,\mathcal{R}_{23}(x)\,\mathcal{R}_{34}(x)\,[\hat{p}_4]^{x-s}\,.
\end{multline*}
After evident cancellations from the left and right we move $\mathcal{R}_{43}(x)$ rightwards and \\
$[\hat{p}_3]^{s-y}\,\mathcal{R}_{34}(y)\,[\hat{p}_4]^{y-s}$ leftwards
\begin{multline*}
	Q_4(x)\,Q_4(y) =  
	\mathcal{R}_{43}(y)\,\mathcal{R}_{32}(y)\,\mathcal{R}_{21}(y)\,
	\mathcal{K}_1(y)\,\mathcal{R}_{12}(y)\,\mathcal{R}_{23}(y)\,\\ 
	\times \mathcal{R}_{43}(x)\,[\hat{p}_3]^{y-s}\,
	\mathcal{R}_{32}(x)\,[\hat{p}_3]^{s-y}\,
	\mathcal{R}_{34}(y)\,[\hat{p}_4]^{y-s}\,
	\mathcal{R}_{21}(x)\,
	\mathcal{K}_1(x)\,\mathcal{R}_{12}(x)\,
	\mathcal{R}_{23}(x)\,\mathcal{R}_{34}(x)\,[\hat{p}_4]^{x-s} \,,
\end{multline*}
and, eventually, it remains to use \eqref{I} with $x\rightleftarrows y$ in the opposite way
\begin{align*} 
	\mathcal{R}_{43}(x)\,[\hat{p}_3]^{y-s}\,
	\mathcal{R}_{32}(x)\,[\hat{p}_3]^{s-y}\,
	\mathcal{R}_{34}(y)\,[\hat{p}_4]^{y-s} = 
	\mathcal{R}_{34}(y)\,[\hat{p}_4]^{y-s}\,
	\mathcal{R}_{43}(x)\,\mathcal{R}_{32}(x)
\end{align*}
to derive 
\begin{multline*}
	Q_4(x)\,Q_4(y) =  
	\mathcal{R}_{43}(y)\,\mathcal{R}_{32}(y)\,\mathcal{R}_{21}(y)\,
	\mathcal{K}_1(y)\,\mathcal{R}_{12}(y)\,\mathcal{R}_{23}(y)\, 
	\mathcal{R}_{34}(y)\,[\hat{p}_4]^{y-s}\,\\
	\times \mathcal{R}_{43}(x)\,\mathcal{R}_{32}(x)\,
	\mathcal{R}_{21}(x)\,
	\mathcal{K}_1(x)\,\mathcal{R}_{12}(x)\,
	\mathcal{R}_{23}(x)\,\mathcal{R}_{34}(x)\,[\hat{p}_4]^{x-s} = 
	Q_4(y)\,Q_4(x) \,.
\end{multline*}

\subsection{Baxter equation}

Now we are going to the derivation of the Baxter equation \eqref{BaxB}. 
Everything is based on the key commutation 
relations~\eqref{defR} and~\eqref{equiv} for $R$-operator and $L$-operators.
It is useful to represent the operator $B_n(u)$ is the following form
\begin{align*}
B_n(u) = 
	\langle \uparrow |\,
	L_n(u_1,u_2) \cdots L_1(u_1,u_2) K(u) L_1(u_1,u_2) \cdots L_n(u_1,u_2)\,
	|\downarrow\rangle 
\end{align*}
where 
\begin{align*}
	|\uparrow\rangle = \left(\begin{array}{cc}
		1  \\
		0 \end{array} \right )\ \,, \ 
	|\downarrow\rangle = \left(\begin{array}{cc}
		0  \\
		1 \end{array} \right )\ \,, \ 
	\langle \uparrow | = \left(\begin{array}{cc}
		1  &  0 \end{array} \right )\ \,, \
	\langle \downarrow | = \left(\begin{array}{cc}
		0  &  1 \end{array} \right )
\end{align*}
Operator $B(u)$ is the matrix element in the first row and second column of 
the monodromy matrix $T(u)$.
The operation $\langle \uparrow |\cdots|\downarrow\rangle$ extracts the needed matrix element or equivalently the first row of the left matrix $L_n(u)$ and the second 
column of the right matrix $L_n(u)$.

We have
\begin{align*}
	&Q(x)\,B(u) =   
	\Lambda_n(x)\,[\hat{p}_n]^{x-s}\,
	\langle \uparrow |\,
	L_n(u_1,u_2) \cdots L_1(u_1,u_2) K(u) L_1(u_1,u_2) \cdots L_n(u_1,u_2)\,
	|\downarrow\rangle = \\
	&\Lambda_n(x)\,
	\langle \uparrow |\,L_n(u+x-1,u_2) \cdots L_1(u_1,u_2) K(u) L_1(u_1,u_2) \cdots L_n(u_1,u-x)\, |\downarrow\rangle \,[\hat{p}_n]^{x-s}
\end{align*}
where we used the similarity transformation \eqref{sim}.
Using the commutation relation \eqref{equiv} it is possible to 
derive the following formula 
\begin{align*}
	&\mathcal{R}_{12}(x)\cdots \mathcal{R}_{n-1\,n}(x)\,
	L_{n}\left(u+x-1,u_{2}\right)\cdots
	\,L_{1}\left(u_{1},u_2\right)\,K(u)\,
	L_{1}\left(u_{1},u_{2}\right)\cdots L_{n}\left(u_{1},u-x\right) = \\
	&L_{n}\left(u_{1},u_2\right)\cdots
	L_{1}\left(u+x-1,u_{2}\right)K(u)
	\mathcal{R}_{12}(x)L_{1}\left(u_{1},u_2\right)\cdots
	\mathcal{R}_{n-1\,n}(x)L_{n-1}\left(u_{1},u_{2}\right)
	L_{n}\left(u_{1},u-x\right)
\end{align*}
In this relation, we swapped the product of $R$-operators with 
the left product of $L$-operators, which caused the parameter 
$u-x+1$ to move from left to right.
Next we multiply the last relation to the remaining part of the 
$Q$-operator and regroup the $R$-operators 
\begin{multline*}
	Q(x)\,B(u) =  
	\langle \uparrow |\, \mathcal{R}_{n\,n-1}(x)L_{n}\left(u_{1},u_2\right)\cdots
	\mathcal{R}_{2\,1}(x)L_{2}\left(u_{1},u_2\right)
	\mathcal{K}_1(s,x)\,L_{1}\left(u+x-1,u_{2}\right)K(u)\,\\
	\mathcal{R}_{12}(x)L_{1}\left(u_{1},u_2\right)\cdots
	\mathcal{R}_{n-1\,n}(x)L_{n-1}\left(u_{1},u_{2}\right)\,
	L_{n}\left(u_{1},u-x\right)\,|\downarrow\rangle \,[\hat{p}_n]^{x-s}
\end{multline*}
In complete analogy with the derivation of the Baxter equation for the 
case of a spin chain of A-type \cite{ADV1}, in the next step we need to 
use the key defining relation~(\ref{defR}), 
rewritten in the following equivalent form
\begin{multline*}
	Z_1^{-1}\,
	\mathcal{R}_{12}(x)\,L_1(u_1,u_2)\,Z_2 = \\ 
	\left(%
	\begin{array}{cc}
		\imath\mathcal{R}_{12}(x+1,\bar{x}) +
		(u-x)\mathcal{R}_{12}(x,\bar{x}) & -\mathcal{R}_{12}(x,\bar{x})\,\partial_{1} \\
		-(u-x)\,z_{12}\,\mathcal{R}_{12}(x,\bar{x}) & -\imath(x+s-1)(x-s)
		\mathcal{R}_{12}(x-1,\bar{x})+(u-x)\mathcal{R}_{12}(x,\bar{x})\\
	\end{array}%
	\right)
\end{multline*}
where $Z_k = \left(%
\begin{array}{cc}
	1 & 0 \\
	z_k & 1 \\
\end{array}%
\right)$. 
For $x=u$ matrix is triangular so that one immediately obtains 
\begin{multline*}
	Z_1^{-1}\,\mathcal{R}_{12}(u)L_1(u_1\,,u_2)\cdots 
	\mathcal{R}_{n-1\,n}(u)\,L_{n-1}(u_1\,,u_2)\,Z_n = \\
	\medskip
	= \left(%
	\begin{array}{cc}
		\imath\mathcal{R}_{12}(u+1,\bar{u}) & \ldots \\
		0 & \ldots \\
	\end{array}%
	\right)\cdots\left(%
	\begin{array}{cc}
		\imath\mathcal{R}_{n-1\,n}(u+1,\bar{u}) & \ldots \\
		0 & \ldots \\
	\end{array}%
	\right)\,
\end{multline*}
and similar relation for the product in opposite order 
\begin{multline*}
	Z_n^{-1}\,\mathcal{R}_{n\,n-1}(u)L_n(u_1\,,u_2)\cdots 
	\mathcal{R}_{2\,1}(u)\,L_{2}(u_1\,,u_2)\,Z_1 = \\
	\medskip
	= \left(%
	\begin{array}{cc}
		\imath\mathcal{R}_{n\,n-1}(u+1,\bar{u}) & \ldots \\
		0 & \ldots \\
	\end{array}%
	\right)\cdots\left(%
	\begin{array}{cc}
		\imath\mathcal{R}_{2\,1}(u+1,\bar{u}) & \ldots \\
		0 & \ldots \\
	\end{array}%
	\right)\,
\end{multline*}
Due to this triangularity at the point $x=u$ we have 
\begin{multline*}
	Q(u)\,B(u) =   
	\langle \uparrow|\, Z_n\,\left(%
	\begin{array}{cc}
		\imath^{n-1}\mathcal{R}_{n\,n-1}(u+1,\bar{u})\cdots
		\mathcal{R}_{21}(u+1,\bar{u}) & \ldots \\
		0 & \ldots \\
	\end{array}%
	\right) \\ 
	\left.
	Z^{-1}_1\,\mathcal{K}_1(s,x)\,L_{1}\left(u+x-1,u_{2}\right)\,
	K(u)\,Z_1\right|_{x=u}\\
	\left(
	\begin{array}{cc}
		\imath^{n-1}\mathcal{R}_{12}(u+1,\bar{u})\cdots
		\mathcal{R}_{n-1\,n}(u+1,\bar{u}) & \ldots \\
		0 & \ldots \\
	\end{array}%
	\right)\,
	\left.Z_{n}^{-1}\,
	L_{n}\left(u_{1},u-x\right)\,|\downarrow\rangle\,[\hat{p}_n]^{x-s}\right|_{x=u}
\end{multline*}
The right matrix in explicit form looks as follows 
($\partial_{n} = \imath\hat{p}_n$) 
\begin{align*}
	Z_{n}^{-1}\,L_n\left(u_{1},u-x\right)\,|\downarrow\rangle = \left(%
	\begin{array}{cc}
		u_1 & -\imath\hat{p}_n \\
		0 & u-x \\
	\end{array}%
	\right)\, \left(%
	\begin{array}{cc}
		0 \\ 1 \\
	\end{array}%
	\right) = \left(%
	\begin{array}{cc}
		-\imath\hat{p}_n \\
		u-x \\
	\end{array}%
	\right)
\end{align*}
so that at the point $x=u$ we obtain 
\begin{align*}
	\left.Z_{n}^{-1}\,
	L_{n}\left(u_{1},u-x\right)
	\,|\downarrow\rangle \,[\hat{p}_n]^{x-s}\right|_{x=u} =  
	|\uparrow\rangle \,(-\imath\hat{p}_n)\,[\hat{p}_n]^{u-s}
\end{align*}
This leads to simplification of the whole relation 
\begin{multline*}
	Q(u)\,B(u) =   
	\langle \uparrow|\,Z_n\,\left(%
	\begin{array}{cc}
		\imath^{n-1}\mathcal{R}_{n\,n-1}(u+1,\bar{u})\cdots
		\mathcal{R}_{21}(u+1,\bar{u}) & \ldots \\
		0 & \ldots \\
	\end{array}%
	\right) \\ 
	\left.
	Z^{-1}_1\,\mathcal{K}_1(s,x)\,L_{1}\left(u+x-1,u_{2}\right)\,
	K(u)\,Z_1\right|_{x=u}\,|\uparrow\rangle
	\\
	\imath^{n-1}\mathcal{R}_{12}(u+1,\bar{u})\cdots
	\mathcal{R}_{n-1\,n}(u+1,\bar{u})\,
	(-\imath\hat{p}_n)\,[\hat{p}_n]^{u-s}
\end{multline*}
The middle expression in explicit form looks as follows 
($\partial_{z} = \imath\hat{p}$)
\begin{align*}
	&Z^{-1}\,\mathcal{K}(s,x)\,L\left(u+x-1,u_{2}\right)\,
	K(u)\,Z\,|\uparrow\rangle  =   
	z_{+}^{g-s}\,z_{-}^{1-s-g}\,
	\left(%
	\begin{array}{cc}
		1 & 0 \\
		-z& 1 \\
	\end{array}%
	\right)\,
	\hat{p}^{x-s}
	\left(%
	\begin{array}{cc}
		1 & 0 \\
		z & 1 \\
	\end{array}%
	\right)\, \\
	&\left(%
	\begin{array}{cc}
		\imath(u-\frac{1}{2})\,\hat{p}\,z_{+}^{x-g+1}\,z_{-}^{x+g} +
		(u-x)\,\bigr((u-\frac{1}{2})z-(g-\frac{1}{2})\gamma\bigl)\,
		z_{+}^{x-g}\,z_{-}^{x+g-1}\\ 
		-(u-s)(u-\frac{1}{2})\,z_{+}^{x-g+1}\,z_{-}^{x+g} \\
	\end{array}%
	\right) 
\end{align*}
where we used the explicit expression \eqref{Kop-d} for the reflection operator 
\begin{align*}
	\mathcal{K}(s,x) =
	[z_+]^{g-s} \,[z_-]^{1-s-g} \,
	[\hat{p}]^{x-s} \,[z_+]^{x-g} \,
	[z_-]^{x+g-1}.
\end{align*}
For simplicity we skip the index $z_1 \to z$ 
and use compact notations $z_{\pm} = z\pm\gamma$.
Note that we show the holomorphic part only and 
antiholomorphic part of $\mathcal{K}(s,x)$ is unchanged.

At the point $x=u$ one obtains 
\begin{multline*}
	\left.Z^{-1}\,\mathcal{K}(s,x)\,L\left(u+x-1,u_{2}\right)\,
	K(u)\,Z\,|\uparrow\rangle\right|_{x=u}  = \\  
	{\textstyle \left(u-\frac{1}{2}\right)}\,
	z_{+}^{g-s}\,z_{-}^{1-s-g}\,
	\left(%
	\begin{array}{cc}
		1 & 0 \\
		-\imath(u-s)\,\hat{p}^{-1} & 1 
	\end{array}%
	\right)\,\left(%
	\begin{array}{cc}
		\imath\hat{p}
		\\ 
		-(u-s) \\
	\end{array}%
	\right)\,\hat{p}^{u-s}\,z_{+}^{u-g+1}\,z_{-}^{u+g} = \\ 
	\left(%
	\begin{array}{cc}
		1\\ 
		0
	\end{array}%
	\right)
	\imath{\textstyle \left(u-\frac{1}{2}\right)}\,
	z_{+}^{g-s}\,z_{-}^{1-s-g}\,\hat{p}^{u-s+1}\,z_{+}^{u-g+1}\,z_{-}^{u+g} 
\end{multline*}
so that in the right hand side appears the reflection operator 
with shifted spectral parameter $u \to u+1$
\begin{align*}
	\left.Z^{-1}\,\mathcal{K}(s,x)\,L\left(u+x-1,u_{2}\right)\,
	K(u)\,Z\,\right|_{x=u}\,|\uparrow\rangle = \imath {\textstyle \left(u-\frac{1}{2}\right)}\,|\uparrow\rangle
	\,\mathcal{K}(s,u+1\,,\bar{u})
\end{align*}
Collecting everything together we obtain the Baxter equation 
\begin{multline*}
	Q(u)\,B(u) =   
	\imath^{2(n-1)}\,{\textstyle \left(u-\frac{1}{2}\right)}\,\langle \uparrow|Z_n\,\left(%
	\begin{array}{cc}
		\mathcal{R}_{n\,n-1}(u+1,\bar{u})\cdots
		\mathcal{R}_{21}(u+1,\bar{u}) & \ldots \\
		0 & \ldots \\
	\end{array}%
	\right)\,|\uparrow\rangle \\ 
	\mathcal{K}_1(s,u+1\,,\bar{u})\,
	\mathcal{R}_{12}(u+1,\bar{u})\cdots
	\mathcal{R}_{n-1\,n}(u+1,\bar{u})\,\hat{p}_n\,[\hat{p}_n]^{u-s} =  \\ 
	\imath^{2(n-1)}\,{\textstyle \left(u-\frac{1}{2}\right)}
	\mathcal{R}_{n\,n-1}(u+1,\bar{u})\cdots
	\mathcal{R}_{21}(u+1,\bar{u})\,
	\mathcal{K}_1(s,u+1\,,\bar{u})\,\\
	\mathcal{R}_{12}(u+1,\bar{u})\cdots
	\mathcal{R}_{n-1\,n}(u+1,\bar{u})\,\hat{p}_n\,[\hat{p}_n]^{u-s} = 
	\imath^{2(n-1)}\,\,{\textstyle \left(u-\frac{1}{2}\right)}\,
	Q(u+1\,,\bar{u})\
\end{multline*}

\section{Eigenfunctions} \label{sect:Eigenunc}

In this section we are going to prove by 
direct calculation that functions $\Psi_{\bm x_n}(\bm z_n)$
are eigenfunctions of the $Q$-operator and to derive the explicit 
expression for the corresponding eigenvalues. 

Next we investigate the behaviour of eigenfunctions 
$\Psi_{\bm x_n}(\bm z_n)$ with respect to 
permutations of $x_1, \ldots, x_n$ and reflections $x_k \to -x_k$.

The model under consideration contains two external parameters -- 
spin $s$ which labels the principal series representation and 
parameter $g$ which enters in reflection matrix $K(u)$.
In the last part of this section we investigate the behaviour 
of eigenfunctions $\Psi_{\bm x_n}(\bm z_n)$ with respect to 
transformation $s \to 1-s \,, g \to 1-g$.

\subsection{Eigenfunctions of $Q$-operator}

The operators $B_n(u)$ and $Q_n(v)$ commute and
therefore have the common system of eigenfunctions. 
The explicit proof of the fact that
$\Psi_{\bm x_n}(\bm z_n)$ is an eigenfunction
of the operator $Q_n(u)$ and calculation of the corresponding eigenvalue 
are grounded on the following exchange relation valid for $n=2,\,3,\ldots$
\begin{eqnarray}
\label{exchQA1}
Q_n(u)\,\Lambda_n(x)\,\Psi(\bm{z}_{n-1})
= q(u\,,x)\,\Lambda_n(x)\,Q_{n-1}(u)\,\Psi(\bm{z}_{n-1}) \,,
\end{eqnarray}
where 
\begin{equation} \nonumber
q(u\,,x) =  
[-1]^{u-s}\,\frac{\mathbf{\Gamma}(u\pm x)}{\mathbf{\Gamma}(s\pm x)}
\end{equation}
and $\Psi(\bm{z}_{n-1})$ is an arbitrary function which depends on $z_1,\dots,z_{n-1}$.
We will use the compact notations 
\begin{align}\label{pm}
\mathbf{\Gamma}(u\pm x) = \mathbf{\Gamma}(u+x)\mathbf{\Gamma}(u-x)
\end{align}
Using this exchange relation in the expression
\begin{equation} \nonumber
	Q_n(u)\,\Psi_{\bm x_n}(\bm z_n) =
	Q_n(u) \, \Lambda_n(x_n)\,
	\Lambda_{n-1}(x_{n-1})\cdots 
	\Lambda_{2}(x_{2})\,\Lambda_{1}(x_{1})\cdot 1
\end{equation}
we move the $Q$-operator to the right and obtain
\begin{align}
	\label{QPsi}
	Q_n(u)\,\Psi_{\bm x_n}(\bm z_n)
	= q(u\,,\bm x_{n})
	\, \Psi_{\bm x_n}(\bm z_n) \,,
\end{align}
where 
\begin{align}\label{qn}
q(u\,,\bm x_{n}) = [-1]^{s-u}\,q(u\,,x_{1})\,q(u\,,x_{2})\cdots q(u\,,x_{n}) = [-1]^{(n-1)(u-s)} \prod_{k=1}^{n} \frac{\mathbf{\Gamma}(u\pm x_k)}
{\mathbf{\Gamma}(s\pm x_k)} \,.
\end{align}
On the last step we used the basic relation
\begin{equation} \nonumber
Q_1(u) \,\Lambda_{1}(x_{1})\cdot 1  = 
\frac{\mathbf{\Gamma}(u\pm x_1)}{\mathbf{\Gamma}(s\pm x_1)}\,
\Lambda_{1}(x_{1})\cdot 1
= [-1]^{s-u}q(u,x_1)\,\Lambda_{1}(x_{1})\cdot 1 \,.
\end{equation}
This identity is a consequence of the commutativity relation 
for $Q$-operators \eqref{QQ1} and formula \eqref{star-tr1} from Appendix. 
To deduce it we first of all rewrite \eqref{QQ1} with $x=u,\,y=x_1$
\begin{align*}
\mathcal{K}(u)\,[\hat{p}]^{u-s}\,
\mathcal{K}(x_1)\,[\hat{p}]^{x_1-s} = 
\mathcal{K}(x_1)\,[\hat{p}]^{x_1-s}\,
\mathcal{K}(u)\,[\hat{p}]^{u-s}
\end{align*}
in equivalent form 
\begin{align*}
	\mathcal{K}(u)\,[\hat{p}]^{u-s}\,
	\mathcal{K}(x_1) = 
	\mathcal{K}(x_1)\,[\hat{p}]^{x_1-s}\,\mathcal{K}(u)\,[\hat{p}]^{u-x_1}
\end{align*}
and then apply it to the constant function $\psi(z) = 1$. In terms of $Q$- and $\Lambda$-operators the resulting expression reads
\begin{align*}
Q_1(u)\,\Lambda_1(x_1)\cdot 1 = 
\Lambda_1(x_1)\,[\hat{p}]^{x_1-s}\,\mathcal{K}(u)\,[\hat{p}]^{u-x_1} \cdot 1 \,.
\end{align*}
It remains to calculate $[\hat{p}]^{x_1-s}\,\mathcal{K}(u)\,[\hat{p}]^{u-x_1} \cdot 1$, in order to do this we use the explicit formula \eqref{Kop-d} for reflection operator and insert mutually inverse factors $[\hat{p}]^{g-x_1}\, [\hat{p}]^{x_1-g}$, $[\hat{p}]^{1-g-x_1}\,
[\hat{p}]^{x_1+g-1}$ getting
\begin{align*}
[\hat{p}]^{x_1-s}\,\mathcal{K}(u)\,[\hat{p}]^{u-x_1} \cdot 1 & = 
[\hat{p}]^{x_1-s}\,[z_{+}]^{g-s}\,[\hat{p}]^{g-x_1}\, [\hat{p}]^{x_1-g}\,[z_{-}]^{1-s-g}\,[\hat{p}]^{1-s-x_1}\,\\
& \times
[\hat{p}]^{u+x_1-1}\,[z_{+}]^{u-g}\,[\hat{p}]^{1-g-x_1}\,
[\hat{p}]^{x_1+g-1}\,[z_{-}]^{u+g-1}\,[\hat{p}]^{u-x_1}\cdot 1 \,.
\end{align*}
Using \eqref{star-tr1} four times one obtains
\begin{equation} \nonumber
	[\hat{p}]^{x_1-s}\,\mathcal{K}(u)\,[\hat{p}]^{u-x_1} \cdot 1 =
	\frac{\mathbf{\Gamma}(u\pm x_1)}{\mathbf{\Gamma}(s\pm x_1)} \,.
\end{equation}

Let us return to the proof of the exchange relation \eqref{exchQA1}.
The operators $Q_n(x)$ and $\Lambda_n(x)$ are closed relatives and comparison of the operator expressions 
\begin{align}
\label{Qex}
Q_n( x) & = \mathcal{R}_{n\,,n-1}(x)\cdots 
\mathcal{R}_{21}(x)\,\mathcal{K}_{1}(x)\,
\mathcal{R}_{12}(x)\cdots\mathcal{R}_{n-1\,n}(x)\,
[\hat{p}_n]^{x-s}\,, \\
\nonumber
\Lambda_n( x) & = \mathcal{R}_{n\,,n-1}(x)\cdots 
\mathcal{R}_{21}(x)\,\mathcal{K}_{1}(x)\,
\mathcal{R}_{12}(x)\cdots\mathcal{R}_{n-1\,n}(x)
\end{align}
immediately shows that 
\begin{align}\label{QL1}
Q_n( x) = \Lambda_n(x)\,[\hat{p}_n]^{x-s} \,.
\end{align}
In addition, from \eqref{Qex} follows the recurrent formula
\begin{align} \label{QnQn-1}
Q_n(u) = \mathcal{R}_{n\,,n-1}(u)\, Q_{n-1}(u)\,[\hat{p}_{n-1}]^{s-u}\,
\mathcal{R}_{n-1\,n}(u)\,[\hat{p}_n]^{u-s} \,.
\end{align}
Applying \eqref{QL1} and \eqref{QnQn-1} we rewrite the commutativity 
relation $Q_n(u)\,Q_n(x) = Q_n(x)\,Q_n(u)$ in equivalent form  
\begin{align*} 
Q_n(u)\,\Lambda_n(x)\,[\hat{p}_n]^{x-s} = 
\Lambda_n(x)\,[\hat{p}_n]^{x-s}\,\mathcal{R}_{n\,,n-1}(u)\,
Q_{n-1}(u)\,[\hat{p}_{n-1}]^{s-u}\,
\mathcal{R}_{n-1\,n}(u)\,[\hat{p}_n]^{u-s} \,.
\end{align*}
Multiplying both sides of the last equation from the right by $[\hat{p}_n]^{s-x}$ and using the explicit formula for $\mathcal{R}$-operators we transform it to the form which is needed for our purpose
\begin{align}
\nonumber
Q_n(u)\,\Lambda_n(x) & = 
[-1]^{u-s}\Lambda_n(x)\,
[\hat{p}_n]^{u+x-1}\,[z_{n\,n-1}]^{u-s}\,[\hat{p}_n]^{1-x-s}\,
Q_{n-1}(u)  \\
& \label{QL2}
\times [\hat{p}_{n-1}]^{2s-1}\,
[\hat{p}_n]^{x-s}\,[z_{n\,n-1}]^{u-s}\,[\hat{p}_n]^{u-x}\,
[\hat{p}_{n-1}]^{1-2s} \,.
\end{align}
Eventually, we act by both sides of \eqref{QL2} on the function $\Psi(\bm z_{n-1})$, which does not depend on variable $z_n$. Using this property we employ the formula \eqref{star-tr1} twice in the right hand side
\begin{align*}
&
[\hat{p}_n]^{x-s}\,[z_{n\,n-1}]^{u-s}\,[\hat{p}_n]^{u-x}\cdot 1 = 
 \frac{[\imath]^{u-s+1}\,[-1]^{x-u}}{\mathbf{\Gamma}(s-x,1-u+x)} \,,\\
 &
[\hat{p}_n]^{u+x-1}\,[z_{n\,n-1}]^{u-s}\,[\hat{p}_n]^{1-x-s}\cdot 1 = 
\frac{[\imath]^{u-s+1}\,[-1]^{x+s-1}}{\mathbf{\Gamma}(1-u-x,x+s)}  
\end{align*}
and finally obtain \eqref{exchQA1}.

\subsection{Symmetry} \label{sect:Symmetry}

The eigenvalues of operators $B_n$ and $\bar{B}_n$ in the 
definition of eigenfunctions \eqref{BPsi} are invariant under 
permutations of $x_1, \ldots, x_n$ and reflections $x_k \to -x_k$. 
We expect that the spectrum is simple so that for any 
permutation $\tau \in \mathfrak{S}_n$ and reflection $\sigma_k$ 
the eigenfunctions $\Psi_{\tau \bm x_n}$, $\Psi_{\sigma_k \bm x_n}$ 
and $\Psi_{\bm x_n}$ can differ by multiplication on the constant only 
\begin{align*} 
\Psi_{\tau \bm x_n}(\bm z_n) = 
c(\tau,\bm x_n)\,\Psi_{\bm x_n}(\bm z_n)\, ; \\ 
\Psi_{\sigma_k \bm x_n}(\bm z_n) = 
c_k(\bm x_n)\,\Psi_{\bm x_n}(\bm z_n)\,,
\end{align*}
where $\tau \bm x_n = (x_{\tau(1)},\bar{x}_{\tau(1)}, \ldots , x_{\tau(n)},\bar{x}_{\tau(n)})$ and 
$\sigma_k \bm x_n = (x_{1},\bar{x}_1,\ldots , -x_{k},-\bar{x}_{k}, 
\ldots , x_{n},\bar{x}_{n})$.
It is indeed the case and the behaviour under permutations of $x_1, \ldots, x_n$ is governed by the following 
commutation relation between $\Lambda$-operators ($n=2\,,3\,,\ldots$)
\begin{multline} \label{exchLL1}
[-1]^{y-s}\,\mathbf{\Gamma}^2(s+y)\,\Lambda_n(x)\,
\Lambda_{n-1}(y)\,\Psi(\bm{z}_{n-2}) = \\
[-1]^{x-s}\,\mathbf{\Gamma}^2(s+x)\,\Lambda_n(y)\,\Lambda_{n-1}(x)\,\Psi(\bm{z}_{n-2})
\end{multline}
and the formula ($c(\alpha)$ is defined in \eqref{c})
\begin{multline} \label{-x}
[-2\imath\gamma]^{x-s}\,\left(c(x-s)\right)^{2(n-1)}\,
\mathbf{\Gamma}(g+x)\,\Lambda_n(-x)\,\Psi(\bm{z}_{n-1}) \\
= [-2\imath\gamma]^{-x-s}\,\left(c(-x-s)\right)^{2(n-1)}\,
\mathbf{\Gamma}(g-x)\,\Lambda_n(x)\,\Psi(\bm{z}_{n-1}) \,,
\end{multline}
where $\Psi(\bm{z}_k)$ denotes a function which depends only on $z_1,\dots,z_k$.

Let us start from derivation of the commutation relation for $\Lambda$-operators \eqref{exchLL1} from the commutation relation \eqref{exchQA1} for $Q$- and $\Lambda$-operators 
\begin{align} \label{QL3}
Q_n( x)\,\Lambda_{n}( y) = 
q(x,y)\,\Lambda_{n}( y)\,Q_{n-1}(x) \,. 
\end{align}
Recall that the identity \eqref{QL3} holds true in the domain of functions which do not depend on $z_n$.
We have 
\begin{align*}
&Q_n( x) = \Lambda_n(x)\,[\hat{p}_n]^{x-s} \,, \quad
Q_{n-1}( x) = \Lambda_{n-1}(x)\,[\hat{p}_{n-1}]^{x-s} \,, \\ 
&\Lambda_{n}( y) = \mathcal{R}_{n\,,n-1}(y)\cdots 
\mathcal{R}_{21}(y)\,\mathcal{K}_{1}(y)\,
\mathcal{R}_{12}(y)\cdots\mathcal{R}_{n-1\,n}(y) = 
\mathcal{R}_{n\,,n-1}(y)\,
\Lambda_{n-1}( y)\,\mathcal{R}_{n-1\,n}(y)
\end{align*}
so that 
\begin{align*}
\Lambda_n(x)\,[\hat{p}_n]^{x-s}\,
\mathcal{R}_{n\,,n-1}(y)\,
\Lambda_{n-1}( y)\,\mathcal{R}_{n-1\,n}(y) = 
q(x,y)\,\Lambda_{n}( y)\,
\Lambda_{n-1}(x)\,[\hat{p}_{n-1}]^{x-s} \,.
\end{align*}
Multiplying both sides of this equation from the right by $[\hat{p}_{n-1}]^{s-x}$ and using the explicit formula \eqref{R1} for $\mathcal{R}$-operators one arrives at
\begin{multline*}
	\Lambda_n(x)\,[\hat{p}_n]^{x+y-1}\,[z_{n\,,n-1}]^{y-s}\,
	[\hat{p}_n]^{1-s-x}\,
	\Lambda_{n-1}( y)\,
	[\hat{p}_{n-1}]^{s+y-1}\,[\hat{p}_n]^{x-s}\,[z_{n-1\,n}]^{y-s}\,
	[\hat{p}_{n-1}]^{1-s-x} \\
	= q(x,y)\,\Lambda_{n}( y)\,\Lambda_{n-1}(x) \,.
\end{multline*}
To obtain \eqref{exchLL1} we apply this operator identity to the function 
$\Psi(\bm z_{n-2})$, which does not depend on $z_n$ and $z_{n-1}$, and then transform the left hand side with the help of chain relation~\eqref{Chain1} 
\begin{equation*}
[\hat{p}_n]^{x-s}\,[z_{n-1\,n}]^{y-s} =
\frac{[\imath]^{x-s+1}\,[-1]^{y-s}}
{\mathbf{\Gamma}(s-y,1-x+y)}\,
[z_{n-1\,n}]^{y-x},
\end{equation*}
and the formula \eqref{star-tr1}
\begin{align*}
&
[\hat{p}_{n-1}]^{s+y-1}\,[z_{n-1\,n}]^{y-x}\,[\hat{p}_{n-1}]^{1-s-x}\cdot 1 = 
\frac{[\imath]^{y-x+1}\,[-1]^{s+x-1}}{\mathbf{\Gamma}(1-s-y,s+x)} \,, \\
&
[\hat{p}_n]^{x+y-1}\,[z_{n\,,n-1}]^{y-s}\,[\hat{p}_n]^{1-s-x}\cdot 1 =  
\frac{[\imath]^{y-s+1}\,[-1]^{s+x-1}}{\mathbf{\Gamma}(1-x-y,s+x)} \,.
\end{align*}

Now we derive the relation \eqref{-x}. 
In the simplest case $n=1$ this relation is proven in \cite{ABDV},
therefore we start from the case $n=2$ and transform the initial expression
\begin{align*}
\Lambda_{2}( x)\,\Psi(z_{1}) = \mathcal{R}_{21}(x)\,\mathcal{K}_{1}(x)\,
\mathcal{R}_{12}(x)\,\Psi(z_{1}) = a(x)\,
\mathcal{R}_{21}(x)\,\mathcal{K}_{1}(x)\,
\mathcal{R}_{21}(-x)\,\Psi(z_{2})
\end{align*}
using the key relation 
\begin{align} \label{key0}
\mathcal{R}_{12}(x)\,\Psi(z_{1}) = 
a(x)\,\mathcal{R}_{21}(-x)\,\Psi(z_{2})
\end{align}
where 
\begin{align} \label{a}
a(x) = \frac{c(x-s)}{c(-x-s)} \,.
\end{align}
Next we use the explicit form \eqref{Kop-d} of the operator $\mathcal{K}_{1}(x)$ and obtain
\begin{align} \label{L2Psi}
&\Lambda_{2}( x)\,\Psi(z_{1})
= a(x)\, [z_{1+}]^{g-s}\, [z_{1-}]^{1-s-g}\,
\mathcal{R}_{21}(x)\,
[\hat{p}_1]^{x-s}\,
\mathcal{R}_{21}(-x)\,[z_{1+}]^{x-g}\,
[z_{1-}]^{x+g-1}\,\Psi(z_{2}) \,.
\end{align}
Using the explicit expression for $\mathcal{R}$-operator and star-triangle identity \eqref{star-tr} one can derive the commutation relation
\begin{align*}
	\mathcal{R}_{21}(x)\,[\hat{p}_1]^{x-s}\,
	\mathcal{R}_{21}(y)\,[\hat{p}_1]^{y-s} = 
	\mathcal{R}_{21}(y)\,[\hat{p}_1]^{y-s}\,
	\mathcal{R}_{21}(x)\,[\hat{p}_1]^{x-s} \,,
\end{align*}
which is the simplest case of \eqref{comm}. This formula allows to transform the product of $\mathcal{R}$-operators in \eqref{L2Psi}
\begin{multline*}
\mathcal{R}_{21}(x)\,
[\hat{p}_1]^{x-s}\,
\mathcal{R}_{21}(-x) = \mathcal{R}_{21}(x)\,
[\hat{p}_1]^{x-s}\,
\mathcal{R}_{21}(-x)\,
[\hat{p}_1]^{-x-s}\,
[\hat{p}_1]^{x+s} \\
= \mathcal{R}_{21}(-x)\,
[\hat{p}_1]^{-x-s}\,
\mathcal{R}_{21}(x)\,
[\hat{p}_1]^{2x} \,,
\end{multline*}
so that we have 
\begin{align*}
&
\mathcal{R}_{21}(x)\,
[\hat{p}_1]^{x-s}\,
\mathcal{R}_{21}(-x)\,[z_{1+}]^{x-g}\,
[z_{1-}]^{x+g-1}\,\Psi(z_{2}) \\
&
= \mathcal{R}_{21}(-x)\,
[\hat{p}_1]^{-x-s}\,
\mathcal{R}_{21}(x)\,
[\hat{p}_1]^{2x}\,[z_{1+}]^{x-g}\,
[z_{1-}]^{x+g-1}\,\Psi(z_{2}) \\
&
= [-2\imath\gamma]^{2x}\,\frac{\mathbf{\Gamma}(g+x)}{\mathbf{\Gamma}(g-x)}\,
\mathcal{R}_{21}(-x)\,
[\hat{p}_1]^{-x-s}\,
\mathcal{R}_{21}(x)\,[z_{1+}]^{-x-g}\,
[z_{1-}]^{-x+g-1}\,\Psi(z_{2})\,,
\end{align*}
where on the last step we used the star-triangle relation in the form \eqref{Star1}
\begin{align} \label{p1z1z1}
[\hat{p}_1]^{2x}\,
[z_{1+}]^{x-g}\,
[z_{1-}]^{x+g-1} = 
[-2\imath\gamma]^{2x}\,\frac{\mathbf{\Gamma}(g+x)}{\mathbf{\Gamma}(g-x)}\,
[z_{1+}]^{-x-g}\,[z_{1-}]^{-x+g-1} \,.
\end{align}
After this we move the operator $\mathcal{R}_{21}(x)$ to the right and use the formula \eqref{key0} again
\begin{equation} \nonumber
	\mathcal{R}_{21}(x)\,\Psi(z_{2}) = a(x)\,\mathcal{R}_{12}(-x)\,\Psi(z_1) \,.
\end{equation}
Collecting everything together and joining certain factors to form the reflection operator
\begin{equation} \nonumber
	[z_{1+}]^{g-s}\, [z_{1-}]^{1-s-g}[\hat{p}_1]^{-x-s}[z_{1+}]^{-x-g}\,
	[z_{1-}]^{-x+g-1} = \mathcal{K}_1(-x)
\end{equation}
we obtain from \eqref{L2Psi} the needed result \eqref{-x} for $n=2$
\begin{align*}
&\Lambda_{2}( x)\,\Psi(z_{1}) = [-2\imath\gamma]^{2x}\,a^2(x)\,\frac{\mathbf{\Gamma}(g+x)}{\mathbf{\Gamma}(g-x)}\,
\mathcal{R}_{21}(-x)\,\mathcal{K}_1(-x)\,\mathcal{R}_{12}(-x)\,\Psi(z_{2}) \\ 
&
= \frac{[-2\imath\gamma]^{x-s}\,c^2(x-s)\,\mathbf{\Gamma}(g+x)}{[-2\imath\gamma]^{-x-s}\,c^2(-x-s)\,\mathbf{\Gamma}(g-x)}\,
\Lambda_2(-x) \Psi(z_{2}) \,.
\end{align*}

In the case $n=3$ everything is very similar. With the help of explicit formulas for $\Lambda$- and $\mathcal{K}$-operators one gets the following expression
\begin{align*}
	& \Lambda_3(x)\,\Psi(z_1,z_2) \\
	& = [z_{1+}]^{g-s}\, [z_{1-}]^{1-s-g}\mathcal{R}_{32}(x)\,\mathcal{R}_{21}(x)\,[\hat{p}_1]^{x-s}[z_{1+}]^{x-g}\, [z_{1-}]^{x+g-1}\,\mathcal{R}_{12}(x)\,\mathcal{R}_{23}(x)\,\Psi(z_{1},z_2)
\end{align*} 
There are two essential steps: first of all we apply the key relation  
\begin{align} \label{key2}
\mathcal{R}_{12}(x)\,\mathcal{R}_{23}(x)\,\Psi(z_{1},z_2) = 
a^2(x)\,\mathcal{R}_{32}(-x)\,\mathcal{R}_{21}(-x)\,\Psi(z_{2},z_3) \,,
\end{align}
then the commutation relation \eqref{comm} for $k=3$
\begin{align*}
\mathcal{R}_{32}(y)\,\mathcal{R}_{21}(y)\,[\hat{p}_1]^{y-s}\,
\mathcal{R}_{32}(x)\,\mathcal{R}_{21}(x)\,[\hat{p}_1]^{x-s} = 
\mathcal{R}_{32}(x)\,\mathcal{R}_{21}(x)\,[\hat{p}_1]^{x-s}\,
\mathcal{R}_{32}(y)\,\mathcal{R}_{21}(y)\,[\hat{p}_1]^{y-s}
\end{align*}
allows to perform the main transformation  
\begin{align*}
\mathcal{R}_{32}(x)\,\mathcal{R}_{21}(x)\,
[\hat{p}_1]^{x-s}\,
\mathcal{R}_{32}(-x)\,\mathcal{R}_{21}(-x) = 
\mathcal{R}_{32}(-x)\,\mathcal{R}_{21}(-x)\,
[\hat{p}_1]^{-x-s}\,
\mathcal{R}_{32}(x)\,\mathcal{R}_{21}(x)
\,[\hat{p}_1]^{2x} \,.
\end{align*}
As the result we arrive at
\begin{multline*}
	\Lambda_3(x)\,\Psi(z_1,z_2) = a^2(x)\,[z_{1+}]^{g-s}\,
	[z_{1-}]^{1-s-g} \\
	\times
	\mathcal{R}_{32}(-x)\,\mathcal{R}_{21}(-x)\,
	[\hat{p}_1]^{-x-s}\,
	\mathcal{R}_{32}(x)\,\mathcal{R}_{21}(x)\,
	[\hat{p}_1]^{2x}\,
	[z_{1+}]^{x-g}\,
	[z_{1-}]^{x+g-1}\,\Psi(z_{2}\,,z_3) \,.
\end{multline*}
Then applying the operator $[\hat{p}_1]^{2x}$ to the function $[z_{1+}]^{x-g}\,
[z_{1-}]^{x+g-1}$ by means of \eqref{p1z1z1} and using \eqref{key2} again
\begin{equation} \nonumber
	\mathcal{R}_{32}(x)\,\mathcal{R}_{21}(x)\,\Psi(z_2,z_3) =
	a^2(x)\,\mathcal{R}_{12}(-x)\,\mathcal{R}_{23}(-x)\,\Psi(z_1,z_2)
\end{equation}
we get the desired relation \eqref{-x} for $n=3$
\begin{align*}
	&\Lambda_3( x)\,\Psi(z_1,z_2) \\
	& = [-2\imath\gamma]^{2x}\,a^4(x)\,\frac{\mathbf{\Gamma}(g+x)}{\mathbf{\Gamma}(g-x)}\,
	\mathcal{R}_{32}(-x)\,\mathcal{R}_{21}(-x)\,\mathcal{K}_1(-x)\, \mathcal{R}_{12}(-x)\,\mathcal{R}_{23}(-x)\,\Psi(z_1,z_2) \\ 
	&
	= \frac{[-2\imath\gamma]^{x-s}\,c^4(x-s)\,\mathbf{\Gamma}(g+x)}{[-2\imath\gamma]^{-x-s}\,c^4(-x-s)\,\mathbf{\Gamma}(g-x)}\,
	\Lambda_3(-x)\,\Psi(z_1,z_2) \,.
\end{align*}

We hope that generalization to the case of arbitrary $n$ is clear.
Everything is based on the key relation  
\begin{align}\label{key}
	&\mathcal{R}_{12}(x)\cdots\mathcal{R}_{n-1\,,n}(x)\,\Psi(z_{1},\ldots,z_{n-1}) = 
	\\
	\nonumber
	&a^{n-1}(x)\,\mathcal{R}_{n\,,n-1}(-x)\cdots\mathcal{R}_{21}(-x)\,\Psi(z_{2},\ldots,z_n)
\end{align}
and the commutation relation \eqref{comm} in the form
\begin{align} \nonumber
	&\mathcal{R}_{n\,,n-1}(y)\cdots\mathcal{R}_{21}(y)\,[\hat{p}_1]^{y-s}\,
	\mathcal{R}_{n\,,n-1}(x)\cdots\mathcal{R}_{21}(x)\,[\hat{p}_1]^{x-s} = 
	\\
	\nonumber
	&\mathcal{R}_{n\,,n-1}(x)\cdots\mathcal{R}_{21}(x)\,[\hat{p}_1]^{x-s}\,
	\mathcal{R}_{n\,,n-1}(y)\cdots\mathcal{R}_{21}(y)\,[\hat{p}_1]^{y-s} \,.
\end{align}
The identity \eqref{comm} is proven in appendix. To prove \eqref{key} we rewrite both sides in an explicit integral form. We consider the the simplest example when $n=1$
\begin{align}\label{key1}
	\mathcal{R}_{12}(x)\,\Psi(z_{1}) = 
	a(x)\,\mathcal{R}_{21}(-x)\,\Psi(z_{2}) \,,
\end{align}
and the generalization to the case of arbitrary $n$ is straightforward. 

We have the following explicit expressions for 
integral operator in the left hand side of~\eqref{key1}
\begin{multline*}
	\mathcal{R}_{12}(x)\,\Psi(z_{1}) = 
	[z_{12}]^{1-2s}\,[\hat{p}_1]^{x-s}\,[z_{12}]^{s+x-1}\,\Psi(z_{1}) \\
	= \frac{c(x-s)}{[z_{12}]^{2s-1}}\,
	\int \mathrm{d}^2 w \,
	\frac{\Psi(w)}{[z_1-w]^{1-s+x} \, [w-z_2]^{1-s-x}} 
\end{multline*}
and for the integral operator in the right hand side
\begin{multline*} 
	\mathcal{R}_{21}(-x)\,\Psi(z_{2}) = 
	[z_{21}]^{1-2s}\,[\hat{p}_2]^{-x-s}\,[z_{21}]^{s-x-1}\,\Psi(z_{2}) \\
	= \frac{c(-x-s)}{[z_{21}]^{2s-1}}\,
	\int \mathrm{d}^2 w \,
	\frac{\Psi(w)}{[z_2-w]^{1-s-x} \, [w-z_1]^{1-s-x}}
\end{multline*}
The integral kernels are the same 
(the overall sign factors cancel) and 
there exists only one difference in the 
normalization factors which produces $a(x)$ in \eqref{key1}.

\subsection{The symmetry $s,\,g \to 1-s,\,1-g$}

As we mentioned in section~\ref{sect:Rep}, principal series representations corresponding to spins $(s,\bar{s})$ and $(1-s,1-\bar{s})$ are equivalent. In the case of A-type open spin chain this yields the simple transformation rule for eigenfunctions: the functions for spins $s$ and $1-s$ corresponding to the same eigenvalue differ by a product of power functions \cite[Section~4.3]{ADV1}.

In this section we derive a similar relation for open BC-type spin chain. In addition to spin $s$, this model has parameter $g$ of the same form, which also needs to be taken into consideration. To show the dependence of eigenfunctions on these parameters we use the notation $\Psi_{\bm{x}_n}(s,g;\bm{z}_n)$, and the same notations will be used for $B$-, $Q$- and $\Lambda$-operators. For brevity we do not display antiholomorphic parameters $\bar{s}, \bar{g}$ implying that, for example, the argument $s$ should be understood as $(s,\bar{s})$ and $1-s$ -- as $(1-s, 1-\bar{s})$.

It comes out that eigenfunctions corresponding to parameters $s,g$ and $1-s,1-g$ differ by a factor
\begin{equation} \label{Psi1-s}
	\Psi_{\bm{x}_n}(1-s,1-g;\bm{z}_n) = c(\bm{x}_n)\,S_n^{-1}\Psi_{\bm{x}_n}(s,g;\bm{z}_n) \,.
\end{equation}
Here $S_n$ is the operator of multiplication by the function
\begin{align}\label{Sn1}
	S_n = [z_{12}]^{1-2s}\,[z_{23}]^{1-2s} 
	\cdots [z_{n-1\,n}]^{1-2s}\,[z_{1+}]^{g-s}\,
	[z_{1-}]^{1-g-s} \,,
\end{align} 
where $z_{1\pm} = z_1\pm\gamma$, and the coefficient $c(\bm{x}_n)$ has the form
\begin{align}
	\label{cx}
	c(\bm{x}_n) & = [\imath]^{2sn}\,[-1]^{\underline{\bm{x}_n}+ns}\,\mathbf{\Gamma}(s,\pm\bm{x}_n)\,
	\prod\limits_{k=1}^{n-1} q^{-1}(1-s;\bm{x}_k) \\
	\nonumber
	& =
	[\imath]^{2ns}\,[-1]^{\underline{\bm{x}_n}+(2-n)s}
	\prod\limits_{k=1}^{n}\mathbf{\Gamma}^{2(n-k)+1}(s\pm x_k) \,,
\end{align}
where we use the notations \eqref{pm}, \eqref{nota1} for gamma functions.
This property follows from the fact that operators corresponding to parameters $s,g$ and $1-s, 1-g$ are connected by similarity transformations
\begin{align}\label{BS1}
&S^{-1}_n B_n(s,g;u)\, S_n = B_n(1-s,1-g;u) \,, \\ 
\label{SQ1}
&S_n\,Q_n(1-s,1-g; x)\,S^{-1}_{n} = Q_n(s,g; x)\,Q^{-1}_{n}(s,g;1-s) \,, \\
& \label{SL1}
\begin{aligned} 
	S_n\,\Lambda_n(1-s,1-g;x)\,S^{-1}_{n-1}\,\Psi(\bm{z}_{n-1}) & \\
	= [\imath]^{2s}\,[-1]^{x+s} &\,\mathbf{\Gamma}(s\pm x)\,
	\Lambda_n(s,g;x)\,Q^{-1}_{n-1}(s,g;1-s)\,\Psi(\bm{z}_{n-1}) \,,
\end{aligned}
\end{align}
where $\Psi(\bm{z}_{n-1})$ is an arbitrary function which does not depend on $z_n$. There are two things worth noting. First, there exists the antiholomorphic analog of relation \eqref{BS1}, which has the same form (and can be proven in the similar way), one just needs to replace operator $B_n$ with its antiholomorphic counterpart $\bar{B}_n$ which depends on parameters $\bar{s}, \bar{g}, \bar{u}$. Second, for $n=1$ the equation \eqref{SL1} degenerates into
\begin{equation} \label{SL1deg}
	S_1\,\Lambda_{1}(1-s,1-g;x)\cdot 1 =
	[\imath]^{2s}\,[-1]^{x+s}\,\Lambda_1(s,g;x)\cdot 1 \,.
\end{equation}

One may come to relation \eqref{Psi1-s} for eigenfunctions by the following argumentation. Since $\Psi_{\bm{x}_n}(s,g;\bm{z}_n)$ is an eigenfunction of the operator $B_n(s,g;u)$, it follows from the identity \eqref{BS1} that $S_n^{-1}\Psi_{\bm{x}_n}(s,g;\bm{z}_n)$ is the eigenfunction of $B_n(1-s,1-g;u)$ corresponding to the same eigenvalue. Therefore, $S_n^{-1}\Psi_{\bm{x}_n}(s,g;\bm{z}_n)$ should coincide with $\Psi_{\bm{x}_n}(1-s,1-g;\bm{z}_n)$ up to some constant $c(\bm{x}_n)$.

We prove \eqref{Psi1-s} by induction.
For $n=1$ we have $\Psi_{x_1}(s,g;z_1)=\Lambda_1(s,g;x_1)\cdot 1$, and in this case \eqref{Psi1-s} is equivalent to \eqref{SL1deg}. Using the latter relation one finds the constant $c(x_1) = [\imath]^{2s}\,[-1]^{x_1+s}\,\mathbf{\Gamma}(s\pm x_1)$.
For the induction step $n-1\to n$ we need the recurrent formula for eigenfunctions 
\begin{equation} \nonumber
	\Psi_{\bm{x}_n}(1-s,1-g;\bm{z}_n) =
	\Lambda_n(1-s,1-g;x_n)\,\Psi_{\bm{x}_{n-1}}(1-s,1-g;\bm{z}_{n-1}) \,.
\end{equation}
We act by $S_n$ on both sides and insert $S_{n-1}^{-1}S_{n-1}$ between $\Lambda_n$ and $\Psi_{\bm{x}_{n-1}}$
\begin{equation} \nonumber
	S_n\,\Psi_{\bm{x}_n}(1-s,1-g;\bm{z}_n) =
	S_n\,\Lambda_n(1-s,1-g;x_n)\,S_{n-1}^{-1}S_{n-1}\,\Psi_{\bm{x}_{n-1}}(1-s,1-g;\bm{z}_{n-1}) \,.
\end{equation}
Then one can use the induction proposition
\begin{equation*}
	S_{n-1}\,\Psi_{\bm{x}_{n-1}}(1-s,1-g;\bm{z}_{n-1}) =
	c(\bm{x}_{n-1})\,\Psi_{\bm{x}_{n-1}}(s,g;\bm{z}_{n-1})
\end{equation*}
and the formula \eqref{SL1} together with the eigenvalue equation \eqref{QPsi} for $Q$-operator
\begin{multline} \nonumber
	S_n\,\Lambda_n(1-s,1-g;x_n)\,S^{-1}_{n-1}\,\Psi_{\bm{x}_{n-1}}(s,g;\bm{z}_{n-1}) \\
	= [\imath]^{2s}\,[-1]^{x_n+s} \,\mathbf{\Gamma}(s\pm x_n)\,
	\Lambda_n(s,g;x_n)\,Q^{-1}_{n-1}(s,g;1-s)\,\Psi_{\bm{x}_{n-1}}(s,g;\bm{z}_{n-1}) \\
	= [\imath]^{2s}\,[-1]^{x_n+s} \,\mathbf{\Gamma}(s\pm x_n)\,q^{-1}(1-s;\bm{x}_{n-1})\,\Psi_{\bm{x}_n}(s,g;\bm{z}_{n-1}) \,.
\end{multline}
As the result we obtain the needed identity
\begin{equation} \nonumber
	S_n\Psi_{\bm{x}_n}(1-s,1-g;\bm{z}_n) = c(\bm{x}_n)\,\Psi_{\bm{x}_n}(s,g;\bm{z}_n) \,,
\end{equation}
where
\begin{equation} \label{crec}
	c(\bm{x}_n) = [\imath]^{2s}\,[-1]^{x_n+s} \,\mathbf{\Gamma}(s\pm x_n)\,q^{-1}(1-s;\bm{x}_{n-1})\,c(\bm{x}_{n-1}) \,.
\end{equation}
Solving the recurrence relation \eqref{crec} for $c(\bm{x}_n)$ we obtain the explicit expression \eqref{cx} for this coefficient.

Now we turn to the derivation of the formula \eqref{BS1} for similarity transformation of $B$-operator, let us write it again for convenience
\begin{equation} \label{BS2}
	S^{-1}_n\,B_n(s,g;u)\,S_n = B_n(1-s,1-g;u) \,.
\end{equation}
We recall that this operator is the element $(1,2)$ of the monodromy matrix~\eqref{Top}
\begin{align}
	\label{T1}
	T_n(s,g;u) & = L_n(u_1\,,u_2) \cdots L_1(u_1\,,u_2)\,K(g;u)\,L_1(u_1\,,u_2) \cdots L_n(u_1\,,u_2) \\
	\label{T2}
	& =
	\begin{pmatrix}
		A_n(s,g;u) & B_n(s,g;u) \\
		C_n(s,g;u) & D_n(s,g;u)
	\end{pmatrix} ,
\end{align}
where $u_1 \,, u_2$ are two alternative parameters of the $L$-operator, which are connected to $u, s$ by the following change of variables
\begin{equation} \label{u12}
	u_1 = u+s-1 \,, \quad u_2 = u-s \,.
\end{equation}
Formula \eqref{BS2} is the consequence of the following matrix relation
\begin{equation} \label{TS}
	S^{-1}_n\,T_n(s,g;u)\,S_n = 
	(u_1\,u_2)^{-1}
	\left(%
	\begin{array}{cc}
		u_1 & 0 \\
		z_n\,u_{12} & u_2 \\
	\end{array}%
	\right)
	T_n(1-s,1-g;u)
	\left(%
	\begin{array}{cc}
		u_1 & 0 \\
		z_n\,u_{12} & u_2 \\
	\end{array}%
	\right) ,
\end{equation}
where $u_{12}=u_1-u_2$. Indeed, substituting the expression \eqref{T2} for monodromy matrix into the last identity and comparing matrix elements $(1,2)$ in both sides one obtains \eqref{BS2}.

Thus, it remains to derive \eqref{TS} and we
will apply a method similar to that used in~\cite{BDM}.
We will use the factorized expression for the $L$-operator
\begin{equation} \label{Lfact1}
	L(u_1\,,u_2) =
	\left(%
	\begin{array}{cc}
		1 & 0 \\
		z & u_2 \\
	\end{array}%
	\right)\left(%
	\begin{array}{cc}
		1 & -\partial \\
		0 & 1 \\
	\end{array}%
	\right)\left(%
	\begin{array}{cc}
		u_1 & 0 \\
		-z & 1 \\
	\end{array}%
	\right)\,,
\end{equation}
which is equivalent to \eqref{Lfact}.
For our purposes it is also convenient to represent the operator $S_n$ as $S_n = S_{0\,n}\,S_{\pm\, 1}$, where
\begin{align*}
	S_{0\,n} = [z_{12}]^{1-2s}\,[z_{23}]^{1-2s} 
	\cdots [z_{n-1\,n}]^{1-2s} \,, \qquad
	S_{\pm\, 1} = [z_{1+}]^{g-s}\,
	[z_{1-}]^{1-g-s} \,.
\end{align*}

Let us start from the case $n=1$.
By direct calculation one can verify that
\begin{multline} \label{TS1}
	S^{-1}_1\,T_1(s,g;u)\,S_1 \equiv
	S_{\pm\, 1}^{-1}\,L_1(u_1,u_2)\,K(g;u)\,L_1(u_1,u_2)\,S_{\pm\, 1} \\
	 = L_1(u_2,u_2)\,K(1-g;u)\,L_1(u_1,u_1) \,.
\end{multline}
Now we need to rewrite the last expression in \eqref{TS1} in terms of $T_1(1-s,1-g;u)$. First, one can see that in the right hand side of \eqref{TS1} the parameter $g$ of the reflection matrix changed to $1-g$. According to the definition \eqref{T1} and the explicit expression \eqref{u12} for $u_1\,,u_2$, the interchange $s\leftrightarrows 1-s$ in the monodromy matrix is equivalent to the permutation of arguments $(u_1,u_2)\to (u_2,u_1)$ in every $L$-operator, so that
\begin{equation} \nonumber
	T_1(1-s,1-g;u) = L_1(u_2,u_1)\,K(1-g;u)\,L_1(u_2,u_1) \,.
\end{equation}
And to obtain this matrix in the right hand side of \eqref{TS1} we rewrite the $L$-operators using the following relations
\begin{align}\label{u1u2}
	L(u_2\,,u_2) = \frac{1}{u_1}
	\left(%
	\begin{array}{cc}
		u_1 & 0 \\
		z\,u_{12} & u_2 \\
	\end{array}%
	\right)\,L(u_2\,,u_1) \,, \quad
	L(u_1\,,u_1) =
	\frac{1}{u_2}\,L(u_2\,,u_1)\,
	\left(%
	\begin{array}{cc}
		u_1 & 0 \\
		z\,u_{12} & u_2 \\
	\end{array}%
	\right) ,
\end{align}
which are direct consequences of the factorization \eqref{Lfact1}. This way, we get the needed identity \eqref{TS} for $n=1$
\begin{equation} \nonumber
	S^{-1}_1\,T_1(s,g;u)\,S_1 = 
	(u_1\,u_2)^{-1}
	\left(%
	\begin{array}{cc}
		u_1 & 0 \\
		z_1\,u_{12} & u_2 \\
	\end{array}%
	\right)
	T_1(1-s,1-g;u)
	\left(%
	\begin{array}{cc}
		u_1 & 0 \\
		z_1\,u_{12} & u_2 \\
	\end{array}%
	\right) .
\end{equation}

In the case of arbitrary $n$ we have
\begin{equation} \label{TS2}
	S^{-1}_n\,T_n(s,g;u)\,S_n =
	S_{0\,n}^{-1}\,S_{\pm\, 1}^{-1}\,T_n(s,g;u)\,S_{\pm\, 1}\,S_{0\,n}\,.
\end{equation}
Again, our aim is to express the right hand side of the last expression in terms of the matrix
\begin{equation} \label{T3}
	T_n(1-s,1-g;u) = L_n(u_2\,,u_1) \cdots L_1(u_2\,,u_1)\,K(1-g;u)\, L_1(u_2\,,u_1) \cdots L_n(u_2\,,u_1) \,.
\end{equation}
First, to replace $g$ with $1-g$ we conjugate $T_n(s,g;u)$ by $S_{\pm\, 1} = [z_{1+}]^{g-s}\,
[z_{1-}]^{1-g-s}$. Since this function depends only on $z_1$, everything reduces to the similarity transformation in the middle of the expression for monodromy matrix
\begin{equation} \nonumber
	S_{\pm\, 1}^{-1}\,L_1(u_1\,,u_2) K(g;u) L_1(u_1\,,u_2)\,S_{\pm\, 1} =
	L_1(u_2,u_2)\,K(1-g;u)\,L_1(u_1,u_1) \,,
\end{equation}
where we have used the formula \eqref{TS1}. Thus \eqref{TS2} takes the form
\begin{multline} \label{TnConj}
	S^{-1}_n\,T_n(s,g;u)\,S_n =
	S_{0\,n}^{-1}\,L_n(u_1\,,u_2) \cdots L_2(u_1\,,u_2)\,L_1(u_2\,,u_2) \\
	\times K(1-g;u)\,L_1(u_1\,,u_1)\,L_2(u_1\,,u_2) \cdots L_n(u_1\,,u_2)\,S_{0\,n}\,.
\end{multline}

To obtain the matrix \eqref{T3} in the right hand side of \eqref{TnConj} it remains to make the arguments of every $L$-operator equal to $(u_2\,,u_1)$.
We achieve this by commuting $S_{0\,n}^{-1}$ with $L$-operators using the relation~\cite[Section~3.1]{DM09}
\begin{equation} \label{zLL}
	[z_{ij}]^{\alpha_1-\beta_2}\,L_i(\alpha_1,\alpha_2)\,L_j(\beta_1,\beta_2) =
	L_i(\beta_2,\alpha_2)\,L_j(\beta_1,\alpha_1)\,[z_{ij}]^{\alpha_1-\beta_2} \,,
\end{equation}
which holds true for arbitrary values of parameters $\alpha_1\,,\alpha_2$, $\beta_1\,,\beta_2$. First we commute $S_{0\,n}^{-1}=[z_{12}]^{2s-1} 
\cdots [z_{n-1\,n}]^{2s-1}$ with the product $L_n(u_1\,,u_2) \cdots L_2(u_1\,,u_2)\,L_1(u_2\,,u_2)$. Since $u_1-u_2=2s-1$, we can apply \eqref{zLL} in the form
\begin{equation} \nonumber
	[z_{i\,i+1}]^{2s-1}\,L_{i+1}(u_1\,,u_2)\,L_{i}(u_2\,,u_2) =
	L_{i+1}(u_2\,,u_2)\,L_{i}(u_2\,,u_1)\,[z_{i\,i+1}]^{2s-1}
\end{equation}
first for $i=1$, then for $i=2$, and so on till $i=n-1$. As the result, we obtain
\begin{equation} \nonumber
	S_{0\,n}^{-1}\,L_n(u_1\,,u_2) \cdots L_2(u_1\,,u_2)\,L_1(u_2\,,u_2) =
	L_n(u_2\,,u_2)\,L_{n-1}(u_2\,,u_1) \cdots\,L_1(u_2\,,u_1)\,S_{0\,n}^{-1} \,.
\end{equation}
Since $K(1-g;u)$ is a numerical matrix, it commutes with $S_{0\,n}^{-1}$. So next we need to commute $S_{0\,n}^{-1}$ with the second product of $L$-operators from \eqref{TnConj} using the same relation~\eqref{zLL}
\begin{equation} \nonumber
	S_{0\,n}^{-1}\,L_1(u_1\,,u_1)\,L_2(u_1\,,u_2) \cdots L_n(u_1\,,u_2) =
	L_1(u_2\,,u_1) \cdots L_{n-1}(u_2\,,u_1)\, L_n(u_1\,,u_1)\,	S_{0\,n}^{-1} \,.
\end{equation}

Collecting everything together we obtain from \eqref{TnConj}
\begin{multline} \nonumber
	S^{-1}_n\,T_n(s,g;u)\,S_n =
	L_n(u_2\,,u_2)\,L_{n-1}(u_2\,,u_1) \cdots\,L_1(u_2\,,u_1) \\
	\times K(1-g;u)\,L_1(u_2\,,u_1) \cdots L_{n-1}(u_2\,,u_1)\, L_n(u_1\,,u_1)\,.
\end{multline}
The right hand side of this relation is almost the matrix $T_n(1-s,1-g;u)$, and rewriting the leftmost and rightmost $L$-operators by means of \eqref{u1u2}
\begin{align} \nonumber
	L_n(u_2\,,u_2) = \frac{1}{u_1}
	\left(%
	\begin{array}{cc}
		u_1 & 0 \\
		z_n\,u_{12} & u_2 \\
	\end{array}%
	\right)\,L_n(u_2\,,u_1) \,, \quad
	L_n(u_1\,,u_1) =
	\frac{1}{u_2}\,L_n(u_2\,,u_1)\,
	\left(%
	\begin{array}{cc}
		u_1 & 0 \\
		z_n\,u_{12} & u_2 \\
	\end{array}%
	\right)
\end{align}
we finally derive the desired relation \eqref{TS}.

Next we move on to the relation \eqref{SQ1} for similarity transformation of $Q$-operator and prove it in an equivalent form 
\begin{align} \label{SQ2}
S_n\,Q_n(1-s,1-g; x) = Q_n(s,g; x)\,
\left(S^{-1}_{n}\,Q_{n}(s,g;1-s)\right)^{-1}
\end{align}
by induction.
In the case $n=1$ we have two equivalent representations for $Q$-operator 
\begin{align}
\label{Qrep}
Q_1(s,g;x) = \mathcal{K}(s,g;x)\,[\hat{p}]^{x-s} = 
[z_+]^{g-s}\,
[z_-]^{1-s-g}\,
[\hat{p}]^{x-s}\,
[z_+]^{x-g}\,
[z_-]^{x+g-1}\,[\hat{p}]^{x-s} \\ 
\nonumber
= [\hat{p}]^{x+s-1}\,
[z_+]^{g+x-1}\,
[z_-]^{x-g}\,[\hat{p}]^{x-s}\,
[z_+]^{1-s-g}\,
[z_-]^{g-s}\,[\hat{p}]^{1-2s}
\end{align}
The first explicit expression follows directly from the formula \eqref{Kop-d} for reflection operator, and using the star-triangle relation \eqref{star-tr} one can obtain the second expression.
Replacing $s,g$ with $1-s,1-g$ in the first representation for the $Q$-operator one obtains
\begin{align*}
Q_1(1-s,1-g;x) = 
[z_+]^{s-g}\,
[z_-]^{s+g-1}\,
[\hat{p}]^{x+s-1}\,
[z_+]^{x+g-1}\,
[z_-]^{x-g}\,[\hat{p}]^{x+s-1} \,.
\end{align*}
Multiplying this identity from the left by $S_1=[z_+]^{g-s}\,
[z_-]^{1-s-g}$ we get the following explicit formula for the left hand side of \eqref{SQ2}
\begin{equation} \label{SQ3}
S_1\,Q_1(1-s,1-g; x) =
[\hat{p}]^{x+s-1}\,
[z_+]^{x+g-1}\,
[z_-]^{x-g}\,[\hat{p}]^{x+s-1} \,.
\end{equation}
It remains to show that right hand sides of \eqref{SQ2} and \eqref{SQ3} are equal.
The right hand side of \eqref{SQ3} almost coincides with $x$-dependent part in \eqref{Qrep}, so we easily extract the needed $Q$-operator
\begin{align} \label{pzzp}
[\hat{p}]^{x+s-1}\,
[z_+]^{x+g-1}\,
[z_-]^{x-g}\,[\hat{p}]^{x+s-1} =  
Q_1(s,g;x)\,
[\hat{p}]^{2s-1}\,
[z_+]^{s+g-1}\,
[z_-]^{s-g}\,[\hat{p}]^{2s-1} \,.
\end{align}
Then using the formula \eqref{Qrep} for $x=1-s$ we rewrite the remaining product in terms of $Q$-operator
\begin{equation} \nonumber
	[\hat{p}]^{2s-1}\,
	[z_+]^{s+g-1}\,
	[z_-]^{s-g}\,[\hat{p}]^{2s-1} = \left(S_n^{-1}\,Q(s,g;1-s)\right)^{-1}
\end{equation}
and finally reduce \eqref{SQ3} to the right hand side of \eqref{SQ2}, thus proving \eqref{SQ2} for $n=1$.

In the case $n\geq 2$ the $Q$-operator has the form
\begin{align*}
Q_n(s,g;x) = 
\mathcal{R}_{n\,n-1}(s;x)\cdots 
\mathcal{R}_{21}(s;x)
\mathcal{K}_1(s,g;x)\,
\mathcal{R}_{12}(s;x)
\cdots \mathcal{R}_{n-1\,n}(s;x)\,[\hat{p}_n]^{x-s} 
\end{align*}
To prove the relation \eqref{SQ2} we are going to rewrite it in the explicit form. Since the expression for $Q$-operator contains $\mathcal{R}$- and $\mathcal{K}$-operators, we use the following formulae which relate these operators for spins $s$ and $1-s$
\begin{align}
\label{KK}
&[z_{1+}]^{g-s}\,
[z_{1-}]^{1-s-g}\,\mathcal{K}_1(1-s,1-g;x) = 
\mathcal{K}_1(s,g;x)\,
[\hat{p}_1]^{x+s-1}\,
[z_{1+}]^{s+g-1}\,
[z_{1-}]^{s-g}\,[\hat{p}_1]^{s-x} \,,\\ 
\label{Rk}
&[z_{k+1\,k}]^{1-2s}\,\mathcal{R}_{k+1\,k}(1-s;x)\,[\hat{p}_{k+1}]^{1-2s} = \mathcal{R}_{k+1\,k}(s;x)\,,\\
\label{R(1-s)}
&\mathcal{R}_{k+1\,k}(s;1-s) = [z_{k+1\,k}]^{1-2s}\,[\hat{p}_{k+1}]^{1-2s} \,.
\end{align}
To obtain \eqref{KK} one needs to substitute into \eqref{SQ3} the expression \eqref{pzzp} for the right hand side, and then express $Q$-operators in terms of reflection operators: $Q_1(1-s,1-g; x)=\mathcal{K}_1(1-s,1-g;x)\,[\hat{p}_1]^{x+s-1}$, $Q_1(s,g; x)=\mathcal{K}_1(s,g;x)\,[\hat{p}_1]^{x-s}$. The formula \eqref{Rk} is the direct consequence of the explicit expression \eqref{R1} for $\mathcal{R}$-operator and the star-triangle relation \eqref{star-tr}. And \eqref{R(1-s)} is the particular case of \eqref{R1}.
Using (\ref{KK}--\ref{R(1-s)}) one can derive relations
\begin{multline*}
	S_n\,  
	\mathcal{R}_{n\,n-1}(1-s;x)\cdots 
	\mathcal{R}_{21}(1-s;x)
	\mathcal{K}_1(1-s,1-g;x) = \\
	\mathcal{R}_{n\,n-1}(s;x)\cdots 
	\mathcal{R}_{21}(s;x)
	\mathcal{K}_1(s,g;x)\,
	\left([\hat{p}_{n}]\cdots[\hat{p}_{2}]\right)^{2s-1}\,
	[\hat{p}_1]^{x+s-1}\,[z_{1+}]^{s+g-1}\,
	[z_{1-}]^{s-g}\,[\hat{p}_1]^{s-x} 
\end{multline*}
and 
\begin{multline*}
	S^{-1}_n\,  
	\mathcal{R}_{n\,n-1}(s;1-s)\cdots 
	\mathcal{R}_{21}(s;1-s)
	\mathcal{K}_1(s,g;1-s) = 
	\left([\hat{p}_{n}]\cdots[\hat{p}_{1}]\right)^{1-2s}
	\,[z_{1+}]^{1-s-g}\,[z_{1-}]^{g-s} \,,
\end{multline*}
which allow to reduce the main identity \eqref{SQ2}
\begin{align*}
S_n\,Q_n(1-s,1-g; x) = Q_n(s,g; x)\,
\left(S^{-1}_{n}\,Q_{n}(s,g;1-s)\right)^{-1}
\end{align*}
to the following equivalent and explicit form 
\begin{align} 
& \nonumber
\begin{aligned}
	& \left([\hat{p}_{n}]\cdots[\hat{p}_{2}]\right)^{2s-1}\,
	[\hat{p}_1]^{x+s-1}\,[z_{1+}]^{s+g-1}\,
	[z_{1-}]^{s-g}\,[\hat{p}_1]^{s-x} \, \\
	& \times \mathcal{R}_{12}(1-s;x)
	\cdots \mathcal{R}_{n-1\,n}(1-s;x)\,[\hat{p}_n]^{x+s-1}
\end{aligned}
\\
& \label{equiv1}
\begin{aligned}
	= \mathcal{R}_{12}(s;x)
	\cdots \mathcal{R}_{n-1\,n}(s;x)\,[\hat{p}_n]^{x+s-1}\,
	\mathcal{R}^{-1}_{n-1\,n}(s;1-s)\cdots
	\mathcal{R}^{-1}_{12}(s;1-s)\, \\ 
	\times [z_{1+}]^{s+g-1}\,[z_{1-}]^{s-g}
	\left([\hat{p}_{n}]\cdots[\hat{p}_{1}]\right)^{2s-1} \,.
\end{aligned}
\end{align}

Next we apply the commutation relation \eqref{comm} to transform the product of $\mathcal{R}$-operators in the right hand side of \eqref{equiv1}
\begin{multline} \label{RpRinv}
	\mathcal{R}_{12}(s;x) \cdots \mathcal{R}_{n-1\,n}(s;x)\,[\hat{p}_n]^{x+s-1}\,
	\mathcal{R}^{-1}_{n-1\,n}(s;1-s)\cdots \mathcal{R}^{-1}_{12}(s;1-s) \\
	= [\hat{p}_n]^{2s-1}\,
	\mathcal{R}^{-1}_{n-1\,n}(s;1-s)\cdots \mathcal{R}^{-1}_{12}(s;1-s)\,
	\mathcal{R}_{12}(s;x) \cdots \mathcal{R}_{n-1\,n}(s;x)\,[\hat{p}_n]^{x-s} \\
	= ([\hat{p}_n]\ldots[\hat{p}_1])^{2s-1}([z_{12}]\ldots [z_{n-1\,n}])^{2s-1}
	\mathcal{R}_{12}(s;x) \cdots \mathcal{R}_{n-1\,n}(s;x)\,[\hat{p}_n]^{x-s} \,,
\end{multline}
where on the last step we used the explicit formula for $\mathcal{R}$-operators
\begin{equation} \nonumber
	\mathcal{R}_{k\,k+1}(s;1-s) = [z_{k\,k+1}]^{1-2s}[\hat{p}_k]^{1-2s} \,.
\end{equation}
After substituting \eqref{RpRinv} into \eqref{equiv1} we cancel the product $\left([\hat{p}_{n}]\cdots[\hat{p}_{1}]\right)^{2s-1}$ from the left in both sides of this identity, and the factor $[\hat{p}_n]^{x+s-1}$ cancels from the right. As the result, \eqref{equiv1} reduces to
\begin{multline} \label{equiv2}
	[\hat{p}_1]^{x-s}\,[z_{1+}]^{s+g-1}\,
	[z_{1-}]^{s-g}\,[\hat{p}_1]^{s-x}\,
	\mathcal{R}_{12}(1-s;x)
	\cdots \mathcal{R}_{n-1\,n}(1-s;x) = \\   
	([z_{12}]\ldots [z_{n-1\,n}])^{2s-1}
	\mathcal{R}_{12}(s;x) \cdots \mathcal{R}_{n-1\,n}(s;x)\,
	[z_{1+}]^{s+g-1}\,[z_{1-}]^{s-g}
	\left([\hat{p}_{n-1}]\cdots[\hat{p}_{1}]\right)^{2s-1}\,.
\end{multline}
For $n=2$ the relation \eqref{equiv2} becomes trivial with both sides equal to
\begin{equation} \nonumber
	[\hat{p}_1]^{x-s}\,[z_{1+}]^{s+g-1}\,
	[z_{1-}]^{s-g}\,[z_{12}]^{x+s-1}\,[\hat{p}_1]^{2s-1} \,,
\end{equation}
to see this one just needs to rewrite $\mathcal{R}$-operators with the help of \eqref{R1}
\begin{equation} \nonumber
	\mathcal{R}_{12}(1-s;x) = [\hat{p}_1]^{x-s}\,[z_{12}]^{x+s-1}\,[\hat{p}_1]^{2s-1} \,,
	\quad
	\mathcal{R}_{12}(s;x) = [z_{12}]^{1-2s}[\hat{p}_1]^{x-s}[z_{12}]^{x+s-1} \,.
\end{equation}
If $n>2$, we use the analogue of \eqref{Rk}
\begin{equation} \nonumber
	[z_{n-1\,n}]^{2s-1}\,\mathcal{R}_{n-1\,n}(s;x)\,[\hat{p}_{n-1}]^{2s-1} = \mathcal{R}_{n-1\,n}(1-s;x)
\end{equation}
in the right hand side of \eqref{equiv2} and cancel the factors $\mathcal{R}_{n-1\,n}(1-s;x)$ in both sides of this identity. Thus, \eqref{equiv2} transforms into the similar relation of smaller rank with $n$ replaced by $n-1$
\begin{multline} \nonumber
	[\hat{p}_1]^{x-s}\,[z_{1+}]^{s+g-1}\,
	[z_{1-}]^{s-g}\,[\hat{p}_1]^{s-x}\,
	\mathcal{R}_{12}(1-s;x)
	\cdots \mathcal{R}_{n-2\,n-1}(1-s;x) = \\   
	([z_{12}]\ldots [z_{n-2\,n-1}])^{2s-1}
	\mathcal{R}_{12}(s;x) \cdots \mathcal{R}_{n-2\,n-1}(s;x)\,
	[z_{1+}]^{s+g-1}\,[z_{1-}]^{s-g}
	\left([\hat{p}_{n-2}]\cdots[\hat{p}_{1}]\right)^{2s-1}\,.
\end{multline}
In that way, one can reduce \eqref{equiv2} for arbitrary $n$ to the trivial relation for $n=2$, and we have finally proven \eqref{SQ1}.

The relation \eqref{SL1} can be derived by reduction from the 
relation \eqref{SQ1}. 
Starting from
\begin{align} \label{SLder}
S_n\,Q_n(1-s,1-g; x)\,S^{-1}_{n} = Q_n(s,g; x)\,Q^{-1}_{n}(s,g;1-s)
\end{align}
we rewrite $S_n$ in terms of $S_{n-1}$ with the help of explicit formula \eqref{Sn1}
\begin{equation} \nonumber
	S_n = [z_{n-1\,n}]^{1-2s}\,S_{n-1} \,,
\end{equation}
then express two $Q$-operators in terms of $\Lambda$-operator
\begin{equation} \nonumber
	Q_n(1-s,1-g; x)=\Lambda_n(1-s,1-g; x)\,[\hat{p}_n]^{x+s-1}\,, \quad
	Q_n(s,g; x)=\Lambda_n(s,g; x)\,[\hat{p}_n]^{x-s}
\end{equation}
and apply the recurrent formula \eqref{QnQn-1} for the third $Q$-operator
\begin{align*}
	& Q_{n}(s,g;1-s) =
	\mathcal{R}_{n\,,n-1}(s;1-s)\, Q_{n-1}(s;1-s)\,[\hat{p}_{n-1}]^{2s-1}\,
	\mathcal{R}_{n-1\,n}(s;1-s)\,[\hat{p}_n]^{1-2s} \\
	& = [z_{n\,n-1}]^{1-2s}\,[\hat{p}_{n}]^{1-2s}\,Q_{n-1}(s,g;1-s)\,
	[\hat{p}_{n-1}]^{2s-1}\,[z_{n-1\,n}]^{1-2s}\,
	[\hat{p}_{n-1}]^{1-2s}\,[\hat{p}_n]^{1-2s} \,.
\end{align*}
This way, \eqref{SLder} takes the form
\begin{align*}
&S_n\,\Lambda_n(1-s,1-g; x)\,S^{-1}_{n-1}\,[\hat{p}_n]^{x+s-1}\,[z_{n-1\,n}]^{2s-1} = \\
&\Lambda_n(s,g; x)\,[\hat{p}_n]^{x+s-1}\,[\hat{p}_{n-1}]^{2s-1}\,
[z_{n-1\,n}]^{2s-1}\,[\hat{p}_{n-1}]^{1-2s}\,
Q^{-1}_{n-1}(s,g;1-s)\,[\hat{p}_{n}]^{2s-1}\,
[z_{n\,n-1}]^{2s-1} \,.
\end{align*}
After some evident transformations of this operator identity and application to the function $\Psi(\bm{z}_{n-1})$ which does not depend on the variable $z_n$ we arrive at
\begin{align}
\nonumber
&S_n\,\Lambda_n(1-s,1-g; x)\,S^{-1}_{n-1}\,\Psi(\bm z_{n-1}) \\
\label{SLder1}
& = \Lambda_n(s,g; x)\,[\hat{p}_{n-1}]^{2s-1}\,
{\blue [\hat{p}_n]^{x+s-1}\,[z_{n\,n-1}]^{2s-1}\,[\hat{p}_{n}]^{s-x}}\,
[\hat{p}_{n-1}]^{1-2s}\,Q^{-1}_{n-1}(s,g;1-s)\,\Psi(\bm z_{n-1}) \,. 
\end{align}
Since $\Psi(\bm{z}_{n-1})$ is independent of $z_n$, we can calculate the action of marked operator in the space of functions of variable $z_n$ using \eqref{star-tr1}
\begin{align*}
[\hat{p}_n]^{x+s-1}\,[z_{n\,n-1}]^{2s-1}\,[\hat{p}_{n}]^{s-x}\cdot 1 = 
\frac{[\imath]^{2s-1}\,[-1]^{x-s}}{\mathbf{\Gamma}(1-s-x,1-s+x)} 
= [\imath]^{2s}\,[-1]^{x+s}\,\mathbf{\Gamma}(s\pm x) \,.
\end{align*}
As the result, the operators $[\hat{p}_{n-1}]^{2s-1}$ and $[\hat{p}_{n-1}]^{1-2s}$ in the right hand side of \eqref{SLder1} cancel out, and this identity transforms into the relation \eqref{SL1} 
\begin{multline} \nonumber
	S_n\,\Lambda_n(1-s,1-g; x)\,S^{-1}_{n-1}\,\Psi(\bm z_{n-1}) \\
	= [\imath]^{2s}[-1]^{x+s}\,\mathbf{\Gamma}(s\pm x)\,\Lambda_n(s,g; x)\,Q^{-1}_{n-1}(s,g;1-s)\,\Psi(\bm z_{n-1}) \,,
\end{multline}
which we aimed to prove.

\subsection{Symmetric normalization}
\label{sect:SymmNorm}

In Section~\ref{sect:Symmetry} we discussed the symmetry of eigenfunctions under permutations and reflections of eigenvalue's parameters
\begin{align*} 
	\Psi_{\tau \bm x_n}(\bm z_n) = 
	c(\tau,\bm x_n)\,\Psi_{\bm x_n}(\bm z_n)\, ; \\ 
	\Psi_{\sigma_k \bm x_n}(\bm z_n) = 
	c_k(\bm x_n)\,\Psi_{\bm x_n}(\bm z_n)\,,
\end{align*}
where $\tau \bm x_n = (x_{\tau(1)},\bar{x}_{\tau(1)}, \ldots , x_{\tau(n)},\bar{x}_{\tau(n)})$, $\sigma_k \bm x_n = (x_{1},\bar{x}_1,\ldots , -x_{k},-\bar{x}_{k}, 
\ldots , x_{n},\bar{x}_{n})$ and $c(\tau,\bm x_n), c_k(\bm x_n)$ are some constants. It is natural to choose the normalization of eigenfunctions, such that $c(\tau,\bm x_n)=1, \, c_k(\bm x_n) = 1$.
Do do so, hereinafter we use the new normalization for $\Lambda$-operators
\begin{align*}
	&\Lambda_k(x)= \lambda_k(x)\,\mathcal{R}_{k\,k-1}(x)\,
	\mathcal{R}_{k-1\, k-2}(x)
	\ldots \mathcal{R}_{2 1}(x)\,
	\mathcal{K}_1(s,x)\,
	\mathcal{R}_{1 2}(x) \, \mathcal{R}_{2 3}(x) \ldots 
	\mathcal{R}_{k-1\, k}(x)\,, \\ 
	&\Lambda_1(x) = \lambda_1(x)\,\mathcal{K}_1(s,x)
\end{align*} 
where 
\begin{equation} \label{lambda}
	\lambda_k(x) = \pi^{4k-3}\,[-2\imath\gamma]^{s-x}\,c^{2(k-1)}(-x-s)\,\mathbf{\Gamma}(g-x) \,.
\end{equation}
That is, we multiplied the operator $\Lambda_k(x)$ by the normalization factor $\lambda_k(x)$. The choice of this constant is dictated by the commutation relation \eqref{exchLL1} between $\Lambda$-operators and the reflection symmetry \eqref{-x}. In new normalization these relations take the form
\begin{equation} \nonumber
	\Lambda_n(x)\,\Lambda_{n-1}(y) =  
	\Lambda_n(y)\,\Lambda_{n-1}(x) \,, \qquad
	\Lambda_n(x) = \Lambda_n(-x) \,.
\end{equation}
Consequently, from the definition
\begin{equation} \nonumber
	\Psi_{\bm{x}_n}(\bm z_n) =
	\Lambda_n( x_n)\,
	\Lambda_{n-1}( x_{n-1})\cdots 
	\Lambda_{2}( x_{2})\,\Lambda_{1}\cdot 1
\end{equation}
follows the complete symmetry of eigenfunctions under permutations of $x_1, \ldots, x_n$ and reflections $x_k \to -x_k$
\begin{align*} 
	\Psi_{\tau \bm x_n}(\bm z_n) = 
	\Psi_{\bm x_n}(\bm z_n)\, , \qquad
	\Psi_{\sigma_k \bm x_n}(\bm z_n) = 
	\Psi_{\bm x_n}(\bm z_n)\,.
\end{align*}

\section{Orthogonality}
\label{ort}

In this section we will prove that the eigenfunctions 
$\Psi_{\bm{x}_n}(\bm{z}_n)$ form the orthogonal set with respect 
to the standard scalar product
\begin{equation} \label{orth}
	\langle\Psi_{\bm{y}_n}|\Psi_{\bm{x}_n}\rangle = 
\int \mathrm{d}^2\bm{z}_n \, 
\overline{\Psi_{\bm{y}_n}(\bm{z}_n)} \, \Psi_{\bm{x}_n}(\bm{z}_n) = 
	\mu^{-1}(\bm{x}_n)\, \delta^{(2)}(\bm{x}_n,\bm{y}_n) \,,
\end{equation}
where we recall that $\mathrm{d}^2\bm{z}_n = \prod_{k=1}^n \mathrm{d}^2 z_k$.
The Sklyanin measure $\mu(\bm{x}_n)$ is given by the  
following expression 
\begin{align}\label{mu0}
	\mu(\bm{x}_n) =
	\frac{\prod\limits_{k=1}^n|x_k|^2 \prod\limits_{1\leq i<j\leq n}|x_i^2-x_j^2|^2}{4^n\,\pi^{2n(n+1)}\,n!\,|\gamma|^{2n}} \,,
\end{align}
where $|\alpha|$ is the absolute value of the complex number $\alpha$.

The symmetric delta function $\delta^{(2)}(\bm{x}_n,\bm{y}_n)$ is 
defined in the following way
\begin{align}\label{deltaSym}
	\delta^{(2)}(\bm{x}_n,\bm{y}_n) =
	\frac{1}{2^n\,n!}
	\sum\limits_{
		\begin{smallmatrix}
			\tau\in\mathfrak{S}_n \\ \sigma_1,\ldots,\sigma_n=\pm 1
	\end{smallmatrix}}
	\delta^{(2)}(x_1-\sigma_1\,y_{\tau(1)}) \ldots
	\delta^{(2)}(x_n-\sigma_n\,y_{\tau(n)}) \,,
\end{align}
where $\mathfrak{S}_n$ denotes the symmetric group of degree $n$.
The parameters $x_i, \bar{x}_i, y_j, \bar{y}_j$ have the form~\eqref{x}
\begin{equation*}
	x_i = \tfrac{n_i}{2}+\imath\nu_i, \; 
	\bar{x}_i = -\tfrac{n_i}{2}+\imath\nu_i\,,
	\quad
	y_j=\tfrac{m_j}{2}+\imath\eta_j, \;
	\bar{y}_j=-\tfrac{m_j}{2}+\imath\eta_j \,,
\end{equation*}
where $n_i, m_j \in \mathbb{Z}+\sigma$, and numbers $\nu_i, \eta_j$ are real.
For the pair
\begin{equation*}
	(v,\bar{v}) = \left(\frac{m}{2}+\imath\rho, -\frac{m}{2}+\imath\rho\right) ,
	\qquad m\in\mathbb{Z}, \; \rho\in\mathbb{R}
\end{equation*}
the delta function $\delta^{(2)}(v)$ is defined as
\begin{equation*}
	\delta^{(2)}(v) = \delta_{m,0}\,\delta(\rho) \,,
\end{equation*}
here $\delta_{m,0}$ is the Kronecker symbol and $\delta(\rho)$ is the ordinary one-dimensional delta function.
This way, in the formula \eqref{orth} we have
\begin{equation} \nonumber
	\delta^{(2)}(x_k-\sigma_k\,y_{\tau(k)}) =
	\delta_{n_k,\sigma_k m_{\tau(k)}}\,\delta(\nu_k-\sigma_k\eta_{\tau(k)}) \,.
\end{equation}
First of all we should note that the general form of expression in the 
right hand side of \eqref{orth} is dictated by the expected 
orthogonality of eigenfunctions and its symmetry properties.

As we discussed in Section~\ref{sect:SymmNorm}, eigenfunctions $\Psi_{\bm{x}_n}$ are symmetric with respect to permutations of parameters $x_1,\ldots,x_n$ and reflections $x_k\to -x_k$. Consequently, the scalar product $\langle\Psi_{\bm{y}_n}|\Psi_{\bm{x}_n}\rangle$ is also invariant under such transformations of sets $\bm{y}_n$ and $\bm{x}_n$.  
The sum in \eqref{deltaSym} is the average of the function  
$\delta^{(2)}(x_1-y_{1}) \ldots \delta^{(2)}(x_n-y_{n})$ over the action of the whole group of transformations generated by permutations of $y_1,\ldots,y_n$ and reflections $y_k\to -y_k$. As the result, $\delta^{(2)}(\bm{x}_n,\bm{y}_n)$ is symmetric under the mentioned transformations of the set $\bm{y}_n$ and automatically symmetric with respect to similar transformations of the set $\bm{x}_n$.
Of course the function $\mu(\bm{x}_n)$ must be symmetric but the explicit expression for the function $\mu(\bm{x}_n)$ can not be fixed by the symmetry arguments.

The expression \eqref{mu0} is obtained by direct calculations.
To prove the formulae \eqref{orth} and \eqref{mu0} we 
calculate the scalar product $\langle\Psi_{\bm{y}_n}|\Psi_{\bm{x}_n}\rangle$ 
assuming the conditions 
\begin{equation*}
	y_i \neq \pm x_j \,, \, i \neq j, \quad y_k \neq -x_k \,,
\end{equation*}
and obtain the following result
\begin{align}
	\label{orthIncomplete}
	\langle\Psi_{\bm{y}_n}|\Psi_{\bm{x}_n}\rangle &
	= \frac{2^n\,\pi^{2n(n+1)}\,|\gamma|^{2n}}{\prod\limits_{k=1}^n|x_k|^2 \prod\limits_{1\leq i<j\leq n}|x_i^2-x_j^2|^2}
	\,\delta^{(2)}(x_1-y_1) \ldots \delta^{(2)}(x_n-y_n) \\
	& \nonumber
	= \mu^{-1}(\bm{x}_n)\,\frac{1}{2^n\,n!}\,\delta^{(2)}(x_1-y_1) \ldots \delta^{(2)}(x_n-y_n) \,.
\end{align}
The full expression for the scalar product can be reconstructed 
in a unique way using the symmetry.

The expression \eqref{orthIncomplete} follows from the recurrence relation
\begin{equation} \label{SPrecurOpen}
	\langle\Psi_{\bm{y}_n}|\Psi_{\bm{x}_n}\rangle =
\delta^{(2)}(x_n-y_n)\,
\frac{2\pi^{4n}\,|\gamma|^2}
{|x_n|^2 \prod_{k=1}^{n-1}|x_n^2-x_k^2|^2}
\,\langle\Psi_{\bm{y}_{n-1}}|\Psi_{\bm{x}_{n-1}}\rangle
\end{equation}
which is derived assuming the restrictions 
\begin{equation*}
	y_n \neq \pm x_i \,, \, i = 1,2,\ldots,n-1, \quad y_n \neq -x_n \,.
\end{equation*}
The role of initial condition plays the expression for the 
simplest scalar product 
\begin{equation} \label{SPbaseOpen}
	\langle\Psi_{y_1}|\Psi_{x_1}\rangle =
	\frac{2\pi^4\,|\gamma|^2}{|x_1|^2}\,\delta^{(2)}(x_1-y_1) \,,
\end{equation}
which is valid by condition that $y_1 \neq -x_1$.

Now we are going to the derivation of \eqref{SPrecurOpen}. First we use the recurrence relation for eigenfunctions
\begin{equation} \nonumber
	\Psi_{\bm{y}_n}(\bm{z}_n) = \Lambda_n(y_n)\,\Psi_{\bm{y}_{n-1}}(\bm{z}_{n-1}) \,.
\end{equation}
and obtain
\begin{equation} \nonumber
	\langle\Psi_{\bm{y}_n}|\Psi_{\bm{x}_n}\rangle =
	\langle\Lambda_n(y_n)\,\Psi_{\bm{y}_{n-1}}|\Psi_{\bm{x}_n}\rangle =
	\langle\Psi_{\bm{y}_{n-1}}|\Lambda_n^\dagger(\bm{y}_n)\Psi_{\bm{x}_n}\rangle
\end{equation}
Next, comparing formulas for $\Lambda$ and $Q$ operators
\begin{align}
	\label{LamExpl}
	\Lambda_n( x) & = \lambda_n(x)\,\mathcal{R}_{n\,,n-1}(x)\cdots 
	\mathcal{R}_{21}(x)\,\mathcal{K}_{1}(x)\,
	\mathcal{R}_{12}(x)\cdots\mathcal{R}_{n-1\,n}(x) \,, \\
	\label{QExpl}
	Q_n( x) & = \mathcal{R}_{n\,,n-1}(x)\cdots 
	\mathcal{R}_{21}(x)\,\mathcal{K}_{1}(x)\,
	\mathcal{R}_{12}(x)\cdots\mathcal{R}_{n-1\,n}(x)\,
	[\hat{p}_n]^{x-s}
\end{align}
we express $\Lambda_n(\bm{y}_n)$ in terms of $Q_n(\bm{y}_n)$
\begin{equation} \nonumber
	\Lambda_n(y_n) = \lambda_n(y_n)\,Q_n(y_n)\,[\hat{p}_n]^{s-y_n} \,.
\end{equation}
Applying the conjugation rules \eqref{Qconj} and \eqref{conj} to operators $Q_n(y_n)$ and $[\hat{p}_n]^{s-y_n}$ we arrive at
\begin{align}
	\nonumber
	\langle\Psi_{\bm{y}_n}|\Psi_{\bm{x}_n}\rangle & =
	\lambda_n^\ast(y_n)\,
	\langle\Psi_{\bm{y}_{n-1}}|\,[\hat{p}_n]^{1-s+y_n}Q_n^{-1}(y_n+1)\Psi_{\bm{x}_n}\rangle \\
	& \label{PsiyPsix}
	= \frac{\lambda_n^\ast(y_n)}{q(y_n+1;\bm{x}_n)}\,
	\langle\Psi_{\bm{y}_{n-1}}|\,[\hat{p}_n]^{1-s+y_n}\Psi_{\bm{x}_n}\rangle \,,
\end{align}
where on the last step we used the fact that $\Psi_{\bm{x}_n}$ is an eigenfunction of the $Q$-operator, the $q$-function denotes the corresponding eigenvalue, see equation \eqref{QPsi}.
Now we employ the recurrence formula for $\Psi_{\bm{x}_n}$ in \eqref{PsiyPsix}
\begin{equation} \nonumber
	\Psi_{\bm{x}_n}(\bm{z}_n) = \Lambda_n(x_n)\,\Psi_{\bm{x}_{n-1}}(\bm{z}_{n-1}) \,.
\end{equation}
and rewrite $\Lambda_n(x_n)$ with the help of \eqref{LamExpl}, \eqref{QExpl} in the following form
\begin{equation*}
	\Lambda_n(x_n) = \lambda_n(x_n)\, \mathcal{R}_{n\,n-1}(x_n)\,Q_{n-1}(x_n)\,
[\hat{p}_{n-1}]^{s-x_n}\,\mathcal{R}_{n-1\,n}(x_n)
\end{equation*}
getting
\begin{multline}\label{PsiyPsix1}
\langle\Psi_{\bm{y}_n}|\Psi_{\bm{x}_n}\rangle 
= \frac{\lambda_n^\ast(y_n)\,\lambda_n(x_n)}{q(y_n+1;\bm{x}_n)}\,\\ \langle\Psi_{\bm{y}_{n-1}}|\,[\hat{p}_n]^{1-s+y_n}
\mathcal{R}_{n\,n-1}(x_n)\,Q_{n-1}(x_n)\,[\hat{p}_{n-1}]^{s-x_n}\,
\mathcal{R}_{n-1\,n}(x_n)\,\Psi_{\bm{x}_{n-1}}\rangle
\end{multline}
Using the explicit formula for $\mathcal{R}$-operator we obtain
\begin{equation} \nonumber
	[\hat{p}_n]^{1-s+y_n}\mathcal{R}_{n\,n-1}(x_n) =
	[\hat{p}_n]^{y_n+x_n}[z_{n\,n-1}]^{x_n-s}[\hat{p}_n]^{-s-y_n}[\hat{p}_n]^{1-s+y_n} \,.
\end{equation}
After substituting this expression into \eqref{PsiyPsix1} we perform the following two transformations. First, we move the operator $[\hat{p}_n]^{y_n+x_n}[z_{n\,n-1}]^{x_n-s}[\hat{p}_n]^{-s-y_n}$ to the first argument of scalar product and act by the corresponding hermitian conjugate operator on $\Psi_{\bm{y}_{n-1}}$ using \eqref{star-tr1} since $\Psi_{\bm{y}_{n-1}}$ does not depend on $z_n$
\begin{align}
	& \nonumber
	\left([\hat{p}_n]^{y_n+x_n}[z_{n\,n-1}]^{x_n-s}[\hat{p}_n]^{-s-y_n}\right)^\dagger \Psi_{\bm{y}_{n-1}} =
	[\hat{p}_n]^{y_n+s-1}[z_{n\,n-1}]^{s-x_n-1}[\hat{p}_n]^{-y_n-x_n}\Psi_{\bm{y}_{n-1}} \\
	& \label{pzpPsi}
	= \frac{[\imath]^{s-x_n}\,[-1]^{y_n+x_n}}{\mathbf{\Gamma}(1-s-y_n,1+y_n+x_n)}\,\Psi_{\bm{y}_{n-1}}\,.
\end{align}
Second, we commute the operator $[\hat{p}_n]^{1-s+y_n}$ with $Q_{n-1}(x_n)\,[\hat{p}_{n-1}]^{s-x_n}$ in \eqref{PsiyPsix1} and act by it on the function $\mathcal{R}_{n-1\,n}(x_n)\,\Psi_{\bm{x}_{n-1}}$
\begin{align}
	& \nonumber
	[\hat{p}_n]^{1-s+y_n}\mathcal{R}_{n-1\,n}(x_n)\,\Psi_{\bm{x}_{n-1}} =
	[\hat{p}_{n-1}]^{x_n+s-1}[\hat{p}_n]^{1-s+y_n}[z_{n-1\,n}]^{x_n-s}[\hat{p}_{n-1}]^{1-2s}\,\Psi_{\bm{x}_{n-1}} \\
	& \nonumber
	= \frac{[\imath]^{y_n-s}\,[-1]^{y_n-s}}{\mathbf{\Gamma}(s-x_n,x_n-y_n)}\,
	[\hat{p}_{n-1}]^{x_n+s-1}[z_{n\,n-1}]^{x_n-y_n-1}[\hat{p}_{n-1}]^{1-2s}\,\Psi_{\bm{x}_{n-1}} \,.
\end{align}
Since $\Psi_{\bm{x}_{n-1}}$ does not depend on $z_n$ everything is reduced to the action of $[\hat{p}_n]^{1-s+y_n}$ on the power function $[z_{n-1\,n}]^{x_n-s}$ which is calculated using the formula \eqref{Chain1}.
After all these transformations \eqref{PsiyPsix1} takes the form
\begin{multline} \label{PsiyPsix2}
	\langle\Psi_{\bm{y}_n}|\Psi_{\bm{x}_n}\rangle  = 
\frac{[-1]^{y_n-s}\,\mathbf{\Gamma}(s+x_n)}{\mathbf{\Gamma}(s+y_n)}\,
\frac{[\imath]^{x_n+y_n-2s}\lambda_n^\ast(y_n)\,\lambda_n(x_n)}
{|x_n+y_n|^2\,q(y_n+1;\bm{x}_{n-1})} \\
\langle\Psi_{\bm{y}_{n-1}}|\,Q_{n-1}(x_n)\,
[\hat{p}_{n-1}]^{2s-1}[z_{n\,n-1}]^{x_n-y_n-1}
[\hat{p}_{n-1}]^{1-2s}\,\Psi_{\bm{x}_{n-1}}\rangle \,.
\end{multline}
The resulting coefficient in \eqref{PsiyPsix2} is obtained as follows.
We used \eqref{gamma-conj} to find the complex conjugate of the coefficient from \eqref{pzpPsi} and then extract all factors depending on $x_n$ from 
$q(y_n+1;\bm{x}_{n})$.
Note that coefficient in the right hand side of \eqref{PsiyPsix2} 
is well defined under conditions $y_n \neq -x_n$ and 
$y_n \neq \pm x_k$ for $k=1,\ldots,n-1$.

In the last scalar product only $[z_{n\,n-1}]^{x_n-y_n-1}$ depend on $z_n$, therefore we can integrate over this variable
\begin{equation} \nonumber
	\int \mathrm{d}^2z_n\,[z_{n\,n-1}]^{x_n-y_n-1} = 2\pi^2\,\delta^{(2)}(x_n-y_n) \,.
\end{equation}
As the result, the operators $[\hat{p}_{n-1}]^{2s-1}$ and $[\hat{p}_{n-1}]^{1-2s}$ cancel out and we obtain
\begin{align}
	& \nonumber
	\langle\Psi_{\bm{y}_{n-1}}|\,Q_{n-1}(x_n)\,[\hat{p}_{n-1}]^{2s-1}[z_{n\,n-1}]^{x_n-y_n-1}[\hat{p}_{n-1}]^{1-2s}\,\Psi_{\bm{x}_{n-1}}\rangle \\
	& \nonumber
	= 2\pi^2\,\delta^{(2)}(x_n-y_n) \,
	\langle\Psi_{\bm{y}_{n-1}}|\,Q_{n-1}(x_n)\,\Psi_{\bm{x}_{n-1}}\rangle \,,
\end{align}
where the scalar product in the right hand side is already the integral over $n-1$ variables $z_1, \ldots, z_{n-1}$.
Applying $Q_{n-1}(x_n)$ to its eigenfunction $\Psi_{\bm{x}_{n-1}}$ we get almost the recurrence relation \eqref{SPrecurOpen} which we aimed to derive
\begin{align*}
\langle\Psi_{\bm{y}_n}|\Psi_{\bm{x}_n}\rangle = 
2\pi^2\,\delta^{(2)}(x_n-y_n)\,
\frac{\lambda_n^\ast(y_n)\,\lambda_n(x_n)\,q(x_n;\bm{x}_{n-1})}
{|x_n+y_n|^2\,q(y_n+1;\bm{x}_{n-1})} \,
\langle\Psi_{\bm{y}_{n-1}}|\Psi_{\bm{x}_{n-1}}\rangle \,.
\end{align*}
It remains to transform the fraction in front of the scalar product substituting explicit expressions \eqref{qn}, \eqref{lambda} for functions $q$, $\lambda_n$ and using \eqref{gamma-diff} together with reflection formula \eqref{gamma-refl} for the gamma function. 

To derive the basic expression \eqref{SPbaseOpen} for the scalar product of $\Psi_{y_1}$ and $\Psi_{x_1}$ we employ the explicit formula for these eigenfunctions 
($z_\pm=z\pm\gamma$)
\begin{equation} \label{Psi1expl}
	\Psi_{y_1}(z) = \lambda_1(y_1)\,\mathcal{K}(y_1)\cdot 1 = \lambda_1(y_1)\,[z_+]^{g-s}[z_-]^{1-s-g}[\hat{p}]^{y_1-s}[z_+]^{y_1-g}[z_-]^{y_1+g-1} \,.
\end{equation}
Thus,
\begin{align}
	\nonumber
	\langle\Psi_{y_1}|\Psi_{x_1}\rangle &
	= \lambda_1^\ast(y_1)\,
	\langle[z_+]^{g-s}[z_-]^{1-s-g}[\hat{p}]^{y_1-s}[z_+]^{y_1-g}[z_-]^{y_1+g-1}|\Psi_{x_1}\rangle \\
	& \nonumber
	= \lambda_1^\ast(y_1)\,
	\langle[z_-]^{y_1+g-1}|\,([z_+]^{g-s}[z_-]^{1-s-g}[\hat{p}]^{y_1-s}[z_+]^{y_1-g})^\dagger\,\Psi_{x_1}\rangle \\
	& \nonumber
	= \lambda_1^\ast(y_1)\,\lambda_1(x_1)\,
	\langle[z_-]^{y_1+g-1}|\,[z_+]^{g-y_1-1}[\hat{p}]^{x_1-y_1-1}[z_+]^{x_1-g}[z_-]^{x_1+g-1}\rangle \,,
\end{align}
at the last step we used the hermitian conjugation rules \eqref{conj}, \eqref{zconj} for $[\hat{p}]^\alpha$ and $[z_\pm]^\alpha$, expressed $\Psi_{x_1}$ by \eqref{Psi1expl} and canceled some operator factors. Next we apply the star-triangle relation in the form \eqref{star-tr}
\begin{equation} \nonumber
	[z_+]^{g-y_1-1}[\hat{p}]^{x_1-y_1-1}[z_+]^{x_1-g} = 
	[\hat{p}]^{x_1-g}[z_+]^{x_1-y_1-1}[\hat{p}]^{g-y_1-1}\,.
\end{equation}
and move the operator $[\hat{p}]^{x_1-g}$ to the first argument of the scalar product
\begin{equation} \label{Psiy1Psix12}
	\langle\Psi_{y_1}|\Psi_{x_1}\rangle =
	\lambda_1^\ast(y_1)\,\lambda_1(x_1)\,
	\langle[\hat{p}]^{g-x_1-1}[z_-]^{y_1+g-1}|\,[z_+]^{x_1-y_1-1}[\hat{p}]^{g-y_1-1}[z_-]^{x_1+g-1}\rangle \,.
\end{equation}
The chain relation \eqref{Chain1} gives 
\begin{equation} \label{pz-}
[\hat{p}]^{g-x_1-1}[z_-]^{y_1+g-1} = \frac{[-1]^{x_1+y_1}\,[\imath]^{g-x_1}}
{\mathbf{\Gamma}(1-g-y_1,y_1+x_1+1)}\,[z_-]^{y_1+x_1} \,
\end{equation}
and the corresponding formula for $[\hat{p}]^{g-y_1-1}[z_-]^{x_1+g-1}$ differs from \eqref{pz-} by transposition of $x_1$ and $y_1$.  After substituting these expressions into \eqref{Psiy1Psix12}, all powers of $z_-$ cancel, so we get
\begin{equation} \label{Psiy1Psix13}
	\langle\Psi_{y_1}|\Psi_{x_1}\rangle = \frac{[-1]^{x_1-g}[\imath]^{x_1-y_1}\lambda_1^\ast(y_1)\,\lambda_1(x_1)}
{\mathbf{\Gamma}(g+y_1,1-x_1-y_1,1-g-x_1,1+x_1+y_1)}\,
	\langle 1|\,[z_+]^{x_1-y_1-1}\rangle \,,
\end{equation}
where we additionally used \eqref{gamma-conj} to conjugate the coefficient from \eqref{pz-}. 
Note that coefficient in right hand side of \eqref{Psiy1Psix13}  
is well defined under condition $y_1 \neq -x_1$.

To obtain the \eqref{SPbaseOpen} from \eqref{Psiy1Psix13} it remains to calculate the scalar product in the right hand side
\begin{equation} \nonumber
	\langle 1|\,[z_+]^{x_1-y_1-1}\rangle =
	\int \mathrm{d}^2z\,[z+\gamma]^{x_1-y_1-1} = 2\pi^2\,\delta^{(2)}(x_1-y_1) \,,
\end{equation}
use the explicit formula \eqref{lambda} for $\lambda_1$ and transform the resulting expression with the help of relations \eqref{gamma-diff}, \eqref{gamma-refl} for gamma function.

\section{The Mellin-Barnes representation}
\label{sect:MB}

\subsection{Eigenfunctions for A-type spin chain}

Recall that the monodromy matrix for A-type open $SL(2,\mathbb{C})$ spin chain has the form
\begin{equation} \nonumber
	t_n(u) = L_1(u) \cdots L_n(u) = \begin{pmatrix}
		a_n(u) & b_n(u) \\
		c_n(u) & d_n(u)
	\end{pmatrix} .
\end{equation}
From commutation relations \eqref{tt} it follows that for arbitrary constant $z_0$ the operators
\begin{equation} \label{auz0}
	a_n(u, z_0) = a_n(u)+z_0\,b_n(u) \,, \quad
	\bar{a}_n(\bar{u},\bar{z}_0) = \bar{a}_n(\bar{u})+\bar{z}_0\,\bar{b}_n(\bar{u})
\end{equation}
form the commutative family
\begin{equation} \nonumber
	[a_n(u, z_0), a_n(v, z_0)]=0 \,, \quad
	[\bar{a}_n(\bar{u},\bar{z}_0), \bar{a}_n(\bar{v},\bar{z}_0)]=0 \,, \quad
	[a_n(u, z_0), \bar{a}_n(\bar{v},\bar{z}_0)]=0 \,.
\end{equation}
Therefore, these operators have a common system of eigenfunctions. This system was constructed in \cite{DM14}.
We will use this system for $z_0 = \gamma$, where $\gamma$ is the parameter of $K$-matrix \eqref{K}, to prove the completeness of the set $\{\Psi_{\bm{x}_n}\}$ of eigenfunctions for BC-type $SL(2,\mathbb{C})$ spin chain. 
That is, we will expand $\Psi_{\bm{x}_n}$ over the set of eigenfunctions of operators $a_n(u, \gamma), \bar{a}_n(\bar{u},\bar{\gamma})$ and use the completeness of this system, which was proven in~\cite{M}.
The reason to decompose $\Psi_{\bm{x}_n}$ over this basis is that the kernel of the resulting integral expansion is expressed in the closed form.

In this section we give the expression for eigenfunctions of
$a(u,z_0),\,\bar{a}_n(\bar{u},\bar{z}_0)$ and list all their needed properties.
The operator $a_n(u,z_0)$ is a polynomial of degree $n$ in $u$ \cite{DM14,ADV1}. Since eigenfunctions do not depend on $u$, the eigenvalues are also polynomials of degree $n$ in the spectral parameter. Similarly for $\bar{a}_n(\bar{u}, \bar{z}_0)$. The eigenfunctions $\psi_{\bm{x}_n}$ are labeled by zeroes of these polynomials
\begin{align}
	& \nonumber
	a_n(u, z_0)\,\psi_{\bm{x}_n}(\bm{z}_n) = (u-x_1)\ldots(u-x_n)\,\psi_{\bm{x}_n}(\bm{z}_n) \,, \\
	& \nonumber
	\bar{a}_n(\bar{u},\bar{z}_0)\,\psi_{\bm{x}_n}(\bm{z}_n) =
	(\bar{u}-\bar{x}_1)\ldots(\bar{u}-\bar{x}_n)\,\psi_{\bm{x}_n}(\bm{z}_n) \,,
\end{align}
where $\bm{z}_n$ and $\bm{x}_n$ are the same as in \eqref{xn} and \eqref{x}.
Similarly to the eigenfunctions $\Psi_{\bm{x}_n}$ of BC-type spin chain, $\psi_{\bm{x}_n}$ are constructed iteratively by the use of raising integral operators
\begin{equation} \label{LamCl}
	\widetilde{\Lambda}_k(x) =
	\widetilde{\lambda}_k(x) \,
	\mathcal{R}_{12}(x)\,\mathcal{R}_{23}(x)\ldots\mathcal{R}_{k-1\,k}(x)\,[z_k-z_0]^{-s-x} \,, \quad
	\widetilde{\Lambda}_1(x) = [z_1-z_0]^{-s-x}
\end{equation}
where $\widetilde{\lambda}_k(x)$ is the normalizing coefficient of the form
\begin{equation} \label{lambdaCl}
	\widetilde{\lambda}_k(x) =
	\left(\frac{\pi\,[\imath]^{x-s}}{\mathbf{\Gamma}(s+x)}\right)^{k-1} \,.
\end{equation}
To obtain the eigenfunction one needs to apply the raising operators to the function identically equal to $1$
\begin{equation} \label{reprPsiCl}
	\psi_{\bm{x}_n}(\bm{z}_n) = 
	\widetilde{\Lambda}_n(x_n)\ldots\widetilde{\Lambda}_2(x_2)\,\widetilde{\Lambda}_1(x_1)\cdot 1 \,.
\end{equation}
In the considered normalization $\widetilde{\Lambda}$-operators obey the exchange relation \cite{ADV1,DM14}
\begin{equation} \nonumber
	\widetilde{\Lambda}_k(x)\,\widetilde{\Lambda}_{k-1}(y)\,\Psi(\bm{z}_{k-2}) =
	\widetilde{\Lambda}_k(y)\,\widetilde{\Lambda}_{k-1}(x)\,\Psi(\bm{z}_{k-2}) \,,
\end{equation}
where $\Psi(\bm{z}_{k-2})$ is an arbitrary function depending on variables $z_1,\ldots,z_{k-2}$. In accordance with \eqref{reprPsiCl}, this identity yields the symmetry of $\psi_{\bm{x}_n}$ under permutations of spectral variables
\begin{equation} \nonumber
	\psi_{\tau \bm x_n}(\bm z_n) = 
	\psi_{\bm x_n}(\bm z_n) \,,
\end{equation}
where $\tau \bm x_n = (x_{\tau(1)},\bar{x}_{\tau(1)}, \ldots , x_{\tau(n)},\bar{x}_{\tau(n)})$.

The complete set consists of eigenfunctions $\psi_{\bm{x}_n}$ labeled by parameters of the type~\eqref{x}
\begin{equation} \nonumber
	x_k = \tfrac{n_k}{2}+\imath\nu_k, \; 
	\bar{x}_k = -\tfrac{n_k}{2}+\imath\nu_k\,, \qquad
	n_k \in \mathbb{Z}+\sigma \,, \; \nu_k \in \mathbb{R} \,,
\end{equation}
where the quantity $\sigma$ was introduced in \eqref{sigma}.
The completeness relation reads \cite{M}
\begin{equation} \label{complCl}
	\int \mathcal{D}\bm{x}_n \, \tilde{\mu}(\bm{x}_n) \,
	\psi_{\bm{x}_n}(\bm{z}_n) \, \overline{\psi_{\bm{x}_n}(\bm{w}_n)} =
	\delta^2(z_1-w_1) \ldots \delta^2(z_n-w_n) \,,
\end{equation}
where by $\int \mathcal{D}\bm{x}_n$ we denote the following integral
\begin{equation} \label{intx}
	\int \mathcal{D}\bm{x}_n = \prod\limits_{k=1}^{n}\int\mathcal{D}x_k, \qquad
	\int\mathcal{D}x_k = \sum\limits_{n_k\in\mathbb{Z}+\sigma}\int\limits_{\mathbb{R}}\mathrm{d}\nu_k \,,
\end{equation}
the measure of integration is of the form
\begin{equation} \label{mu}
	\tilde{\mu}(\bm{x}_n) =
	\frac{1}{2^n\,\pi^{n(n+1)}\,n!}
	\prod\limits_{1\leq i<j\leq n}|x_i-x_j|^2 \,,
\end{equation}
and $\delta^2(z)$ is the two-dimensional delta function in the complex plane
\begin{equation} \nonumber
	\delta^2(z) = \delta(\Re z)\,\delta(\Im z) \,.
\end{equation}
In the Dirac notations \eqref{complCl} takes the form
\begin{equation} \label{complClDirac}
	\int \mathcal{D}\bm{x}_n \, \tilde{\mu}(\bm{x}_n) \,
	|\psi_{\bm{x}_n}\rangle \langle\psi_{\bm{x}_n}| = \II \,,
\end{equation}
where $\II$ is the identity operator.

We will also need the $Q$-operator for A-type spin chain
\begin{eqnarray}\label{Qcl}
	\widetilde{Q}_n(x) =
	\mathcal{R}_{12}(x)\,\mathcal{R}_{23}(x)\cdots
	\mathcal{R}_{n-1\,n}(x)\,\mathcal{R}_{n 0}(x)\,.
\end{eqnarray}
The following properties of $\widetilde{Q}_n(x)$ will be used \cite{ADV1,DKM}:
\begin{itemize}
	\item its hermitian conjugation can be expressed in terms of its inverse
	\begin{align*}
		\widetilde{Q}^{\dagger}_n(x) = \widetilde{Q}^{-1}_n(1-\bar{x}^*) \,,
	\end{align*}
	\item it forms commutative family 
	\begin{eqnarray} \nonumber
		\widetilde{Q}_n(x)\,\widetilde{Q}_n(y) = \widetilde{Q}_n(y)\,\widetilde{Q}_n(x) \,,
	\end{eqnarray}
	\item
	functions $\psi_{\bm{y}_n}$ form the set of common eigenfunctions of $a_n(u,z_0), \bar{a}_n(\bar{u},\bar{z}_0)$ and $\widetilde{Q}_n$
	\begin{align} \nonumber 
		\widetilde{Q}_n(u)\,\psi_{\bm x_n}(\bm z_n)
		= \tilde{q}(u\,,\bm x_{n})
		\, \psi_{\bm x_n}(\bm z_n) \,,
	\end{align}
	where the eigenvalue of $Q$-operator reads
	\begin{align} \label{qnCl}
		\tilde{q}(u\,,\bm x_{n}) = [\imath]^{(s-u)n}\,\prod_{k=1}^n\,
		\frac{\mathbf{\Gamma}(u-x_k)}{\mathbf{\Gamma}(s-x_k)} \,.
	\end{align}
	This property is the consequence of the fact that $\widetilde{Q}_n$ commutes with operators $a_n(u,z_0)$ and 
	$\bar{a}_n(\bar{u},\bar{z}_0)$
	\begin{eqnarray} \nonumber
		\widetilde{Q}_n(x)\, a(u,z_0) &=& a(u,z_0) \, \widetilde{Q}_n(x)\,,\\
		\nonumber  
		\widetilde{Q}_n(x)\, \bar{a}(\bar{u},\bar{z}_0) &=& 
		\bar{a}(\bar{u},\bar{z}_0) \, \widetilde{Q}_n(x) \,.
	\end{eqnarray}
\end{itemize}

\subsection{Decomposition over the basis of eigenfunctions for A-type spin chain} \label{sect:Decomp}

In the following we fix $z_0=\gamma$.
Denote by $\Pi_n$ the operator which reverses the order of variables
\begin{equation} \nonumber
\Pi_n\Psi(z_1,\ldots,z_n) = \Psi(z_n,\ldots,z_1) \,
\end{equation} 
and by $\hat{\psi}_{\bm{y}_n}$ the function which is obtained from 
$\psi_{\bm{y}_n}$ by application of $\Pi_n$
\begin{align*}
\hat{\psi}_{\bm{y}_n}(\bm{z}_n) = \left[\Pi_n\psi_{\bm{y}_n}\right](\bm{z}_n) = 
\psi_{\bm{y}_n}(z_n,\ldots,z_1)
\end{align*}
The common eigenfunctions $\psi_{\bm{y}_n}(\bm{z}_n)$ of operators $a_n(u,\gamma), \bar{a}_n(\bar{u},\bar{\gamma})$ \eqref{auz0} were introduced in the previous section, they have the form
\begin{equation} \nonumber
	\psi_{\bm{x}_n}(\bm{z}_n) = 
	\widetilde{\Lambda}_n(x_n)\ldots\widetilde{\Lambda}_2(x_2)\,\widetilde{\Lambda}_1(x_1)\cdot 1 \,,
\end{equation}
where
\begin{equation*}
	\widetilde{\Lambda}_k(x) =
	\widetilde{\lambda}_k(x) \,
	\mathcal{R}_{12}(x)\,\mathcal{R}_{23}(x)\ldots\mathcal{R}_{k-1\,k}(x)\,
[z_{k}-\gamma]^{-s-x} \,, \quad
	\widetilde{\Lambda}_1(x) = [z_{1}-\gamma]^{-s-x}
\end{equation*}
and $\widetilde{\lambda}_k$ is defined in \eqref{lambdaCl}.

In this section we decompose the eigenfunctions $\Psi_{\bm{x}_n}(\bm{z}_n)$ for BC-type spin chain over the basis $\{\hat{\psi}_{\bm{y}_n}(\bm{z}_n)\}$.
The formula for the decomposition reads
\begin{equation} \label{PsiMB}
	\Psi_{\bm{x}_n}(\bm{z}_n) =
	\lim\limits_{\varepsilon\to 0_+}
	\int\mathcal{D}\bm{y}_n\,\tilde{\mu}(\bm{y}_n)\,
	P_\varepsilon(\bm{y}_n,\bm{x}_n)\,
	\hat{\psi}_{\bm{y}_n}(\bm{z}_n) \,,
\end{equation}
where
\begin{equation} \label{Peps}
	P_\varepsilon(\bm{y}_n,\bm{x}_n) =
	\frac{\pi^{\tfrac{(3n+1)n}{2}}\,[-1]^{(n-1)ns}\,[-2\gamma]^{ns+\underline{\bm{y}_n}}\,\mathbf{\Gamma}^n(1-s,\pm\bm{x}_n)\,\mathbf{\Gamma}(\varepsilon-\bm{y}_n,\pm\bm{x}_n)}
	{\mathbf{\Gamma}^n(1-s,\bm{y}_n)\,\mathbf{\Gamma}(g,\bm{y}_n)\,\mathbf{\Gamma}(1-g,\pm\bm{x}_n)\,\prod\limits_{1\leq i<j\leq n}\mathbf{\Gamma}(-y_i-y_j)} \,.
\end{equation}
We use the following notations
\begin{equation} \label{nota}
	\underline{\bm{y}_n}=\sum\limits_{k=1}^n y_k \,, \quad
	\mathbf{\Gamma}(\bm{y}_m,\bm{x}_n) =
	\prod\limits_{i=1}^m\prod\limits_{j=1}^n \mathbf{\Gamma}(y_i-x_j) \,,
	\quad
	\mathbf{\Gamma}(\bm{y}_m, \pm\bm{x}_n) =
	\mathbf{\Gamma}(\bm{y}_m,\bm{x}_n)\,\mathbf{\Gamma}(\bm{y}_m,-\bm{x}_n)
\end{equation}
and $\varepsilon-\bm{y}_n=(\varepsilon-y_1,\varepsilon-\bar{y}_1,\ldots,\varepsilon-y_n,\varepsilon-\bar{y}_n)$. In particular,
\begin{equation} \label{nota1}
	\mathbf{\Gamma}(\lambda,\bm{y}_n) =
	\prod\limits_{k=1}^n \mathbf{\Gamma}(\lambda-y_k) \,, \quad
	\mathbf{\Gamma}(\lambda,\pm\bm{x}_n) =
	\prod\limits_{k=1}^n \mathbf{\Gamma}(\lambda+x_k)\,\mathbf{\Gamma}(\lambda-x_k) \,.
\end{equation}
The choice of $\varepsilon$-prescription is explained in section~\ref{sect:compl}.
It is related to the usability condition for complex analog \eqref{Gust} of BC-type Gustafson integral, which we use in the proof of completeness of the set $\{\Psi_{\bm{x}_n}\}$.

In order to derive \eqref{PsiMB} we use the formula
\begin{equation} \nonumber
\Psi_{\bm{x}_n}(\bm{z}_n) =
\int\mathcal{D}\bm{y}_n\,\tilde{\mu}(\bm{y}_n)\,
\langle\hat{\psi}_{\bm{y}_n}|\Psi_{\bm{x}_n}\rangle\,
\hat{\psi}_{\bm{y}_n}(\bm{z}_n) \,,
\end{equation}
which is obtained by inserting the resolution of unity \eqref{complClDirac}
\begin{equation*}
	\int \mathcal{D}\bm{y}_n \, \tilde{\mu}(\bm{y}_n) \,
	|\hat{\psi}_{\bm{y}_n}\rangle \langle\hat{\psi}_{\bm{y}_n}| = \II \,.
\end{equation*}
Therefore, to get \eqref{PsiMB} one needs to calculate the scalar product
\begin{equation}\label{P}
	P(\bm{y}_n,\bm{x}_n) = \langle\hat{\psi}_{\bm{y}_n}|\Psi_{\bm{x}_n}\rangle = 
\langle \Pi_n \psi_{\bm{y}_n}|\Psi_{\bm{x}_n}\rangle\,.
\end{equation}
It is computed with the help of the following recurrence relation
\begin{align}
	\nonumber
	P(\bm{y}_n,\bm{x}_n) & =
	\frac{\pi\,[-1]^{n(s+y_n)+s+g}\,[\imath]^{x_n-s}\,[2\gamma]^{x_n+y_n}\,
\left(a(x_n)\right)^{n-1}\,\lambda_n(x_n)\,\widetilde{\lambda}_n^\ast(y_n)}
	{\mathbf{\Gamma}(1-s-y_n, g-y_n, 1-g-x_n)} \\
	& \label{SPrecur}
	\times
	\frac{\tilde{q}^\ast(1-s;\bm{y}_{n-1})\,\tilde{q}^\ast(1-y_n;\bm{y}_{n-1})\, q(-y_n;\bm{x}_n)}
	{\tilde{q}^\ast(1+x_n;\bm{y}_n)\,\tilde{q}^\ast(1-x_n;\bm{y}_n)} \,
	P(\bm{y}_{n-1},\bm{x}_{n-1}) \,.
\end{align}
Here the coefficient $a(x_n)$ is defined in \eqref{a}, the normalization coefficient $\lambda_n(x)$ is given in \eqref{lambda}.
The functions $\tilde{q}$ and $q$ are eigenvalues of $Q$-operators \eqref{Qcl} and \eqref{Qop}
\begin{equation} \nonumber
	\widetilde{Q}_{n-1}(u)\,\psi_{\bm y_{n-1}}
	= \tilde{q}(u\,,\bm y_{n-1})
	\, \psi_{\bm y_{n-1}} \,,
	\quad
	Q_{n-1}(u)\,\Psi_{\bm{x}_{n-1}} =
	q(u,\bm{x}_{n-1})\,\Psi_{\bm{x}_{n-1}} \,,
\end{equation}
they are given by expressions \eqref{qnCl} and \eqref{qn}
\begin{equation*}
	\tilde{q}(u\,,\bm y_{n-1}) = [\imath]^{(s-u)(n-1)} \prod_{k=1}^{n-1}
	\frac{\mathbf{\Gamma}(u-x_k)}{\mathbf{\Gamma}(s-x_k)} \,, \quad
	q(u\,,\bm x_{n-1}) = [-1]^{(u-s)n}\prod_{k=1}^{n-1}
	\frac{\mathbf{\Gamma}(u\pm x_k)}{\mathbf{\Gamma}(s\pm x_k)} \,,
\end{equation*}
where we use the notation \eqref{pm} for gamma functions.
Taking into account the initial condition
\begin{equation} \label{SPbase}
	P(y_1, x_1) = \frac{\pi^2\,[-2\gamma]^{s+y_1}\,\mathbf{\Gamma}(-y_1\pm x_1, 1-s\pm x_1)}{\mathbf{\Gamma}(1-s-y_1, g-y_1, 1-g\pm x_1)}
\end{equation}
we obtain from \eqref{SPrecur} the integral kernel \eqref{Peps} which we aimed to calculate
\begin{equation} \nonumber
	P(\bm{y}_n,\bm{x}_n) =
	\frac{\pi^{\tfrac{(3n+1)n}{2}}\,[-1]^{(n-1)ns}\,[-2\gamma]^{ns+\underline{\bm{y}_n}}\,\mathbf{\Gamma}^n(1-s,\pm\bm{x}_n)\,\mathbf{\Gamma}(-\bm{y}_n,\pm\bm{x}_n)}
	{\mathbf{\Gamma}^n(1-s,\bm{y}_n)\,\mathbf{\Gamma}(g,\bm{y}_n)\,\mathbf{\Gamma}(1-g,\pm\bm{x}_n)\,\prod\limits_{1\leq i<j\leq n}\mathbf{\Gamma}(-y_i-y_j)} \,.
\end{equation}

Let us prove the formula \eqref{SPrecur}. 
The proof is based on the following recurrent formula for eigenfunctions of the spin chain of A~type
\begin{align}
	\nonumber
	\Pi_n\psi_{\bm{y}_n} & = \frac{\pi\,[-1]^{n(y_n+s)}\,[2\gamma]^{1-s-y_n}\,\mathbf{\Gamma}(2-g-s)\,\widetilde{\lambda}_n(y_n)\,\tilde{q}(1-s;\bm{y}_{n-1})\,\tilde{q}(1-y_n;\bm{y}_{n-1})}{\mathbf{\Gamma}(s+y_n, 1-g+y_n)} \\
	& \label{Pipsi}
	\times Q_n^{-1}(1-y_n)\,\Pi_n\,\delta^2(z_n-\gamma)\,\psi_{\bm{y}_{n-1}} \,.
\end{align}
Substituting the expression \eqref{Pipsi} into the first argument of the scalar product \eqref{P}, we move the operators $Q_n^{-1}$ and $\Pi_n$ to the second argument by the use of conjugation rule \eqref{Qconj} for the $Q$-operator
and obvious unitarity of $\Pi_n$. Then we apply $Q_n(-y_n)$ to its eigenfunction
\begin{equation} \nonumber
	Q_n(-y_n)\,\Psi_{\bm{x}_n} = q(-y_n;\bm{x}_n)\,\Psi_{\bm{x}_n} \,.
\end{equation}
As the result, we arrive at
\begin{align}
	& \nonumber
	P(\bm{y}_n,\bm{x}_n) =
	\frac{\pi\,[-1]^{n(y_n+s)}\,[2\gamma]^{y_n+s}\,\mathbf{\Gamma}(g+s)\,\widetilde{\lambda}_n^\ast(y_n)}{\mathbf{\Gamma}(1-s-y_n, g-y_n)} \\
	& \label{P1}
	\times \tilde{q}^\ast(1-s;\bm{y}_{n-1})\,\tilde{q}^\ast(1-y_n;\bm{y}_{n-1})\,q(-y_n;\bm{x}_n)\,\langle\delta^2(z_n-\gamma)\,\psi_{\bm{y}_{n-1}}\,|\,\Pi_n\Psi_{\bm{x}_n}\rangle \,,
\end{align}
where we also used the complex conjugation formula \eqref{gamma-conj} for gamma functions.
Next we rewrite the second argument of the scalar product in \eqref{P1} by means of the recurrent relation for the eigenfunction
\begin{equation} \label{PiPsi}
	\Pi_n\Psi_{\bm{x}_n} = \lambda_n(x_n)\,\Pi_n\,\mathcal{R}_{n\,n-1}(x_n)\ldots\mathcal{R}_{21}(x_n)\,\mathcal{K}_1(x_n)\,\mathcal{R}_{12}(x_n)\ldots\mathcal{R}_{n-1\,n}(x_n)\,\Psi_{\bm{x}_{n-1}} \,.
\end{equation}
In the last expression we apply the formula \eqref{key}
\begin{equation} \nonumber
	\mathcal{R}_{12}(x_n)\ldots\mathcal{R}_{n-1\,n}(x_n)\,\Psi_{\bm{x}_{n-1}} = \left(a(x_n)\right)^{n-1}\,\mathcal{R}_{n\,n-1}(-x_n)\ldots\mathcal{R}_{21}(-x_n)\,\Psi_{\bm{x}_{n-1}}(z_2,\ldots,z_n) \,.
\end{equation}
Then we note that since the operator $\Pi_n$ permutes the variables $z_k$ and $z_{n+1-k}$, it obeys the following commutation relations
\begin{equation} \label{PiRK}
	\Pi_n\,\mathcal{R}_{ij}(x)=\mathcal{R}_{n+1-i,\,n+1-j}(x)\,\Pi_n \,, \;
	\Pi_n\,\mathcal{K}_{i}(x)=\mathcal{K}_{n+1-i}(x)\,\Pi_n \,, \;
	\Pi_n\,[\hat{p}_k]^\alpha=[\hat{p}_{n+1-k}]^\alpha\,\Pi_n \,.
\end{equation}
Using these relations we move $\Pi_n$ in \eqref{PiPsi} to the right and, finally, act by this operator on the eigenfunction
\begin{equation} \nonumber
	\Pi_n\,\Psi_{\bm{x}_{n-1}}(z_2,\ldots,z_n) = \Psi_{\bm{x}_{n-1}}(z_{n-1},\ldots,z_1) = \Pi_{n-1}\Psi_{\bm{x}_{n-1}} \,.
\end{equation}
After all these manipulations one obtains
\begin{multline} \nonumber
	\Pi_n\Psi_{\bm{x}_n} = \lambda_n(x_n)\,\left(a(x_n)\right)^{n-1}\,\mathcal{R}_{12}(x_n)\ldots\mathcal{R}_{n-1\,n}(x_n) \\
	\times\mathcal{K}_n(x_n)\,\mathcal{R}_{12}(-x_n)\ldots\mathcal{R}_{n-1\,n}(-x_n)\,\Pi_{n-1}\Psi_{\bm{x}_{n-1}} \,.
\end{multline}
Substituting this result into \eqref{P1} and moving all $\mathcal{R}$- and $\mathcal{K}$-operators to the first argument of the scalar product by means of the corresponding conjugation rules \eqref{Rconj} and \eqref{Kconj} we get
\begin{equation} \label{scPr}
\langle\delta^2(z_n-\gamma)\,\psi_{\bm{y}_{n-1}}\,|\,\Pi_n\Psi_{\bm{x}_n}\rangle = \lambda_n(x_n)\,\left(a(x_n)\right)^{n-1}\,
\langle\phi(\bm{z}_n)\,|\,\Psi_{\bm{x}_{n-1}}\rangle \,,
\end{equation}
where
\begin{align} 
	\nonumber
	\phi(\bm{z}_n) & =
	\Pi_{n-1}[\mathcal{R}_{12}(1-x_n)\ldots\mathcal{R}_{n-1\,n}(1-x_n)]^{-1}\,\mathcal{K}_n^{-1}(1+x_n) \\
	\nonumber &
	\times [\mathcal{R}_{12}(1+x_n)\ldots\mathcal{R}_{n-1\,n}(1+x_n)]^{-1}\delta^2(z_n-\gamma)\,\psi_{\bm{y}_{n-1}} \,.
\end{align}
To transform the expression $\phi$ we use the relation
\begin{align} \label{R_delta_psi}
	\mathcal{R}_{12}(x)\ldots\mathcal{R}_{n-1\,n}(x)\,\delta^2(z_n-\gamma)\,\psi_{\bm{y}_{n-1}} =
	\tilde{q}(x;\bm{y}_{n-1})\,\delta^2(z_n-\gamma)\,\psi_{\bm{y}_{n-1}} \,,
\end{align}
the formula \eqref{Kdelta} for action of reflection operator on delta function
\begin{equation} \label{Kdelta1}
	\mathcal{K}_n(x)\,\delta^2(z_n-\gamma) = \frac{[2\imath\gamma]^{x-s}\,\mathbf{\Gamma}(2-g-s)}{\mathbf{\Gamma}(2-g-x)}\,\delta^2(z_n-\gamma)
\end{equation}
and the fact that $\psi_{\bm{y}_{n-1}}$ does not depend on $z_n$. In other words, $\delta^2(z_n-\gamma)\,\psi_{\bm{y}_{n-1}}$ is the eigenfunction of operators $\mathcal{R}_{12}(x)\ldots\mathcal{R}_{n-1\,n}(x)$ and $\mathcal{K}_n(x)$. Thus, we obtain
\begin{align} \label{phi}
	\phi(\bm{z}_n) = \frac{[2\imath\gamma]^{s-x_n-1}\,\mathbf{\Gamma}(1-g-x_n)}{\mathbf{\Gamma}(2-g-s)\,\tilde{q}(1-x_n;\bm{y}_{n-1})\,\tilde{q}(1+x_n;\bm{y}_{n-1})}\,\delta^2(z_n-\gamma)\,\Pi_{n-1}\,\psi_{\bm{y}_{n-1}} \,.
\end{align}
Substituting \eqref{phi} into \eqref{scPr} and integrating the delta function over $z_n$ we reduce everything to scalar product of functions of $n-1$ variables
\begin{equation} \nonumber
	\langle\delta^2(z_n-\gamma)\,\Pi_{n-1}\,\psi_{\bm{y}_{n-1}}\,|\,\Psi_{\bm{x}_{n-1}}\rangle = \langle\Pi_{n-1}\,\psi_{\bm{y}_{n-1}}\,|\,\Psi_{\bm{x}_{n-1}}\rangle =
	P(\bm{y}_{n-1},\bm{x}_{n-1}) \,,
\end{equation}
that is,
\begin{equation} \label{scPr1}
	\langle\delta^2(z_n-\gamma)\,\psi_{\bm{y}_{n-1}}\,|\,\Pi_n\Psi_{\bm{x}_n}\rangle = 
	\frac{[-2\imath\gamma]^{x_n-s}\,\mathbf{\Gamma}(g+x_n)\,\lambda_n(x_n)\,a^{n-1}(x_n)}{\mathbf{\Gamma}(g+s)\,\tilde{q}^\ast(1-x_n;\bm{y}_{n-1})\,\tilde{q}^\ast(1+x_n;\bm{y}_{n-1})}
	\, P(\bm{y}_{n-1},\bm{x}_{n-1}) \,.
\end{equation}
Substituting \eqref{scPr1} into \eqref{P1} and using the reflection formula \eqref{gamma-refl} for $\mathbf{\Gamma}(g+x_n)$ we finally obtain the needed recurrence relation \eqref{SPrecur}.

Now it is time to prove the relations \eqref{Pipsi} and \eqref{R_delta_psi} which were used in the derivation. To obtain \eqref{R_delta_psi} we rewrite the operator $\mathcal{R}_{n-1\,n}$ in the explicit form
\begin{equation} \nonumber
	\mathcal{R}_{n-1\,n}(x)\,\delta^2(z_n-\gamma)\,\psi_{\bm{y}_{n-1}} =
	[z_{n-1\,n}]^{1-2s}[\hat{p}_{n-1}]^{x-s}[z_{n-1\,n}]^{x+s-1}\,\delta^2(z_n-\gamma)\,\psi_{\bm{y}_{n-1}} \,.
\end{equation}
Since the only operators which act with respect to variable $z_n$ are operators of multiplication by a function, one can substitute $z_n=\gamma$ due to the presence of delta function
\begin{equation} \nonumber
	\mathcal{R}_{n-1\,n}(x)\,\delta^2(z_n-\gamma)\,\psi_{\bm{y}_{n-1}} =
	\mathcal{R}_{n-1,-}(x)\,\delta^2(z_n-\gamma)\,\psi_{\bm{y}_{n-1}} \,,
\end{equation}
where
\begin{equation} \nonumber
\mathcal{R}_{n-1,-}(x) = [z_{n-1}-\gamma]^{1-2s}[\hat{p}_{n-1}]^{x-s}
[z_{n-1}-\gamma]^{x+s-1} \,.
\end{equation}
This way,
\begin{align}
	& \nonumber
	\mathcal{R}_{12}(x)\ldots\mathcal{R}_{n-1\,n}(x)\,\delta^2(z_n-\gamma)\,\psi_{\bm{y}_{n-1}} =
	\mathcal{R}_{12}(x)\ldots\mathcal{R}_{n-2\,n-1}(x)\,\mathcal{R}_{n-1,-}(x)\,\delta^2(z_n-\gamma)\,\psi_{\bm{y}_{n-1}} \\
	& \nonumber
	= \widetilde{Q}_{n-1}(x)\,\delta^2(z_n-\gamma)\,\psi_{\bm{y}_{n-1}}
	= \tilde{q}(x;\bm{y}_{n-1})\,\delta^2(z_n-\gamma)\,\psi_{\bm{y}_{n-1}} \,.
\end{align}
To derive \eqref{Pipsi} we first use the standard recurrence relation for the eigenfunction of A-type spin chain
\begin{equation} \nonumber
	\Pi_n\psi_{\bm{y}_n} = \widetilde{\lambda}_n(y_n)\,\Pi_n\,\mathcal{R}_{12}(y_n)\ldots
\mathcal{R}_{n-1\,n}(y_n)\,[z_{n}-\gamma]^{-s-y_n}\psi_{\bm{y}_{n-1}} \,.
\end{equation}
By means of \eqref{Chain1} we rewrite
\begin{equation} \nonumber
	[z_{n}-\gamma]^{-s-y_n} = \frac{[\imath]^{s-y_n}\,\mathbf{\Gamma}(2s)}{\mathbf{\Gamma}(s+y_n)}\,
[\hat{p}_n]^{y_n-s}\,[z_{n}-\gamma]^{-2s} \,,
\end{equation}
and apply the formula \eqref{RRp}
\begin{align}
	\nonumber
	\mathcal{R}_{12}(y)\ldots\mathcal{R}_{n-1\,n}(y)\,[\hat{p}_n]^{y-s} & =
	[-1]^{(n-1)(y+s)}\,[\hat{p}_1]^{y+s-1}\,\mathcal{R}_{21}^{-1}(1-y)\ldots\mathcal{R}_{n\,n-1}^{-1}(1-y) \\
	& \nonumber
	\times\mathcal{R}_{12}(1-s)\ldots\mathcal{R}_{n-1\,n}(1-s)\,[\hat{p}_n]^{1-2s} \,,
\end{align}
Thus,
\begin{align}
	\nonumber
	\Pi_n\psi_{\bm{y}_n} & = \frac{[-1]^{(n-1)(y_n+s)}\,[\imath]^{s-y_n}\,\mathbf{\Gamma}(2s)\,\widetilde{\lambda}_n(y_n)}{\mathbf{\Gamma}(s+y_n)}\,\Pi_n\,[\hat{p}_1]^{y_n+s-1}\,\mathcal{R}_{21}^{-1}(1-y_n)\ldots\mathcal{R}_{n\,n-1}^{-1}(1-y_n) \\
	& \label{Pipsi1}
	\times\mathcal{R}_{12}(1-s)\ldots\mathcal{R}_{n-1\,n}(1-s)\,
[\hat{p}_n]^{1-2s}\,[z_{n}-\gamma]^{-2s}\,\psi_{\bm{y}_{n-1}} \,.
\end{align}
Then we first extract the $Q$-operator \eqref{Qop} for BC-type spin chain commuting operators with $\Pi_n$ by the use of \eqref{PiRK}
\begin{align}
	& \nonumber
	\Pi_n[\hat{p}_1]^{y_n+s-1}\mathcal{R}_{21}^{-1}(1-y_n)\ldots\mathcal{R}_{n\,n-1}^{-1}(1-y_n) =
	[\hat{p}_n]^{y_n+s-1}\mathcal{R}_{n-1\,n}^{-1}(1-y_n)\ldots\mathcal{R}_{12}^{-1}(1-y_n)\Pi_n \\
	& \nonumber
	= Q_n^{-1}(1-y_n)\,\mathcal{R}_{n\,n-1}(1-y_n)\ldots\mathcal{R}_{21}(1-y_n)\,\mathcal{K}_1(1-y_n)\,\Pi_n \\
	& \nonumber
	= Q_n^{-1}(1-y_n)\,\Pi_n\,\mathcal{R}_{12}(1-y_n)\ldots\mathcal{R}_{n-1\,n}(1-y_n)\,\mathcal{K}_n(1-y_n) \,.
\end{align}
Second, since $\psi_{\bm{y}_{n-1}}$ does not depend on $z_n$, 
it is possible to use \eqref{delta3}
\begin{equation} \nonumber
[\hat{p}_n]^{1-2s}\,[z_n-\gamma]^{-2s} = \frac{\pi\,[\imath]^{-2s}}{\mathbf{\Gamma}(2s)}\,\delta^2(z_n-\gamma) \,.
\end{equation}
After these transformations \eqref{Pipsi1} takes the form
\begin{align}
	\nonumber
	\Pi_n\psi_{\bm{y}_n} & = \frac{\pi\,[-1]^{(n-1)(y_n+s)}\,[\imath]^{-s-y_n}\,\widetilde{\lambda}_n(y_n)}{\mathbf{\Gamma}(s+y_n)}\,Q_n^{-1}(1-y_n)\,\Pi_n\,\mathcal{R}_{12}(1-y_n)\ldots\mathcal{R}_{n-1\,n}(1-y_n) \\
	& \nonumber
	\times\mathcal{K}_n(1-y_n)\,\mathcal{R}_{12}(1-s)\ldots\mathcal{R}_{n-1\,n}(1-s)\,\delta^2(z_n-\gamma)\,\psi_{\bm{y}_{n-1}} \,.
\end{align}
To obtain the desired formula \eqref{Pipsi} it remains to act by operators $\mathcal{R}_{12}\ldots\mathcal{R}_{n-1\,n}$ and $\mathcal{K}_n$ on their eigenfunction $\delta^2(z_n-\gamma)\,\psi_{\bm{y}_{n-1}}$ using relations \eqref{R_delta_psi} and \eqref{Kdelta1}
\begin{align}
	& \nonumber
	\mathcal{R}_{12}(1-s)\ldots\mathcal{R}_{n-1\,n}(1-s)\,\delta^2(z_n-\gamma)\,\psi_{\bm{y}_{n-1}} = \tilde{q}(1-s;\bm{y}_{n-1})\,\delta^2(z_n-\gamma)\,\psi_{\bm{y}_{n-1}} \,, \\
	& \nonumber
	\mathcal{K}_n(1-y_n)\,\delta^2(z_n-\gamma)\,\psi_{\bm{y}_{n-1}} = \frac{[2\imath\gamma]^{1-y_n-s}\mathbf{\Gamma}(2-g-s)}{\mathbf{\Gamma}(1-g+y_n)}\,\delta^2(z_n-\gamma)\,\psi_{\bm{y}_{n-1}} \,, \\
	& \nonumber
	\mathcal{R}_{12}(1-y_n)\ldots\mathcal{R}_{n-1\,n}(1-y_n)\,\delta^2(z_n-\gamma)\,\psi_{\bm{y}_{n-1}} = \tilde{q}(1-y_n;\bm{y}_{n-1})\,\delta^2(z_n-\gamma)\,\psi_{\bm{y}_{n-1}} \,.
\end{align}

We complete this section by deriving the basic formula \eqref{SPbase} for $P(y_1, x_1) =
\langle \psi_{y_1}|\Psi_{x_1}\rangle$. First we express $\psi_{y_1}$ and $\Psi_{x_1}$ by definition and get
\begin{align} \nonumber
	P(y_1, x_1) =
	\lambda_1(x_1)\,\langle [z_-]^{-s-y_1}\,|\,[z_-]^{1-s-g}[z_+]^{g-s}[\hat{p}]^{x_1-s}[z_+]^{x_1-g}[z_-]^{x_1+g-1}\rangle \,,
\end{align}
where $z_{\pm}=z\pm\gamma$. Next we use the conjugation formula \eqref{zconj} for double power
along with the star-triangle relation in the form \eqref{star-tr}
\begin{equation} \nonumber
	[z_+]^{g-s}[\hat{p}]^{x_1-s}[z_+]^{x_1-g} =
	[\hat{p}]^{x_1-g}[z_+]^{x_1-s}[\hat{p}]^{g-s} 
\end{equation}
and arrive at
\begin{align} \nonumber
	P(y_1, x_1) =
	\lambda_1(x_1)\,\langle [z_-]^{g-y_1-1}\,|\,[\hat{p}]^{x_1-g}[z_+]^{x_1-s}[\hat{p}]^{g-s}[z_-]^{x_1+g-1}\rangle \,.
\end{align}
Applying the conjugation rule \eqref{conj} to $[\hat{p}]^{x_1-g}$
and acting by operators on power functions by means of \eqref{Chain1}
\begin{align} 
	& \nonumber
	[\hat{p}]^{g-x_1-1}[z_-]^{g-y_1-1} = \frac{[-1]^{y_1-x_1}[\imath]^{g-x_1}}{\mathbf{\Gamma}(1-g+y_1,1+x_1-y_1)}\,[z_-]^{x_1-y_1}, \\
	& \nonumber
	[\hat{p}]^{g-s}[z_-]^{x_1+g-1} = \frac{[-1]^{s+x_1}[\imath]^{g-s}}{\mathbf{\Gamma}(1-g-x_1,s+x_1)}\,[z_-]^{x_1+s-1}
\end{align}
we obtain
\begin{equation} \nonumber
	P(y_1, x_1) =
	\frac{[-1]^{x_1+y_1+s-g}\,[\imath]^{x_1-s}\,\lambda_1(x_1)}
	{\mathbf{\Gamma}(g-y_1, 1-g-x_1, 1+y_1-x_1, s+x_1)} \,
	\langle[z_-]^{-s-y_1}\,|\,[z_+]^{x_1-s}\rangle \,.
\end{equation}
To calculate the coefficient in the last identity we used the complex conjugation formula~\eqref{gamma-conj} for gamma function.
Eventually, applying the chain relation \eqref{Chain}
\begin{equation} \nonumber
	\langle[z_-]^{-s-y_1}\,|\,[z_+]^{x_1-s}\rangle =
	\int \mathrm{d}^2z\,[z_-]^{s+y_1-1}\,[z_+]^{x_1-s} =
	\frac{\pi\,[-1]^{s-x_1}\,[2\gamma]^{y_1+x_1}}{\mathbf{\Gamma}(1-s-y_1, s-x_1, 1+y_1+x_1)} \,,
\end{equation}
the expression \eqref{lambda} for coefficient $\lambda_1(x_1)$ and the reflection formula~\eqref{gamma-refl} for gamma-function we complete the derivation of \eqref{SPbase}.

\section{Completeness}
\label{sect:compl}

In this section we prove the completeness relation
\begin{equation} \label{compl}
	\int \mathcal{D}\bm{x}_n \, \mu(\bm{x}_n) \,
	\Psi_{\bm{x}_n}(\bm{z}_n) \, \overline{\Psi_{\bm{x}_n}(\bm{w}_n)} =
	\delta^2(z_1-w_1) \ldots \delta^2(z_n-w_n) \,,
\end{equation}
where the spectral variables $x_i, \bar{x}_i$ have the form \eqref{x}, and the integral over variables $\bm{x}_n$ is defined in \eqref{intx} as the integral over continuous components of parameters $x_i$ and the sum over discrete components.
The Sklyanin integration measure reads
\begin{equation} \nonumber
	\mu(\bm{x}_n) =
	\frac{\prod\limits_{k=1}^n|x_k|^2 \prod\limits_{1\leq i<j\leq n}|x_i^2-x_j^2|^2}{4^n\,\pi^{2n(n+1)}\,n!\,|\gamma|^{2n}} \,,
\end{equation}
it follows from the orthogonality relation \eqref{orth}.
We will need the equivalent expression
\begin{equation} \label{measure1}
	\mu(\bm{x}_n) =
	\frac{[-1]^{n(n-1)s}\,\mathbf{\Gamma}(s,\pm\bm{x}_n)\,\mathbf{\Gamma}(1-s,\pm\bm{x}_n)}{4^n\,\pi^{2n(n+1)}\,n!\,[2\gamma]^n
	\prod\limits_{k=1}^n\mathbf{\Gamma}(\pm 2x_k)
	\prod\limits_{1\leq i<j\leq n}\mathbf{\Gamma}(\pm x_i\pm x_j)} \,,
\end{equation}
where we use the notations \eqref{nota} along with
\begin{equation*}
	\mathbf{\Gamma}(\pm 2x_k)=\mathbf{\Gamma}(2x_k)\,\mathbf{\Gamma}(-2x_k), \quad
	\mathbf{\Gamma}(\pm x_i\pm x_j)=\mathbf{\Gamma}(x_i+x_j)\,\mathbf{\Gamma}(x_i-x_j)\,\mathbf{\Gamma}(-x_i+x_j)\,\mathbf{\Gamma}(-x_i-x_j) \,.
\end{equation*}
Formula \eqref{measure1} can be obtained by means of difference equation \eqref{gamma-diff} and reflection relation~\eqref{gamma-refl} for the gamma function.

To derive the identity \eqref{compl} we expand $\Psi_{\bm{x}_n}$ over the basis of eigenfunctions for A-type $SL(2,\mathbb{C})$ spin chain and reduce \eqref{compl} to the proven completeness relation \eqref{complCl} for this basis. The expansion is given by the formula \eqref{PsiMB}
\begin{equation} \label{PsiMB1}
	\Psi_{\bm{x}_n}(\bm{z}_n) =
	\lim\limits_{\varepsilon\to 0_+}
	\int\mathcal{D}\bm{y}_n\,\tilde{\mu}(\bm{y}_n)\,P_\varepsilon(\bm{y}_n,\bm{x}_n)\,
	\hat{\psi}_{\bm{y}_n}(\bm{z}_n) \,,
\end{equation}
where
\begin{align}
	& \nonumber
	P_\varepsilon(\bm{y}_n,\bm{x}_n) =
	\frac{\pi^{\tfrac{(3n+1)n}{2}}\,[-1]^{(n-1)ns}\,[-2\gamma]^{ns+\underline{\bm{y}_n}}\,\tilde{\mu}(\bm{y}_n)\,\mathbf{\Gamma}^n(1-s,\pm\bm{x}_n)\,\mathbf{\Gamma}(\varepsilon-\bm{y}_n,\pm\bm{x}_n)}
	{\mathbf{\Gamma}^n(1-s,\bm{y}_n)\,\mathbf{\Gamma}(g,\bm{y}_n)\,\mathbf{\Gamma}(1-g,\pm\bm{x}_n)\,\prod\limits_{1\leq i<j\leq n}\mathbf{\Gamma}(-y_i-y_j)} \,, \\
	& \nonumber
	\tilde{\mu}(\bm{y}_n) =
	\frac{1}{2^n\,\pi^{n(n+1)}\,n!}
	\prod\limits_{1\leq i<j\leq n}|y_i-y_j|^2 \,.
\end{align}
The variables $\bm{y}_n$ are again of the form \eqref{x}, and the corresponding integral is defined in~\eqref{intx}.

From \eqref{PsiMB1} we can obtain the similar formula for the complex conjugated function $\overline{\Psi_{\bm{x}_n}(\bm{w}_n)}$ applying the conjugation rules \eqref{zconj} and \eqref{gamma-conj} to $[-2\gamma]^{ns+\underline{\bm{y}_n}}$ and gamma functions.
We have
\begin{equation} \label{PsiMB2}
	\overline{\Psi_{\bm{x}_n}(\bm{w}_n)} =
	\lim\limits_{\varepsilon\to 0_+}
	\int\mathcal{D}\bm{y}'_n\,\tilde{\mu}(\bm{y}'_n)\,P_\varepsilon^\ast(\bm{y}'_n,\bm{x}_n)\,
	\overline{\hat{\psi}_{\bm{y}'_n}(\bm{w}_n)} \,,
\end{equation}
where
\begin{equation} \nonumber
	P_\varepsilon^\ast(\bm{y}'_n,\bm{x}_n) =
	\frac{\pi^{\tfrac{(3n+1)n}{2}}\,[-1]^{n(g-s)}\,[-2\gamma]^{n(1-s)-\underline{\bm{y}'_n}}\,\tilde{\mu}(\bm{y}'_n)\,\mathbf{\Gamma}^n(s,\pm\bm{x}_n)\,\mathbf{\Gamma}(\varepsilon+\bm{y}'_n,\pm\bm{x}_n)}
	{\mathbf{\Gamma}^n(s,-\bm{y}'_n)\,\mathbf{\Gamma}(1-g,-\bm{y}'_n)\,\mathbf{\Gamma}(g,\pm\bm{x}_n)\,\prod\limits_{1\leq i<j\leq n}\mathbf{\Gamma}(y'_i+y'_j)} \,.
\end{equation}

Substituting the explicit formulas \eqref{measure1}, \eqref{PsiMB1} and \eqref{PsiMB2}
into the left hand side of the completeness relation \eqref{compl} one obtains the multiple integral
\begin{multline} \label{compl1}
	\int \mathcal{D}\bm{x}_n \, \mu(\bm{x}_n) \,
	\Psi_{\bm{x}_n}(\bm{z}_n) \, \overline{\Psi_{\bm{x}_n}(\bm{w}_n)} \\
	=
	\int \mathcal{D}\bm{y}_n\,\mathcal{D}\bm{y}'_n\, \frac{\pi^{n^2}\,
	[-2\gamma]^{\underline{\bm{y}_n}-\underline{\bm{y}'_n}}\,
	\tilde{\mu}(\bm{y}_n)\,\tilde{\mu}(\bm{y}'_n)\,
	\mathbf{\Gamma}^n(1-s,\bm{y}'_n)\,\mathbf{\Gamma}(g,\bm{y}'_n)\,\hat{\psi}_{\bm{y}_n}(\bm{z}_n)\,
	\overline{\hat{\psi}_{\bm{y}'_n}(\bm{w}_n)}}
	{[-1]^{n(n-1)s+(n+1)\underline{\bm{y}'_n}}\,\mathbf{\Gamma}^n(1-s,\bm{y}_n)\,\mathbf{\Gamma}(g,\bm{y}_n)} \\
	\times \lim\limits_{\varepsilon\to 0_+} I_\varepsilon(\bm{y}_n,\bm{y}'_n)
	 \,,
\end{multline}
where by $I_\varepsilon(\bm{y}_n,\bm{y}'_n)$ we denote the expression which contains the integral over variables $\bm{x}_n$
\begin{multline} \label{Ieps}
	I_\varepsilon(\bm{y}_n,\bm{y}'_n) = \frac{1}{\prod\limits_{1\leq i<j\leq n}\mathbf{\Gamma}(-y_i-y_j, y'_i+y'_j)} \\
	\times
	\int \mathcal{D}\bm{x}_n \,
	\frac{\mathbf{\Gamma}(\varepsilon-\bm{y}_n,\pm\bm{x}_n)\,\mathbf{\Gamma}(\varepsilon+\bm{y}'_n,\pm\bm{x}_n)\,
		\mathbf{\Gamma}(s,\pm\bm{x}_n)\,\mathbf{\Gamma}(1-s-(2n+1)\varepsilon,\pm\bm{x}_n)}
	{(4\pi)^n\,n!\, \prod\limits_{k=1}^n\mathbf{\Gamma}(\pm 2x_k)
		\prod\limits_{1\leq i<j\leq n}\mathbf{\Gamma}(\pm x_i\pm x_j)} \,.
\end{multline}
Note that in the factor $\mathbf{\Gamma}(1-s,\pm\bm{x}_n)$ coming from expression \eqref{measure1} for $\mu(\bm{x}_n)$ we introduced the additional regularization
\begin{equation} \nonumber
	\mathbf{\Gamma}(1-s,\pm\bm{x}_n) =
	\lim\limits_{\varepsilon\to 0_+}\mathbf{\Gamma}(1-s-(2n+1)\varepsilon,\pm\bm{x}_n) \,.
\end{equation}
To calculate $I_\varepsilon(\bm{y}_n,\bm{y}'_n)$ we apply the complex analog of BC-type Gustafson integral~\cite[eq.~(2.3b)]{DM:Gust4},~\cite{N,MN,SS0,SS1,SS2}
\begin{equation} \label{Gust}
	\frac{1}{(4\pi)^n\,n!} \int \mathcal{D}\bm{x}_n \,
	\frac{\prod\limits_{k=1}^n\prod\limits_{j=1}^{2n+2}\mathbf{\Gamma}(\alpha_j\pm x_k)}
	{\prod\limits_{k=1}^n\mathbf{\Gamma}(\pm 2x_k)
	\prod\limits_{1\leq i<j\leq n}\mathbf{\Gamma}(\pm x_i\pm x_j)}
	= \frac{\prod\limits_{1\leq j<k\leq 2n+2}\mathbf{\Gamma}(\alpha_j+\alpha_k)}
	{\mathbf{\Gamma}\left(\sum_{k=1}^{2n+2} \alpha_k\right)} \,.
\end{equation}
The $\varepsilon$-prescription in \eqref{Ieps}, which originates from the Mellin-Barnes representation \eqref{PsiMB}, provides the proper separation of the series of poles which appear due to gamma functions in the numerator, so that the usability conditions \cite[Section~2.1]{DM:Gust4} for the integral \eqref{Gust} are satisfied.
Thus we get
\begin{multline} \label{x_int}
	\lim\limits_{\varepsilon\to 0_+} I_\varepsilon(\bm{y}_n,\bm{y}'_n)
	= [-1]^{3\underline{\bm{y}_n}-\underline{\bm{y}'_n}} \,
	\frac{\mathbf{\Gamma}(s,\bm{y}_n)\,\mathbf{\Gamma}(1-s,\bm{y}_n)}
	{\mathbf{\Gamma}(s,\bm{y}'_n)\,\mathbf{\Gamma}(1-s,\bm{y}'_n)} \\
	\times 
	\lim\limits_{\varepsilon\to 0_+} \mathbf{\Gamma}(1-(2n+1)\varepsilon) \,
	\mathbf{\Gamma}(\underline{\bm{y}_n}-\underline{\bm{y}'_n}+\varepsilon)\,\mathbf{\Gamma}(\bm{y}'_n+2\varepsilon,\bm{y}_n) \,,
\end{multline}
where we additionally used the reflection formula \eqref{gamma-refl} for gamma function.
The limit in the right hand side of \eqref{x_int} is equal to symmetric delta function
\begin{multline} \label{lim}
	\lim\limits_{\varepsilon\to 0_+} \mathbf{\Gamma}(1-(2n+1)\varepsilon) \,
	\mathbf{\Gamma}(\underline{\bm{y}_n}-\underline{\bm{y}'_n}+\varepsilon)\,\mathbf{\Gamma}(\bm{y}'_n+2\varepsilon,\bm{y}_n) \\
	=
	\pi^{-n^2}[-1]^{(n-1)\underline{\bm{y}_n}+n(n-1)s}\,\tilde{\mu}^{-1}(\bm{y}'_n)\, \tilde{\delta}^{(2)}(\bm{y}'_n,\bm{y}_n) \,,
\end{multline}
where
\begin{equation} \nonumber
	\tilde{\delta}^{(2)}(\bm{y}'_n,\bm{y}_n) =
	\frac{1}{n!} \sum\limits_{\tau\in\mathfrak{S}_n}
	\delta^{(2)}(y'_1-y_{\tau(1)}) \ldots \delta^{(2)}(y'_n-y_{\tau(n)}) \,.
\end{equation}
We will prove this formula later, and now let us substitute the results \eqref{x_int} and \eqref{lim} into \eqref{compl1}.
One can see that $\tilde{\mu}(\bm{y}'_n)$ from \eqref{lim} cancels with the same factor from \eqref{compl1}, and the resulting expression does not contain poles on the contours of integration by variables $\bm{y}'_n$.
Therefore, integrating the symmetric delta function with respect to $\bm{y}'_n$ we obtain the following expression
\begin{equation} \nonumber
	\int \mathcal{D}\bm{x}_n \, \mu(\bm{x}_n) \,
	\Psi_{\bm{x}_n}(\bm{z}_n) \, \overline{\Psi_{\bm{x}_n}(\bm{w}_n)}
	= \frac{1}{n!} \sum\limits_{\tau\in\mathfrak{S}_n}
	\int \mathcal{D}\bm{y}_n \, \tilde{\mu}(\bm{y}_n)\,
\hat{\psi}_{\bm{y}_n}(\bm{z}_n)\,
\overline{\hat{\psi}_{\tau\bm{y}_n}(\bm{w}_n)} \,,
\end{equation}
where, as before, $\tau\bm{y}_n=(y_{\tau(1)},\bar{y}_{\tau(1)},\dots,y_{\tau(n)},\bar{y}_{\tau(n)})$.
All terms in the sum are equal thanks to the symmetry of functions $\psi_{\bm{y}_n}$ under permutations of parameters: $\psi_{\tau\bm{y}_n}=\psi_{\bm{y}_n}$. This way, we have
\begin{equation} \nonumber
	\int \mathcal{D}\bm{x}_n \, \mu(\bm{x}_n) \,
	\Psi_{\bm{x}_n}(\bm{z}_n) \, \overline{\Psi_{\bm{x}_n}(\bm{w}_n)}
	= \int \mathcal{D}\bm{y}_n \, \tilde{\mu}(\bm{y}_n)\,
\hat{\psi}_{\bm{y}_n}(\bm{z}_n)\,
\overline{\hat{\psi}_{\bm{y}_n}(\bm{w}_n)} \,,
\end{equation}
that is, the left hand side of \eqref{compl} reduces to the left hand side of completeness relation~\eqref{complCl} for eigenfunctions of A-type spin chain, as it was announced.
Using this relation
\begin{equation} \nonumber
	\int \mathcal{D}\bm{y}_n \, \tilde{\mu}(\bm{y}_n)\,
\hat{\psi}_{\bm{y}_n}(\bm{z}_n)\,
\overline{\hat{\psi}_{\bm{y}_n}(\bm{w}_n)} =
	\delta^2(z_n-w_n) \ldots \delta^2(z_1-w_1)
\end{equation}
we finally obtain \eqref{compl}.

It remains to prove the formula \eqref{lim} for the symmetric delta function.
The nontrivial contribution in the limit $\varepsilon\to 0_+$ is made by singular parts of gamma functions, which we extract by means of the recurrent relation $\Gamma(x+1)=x\,\Gamma(x)$, namely
\begin{align} 
	\nonumber
	& \mathbf{\Gamma}(\underline{\bm{y}_n}-\underline{\bm{y}'_n}+\varepsilon) \to
	\frac{\Gamma(1+\underline{\bm{y}_n}-\underline{\bm{y}'_n})}{\Gamma(1-\underline{\bar{\bm{y}}_n}+\underline{\bar{\bm{y}}'_n})} \,
	\frac{1}{\underline{\bm{y}_n}-\underline{\bm{y}'_n}+\varepsilon} \,, \\
	\label{GG}
	& \mathbf{\Gamma}(\bm{y}'_n+2\varepsilon,\bm{y}_n) \to
	\prod\limits_{i,j=1}^n \frac{\Gamma(1+y'_i-y_j)}{\Gamma(1-\bar{y}'_i+\bar{y}_j)} \,
	\prod\limits_{i,j=1}^n \frac{1}{y'_i-y_j+2\varepsilon} \,,
\end{align}
where $\underline{\bar{\bm{y}}_n} = \sum_{k=1}^n \bar{y}_k$.
We also extract the infinitesimal $\varepsilon$-factor
\begin{equation*}
	\mathbf{\Gamma}(1-(2n+1)\varepsilon) =
	(2n+1)\varepsilon \, \frac{\Gamma(1-(2n+1)\varepsilon)}{\Gamma(1+(2n+1)\varepsilon)}
\end{equation*}
and rewrite the last product in \eqref{GG} with the help of Cauchy determinant formula
\begin{align*}
	\frac{\prod\limits_{i<j}^n (y'_{i}-y'_{j})\,(y_{j}-y_{i})}{\prod\limits_{i,j=1}^{n}(y'_i-y_j+2\varepsilon)} 
	= \det\left(\frac{1}{y'_i-y_{j}+2\varepsilon}\right) =
	\sum_{\tau\in \mathfrak{S}_{n}} (-1)^{s(\tau)}
	\prod_{i=1}^{n}\frac{1}{y'_i-y_{\tau(i)}+2\varepsilon} \,,
\end{align*}
where $s(\tau)$ is the sign of permutation $\tau$.
This way,
\begin{multline} \label{lim1}
	\lim\limits_{\varepsilon\to 0_+} \mathbf{\Gamma}(1-(2n+1)\varepsilon) \,
	\mathbf{\Gamma}(\underline{\bm{y}_n}-\underline{\bm{y}'_n}+\varepsilon)\,\mathbf{\Gamma}(\bm{y}'_n+2\varepsilon,\bm{y}_n) \\
	=
	\frac{\Gamma(1+\underline{\bm{y}_n}-\underline{\bm{y}'_n})}{\Gamma(1-\underline{\bar{\bm{y}}_n}+\underline{\bar{\bm{y}}'_n})} \,
	\frac{1}{\prod_{i<j}^n (y'_{i}-y'_{j})\,(y_{j}-y_{i})}
	\prod\limits_{i,j=1}^n \frac{\Gamma(1+y'_i-y_j)}{\Gamma(1-\bar{y}'_i+\bar{y}_j)} \\
	\times \sum_{\tau\in \mathfrak{S}_{n}} (-1)^{s(\tau)} \,
	\lim\limits_{\varepsilon\to 0_+} 
	\frac{(2n+1)\varepsilon}{(\underline{\bm{y}_n}-\underline{\bm{y}'_n}+\varepsilon)\prod_{i=1}^{n}(y'_i-y_{\tau(i)}+2\varepsilon)} \,.
\end{multline}

To take the limit in every summand in \eqref{lim1} we use the following formula \cite[eq.~(B.1)]{ADV1}
\begin{equation*}
	\lim\limits_{\varepsilon\to 0_+} 
	\frac{(2n+1)\varepsilon}{(\underline{\bm{y}_n}-\underline{\bm{y}'_n}+\varepsilon)\prod_{i=1}^{n}(y'_i-y_i+2\varepsilon)} =
	(2\pi)^n\,\prod\limits_{k=1}^n\delta^{(2)}(y'_k-y_{k})\,,
\end{equation*}
in the term corresponding to permutation $\tau$ the variable $y_k$ must be replaced with $y_{\tau(k)}$.
The sign of permutation $(-1)^{s(\tau)}$ disappears from every term thanks to the presence of the Vandermonde determinant $\prod_{i<j}^n (y_{j}-y_{i})$.
Thus~\eqref{lim1} reduces to
\begin{multline} \label{lim2}
	\lim\limits_{\varepsilon\to 0_+} \mathbf{\Gamma}(1-(2n+1)\varepsilon) \,
	\mathbf{\Gamma}(\underline{\bm{y}_n}-\underline{\bm{y}'_n}+\varepsilon)\,\mathbf{\Gamma}(\bm{y}'_n+2\varepsilon,\bm{y}_n) \\
	=
	\frac{(2\pi)^n}{\prod_{i<j}^n y'_{ij}\,y'_{ji}}
	\prod\limits_{i,j=1}^n \frac{\Gamma(1+y'_{ij})}{\Gamma(1-\bar{y}'_{ij})}
	\sum_{\tau\in \mathfrak{S}_{n}} \prod\limits_{k=1}^n \delta^{(2)}(y'_k-y_{\tau(k)}) \,,
\end{multline}
where $y'_{ij}=y'_i-y'_j$.
It remains to simplify the prefactor by means of the reflection formula \eqref{gamma-refl} for gamma function
\begin{align}
	\nonumber
	& \frac{(2\pi)^n}{\prod_{i<j}^n y'_{ij}\,y'_{ji}}
	\prod\limits_{i,j=1}^n \frac{\Gamma(1+y'_{ij})}{\Gamma(1-\bar{y}'_{ij})}
	= \frac{(2\pi)^n}{\prod_{i<j}^n y'_{ij}\,y'_{ji}}
	\prod\limits_{i<j}^n
	\frac{\Gamma(1+y'_{ij})}{\Gamma(1-\bar{y}'_{ij})}
	\frac{\Gamma(1-y'_{ij})}{\Gamma(1+\bar{y}'_{ij})} \\
	\nonumber
	& = \frac{(2\pi)^n}{\prod_{i<j}^n y'_{ij}\,y'_{ji}}
	\prod\limits_{i<j}^n
	\frac{y'_{ij}}{\bar{y}'_{ij}}\,
	\mathbf{\Gamma}(y'_{ij})\,\mathbf{\Gamma}(1-y'_{ij})
	= \frac{(2\pi)^n}{\prod_{i<j}^n \bar{y}'_{ij}\,y'_{ji}}
	\prod\limits_{i<j}^n [-1]^{y'_{ij}} \\
	\label{pref}
	& = \frac{(2\pi)^n}{\prod_{i<j}^n |y_i'-y_j'|^2}
	\prod\limits_{i<j}^n [-1]^{y'_i-y'_j}
	= \frac{1}{\pi^{n^2}\,n!}\,\mu^{-1}(\bm{y}'_n)\,
	[-1]^{(n-1)\underline{\bm{y}'_n}+n(n-1)s} \,,
\end{align}
where at the last step we applied the explicit formula \eqref{mu} for the measure $\mu(\bm{y}'_n)$ and the identity
\begin{equation*}
	\prod\limits_{i<j}^n [-1]^{y'_i-y'_j} = [-1]^{(n-1)\underline{\bm{y}'_n}+n(n-1)s} \,,
\end{equation*}
which can be checked by direct calculation using the explicit form \eqref{x} of the variables $y'_i, \bar{y}'_i$. Substituting the result \eqref{pref} into \eqref{lim2} we obtain the needed relation \eqref{lim}.

\section{Conclusions} \label{sect:Conclus}

Let us collect the main formulae.
The fundamental building blocks are $\mathcal{R}$-operator \eqref{R} and $\mathcal{K}$-operator \eqref{Kop-d}
\begin{align*}
	& \mathcal{R}_{k j}(x) = 
	[z_{kj}]^{1-2s}\,[\hat{p}_k]^{x-s}\,[z_{kj}]^{s+x-1} \,, \\
	& \mathcal{K}(s,x) =
	[z+\gamma]^{g-s} \,
	[z-\gamma]^{1-s-g} \,
	[\hat{p}]^{x-s} \,
	[z+\gamma]^{x-g} \,
	[z-\gamma]^{x+g-1} \,,
\end{align*}
where $z_{kj}=z_k-z_j$, the double power $[z]^a$ and integral operator $[\hat{p}]^\alpha$ are defined in \eqref{power} and~\eqref{d}, the spin parameter $s$ is given by \eqref{sparam}, and $g$ is the parameter of boundary $K$-matrix~\eqref{K}.
The first building block is determined by the relation \eqref{defR} with $L$-operator~\eqref{Lfact}
\begin{align} \nonumber
	\mathcal{R}_{1 2}(x) \, L_1(u_1, u_2) \, L_2(u_1, u-x) = 
	L_1(u_1, u-x) \, L_2(u_1, u_2) \, \mathcal{R}_{1 2}(x) \,.
\end{align} 
The second one is the solution of the reflection equation \eqref{Kdef}
\begin{multline} \nonumber
	\mathcal{K}(s,x) \, L(u +x - 1, u - s) \, K(u) \, L(u  + s - 1, u - x) \\
	= L(u + s - 1, u - x) \, K(u) \, L(u + x - 1, u - s) \, 
	\mathcal{K}(s,x) \,,
\end{multline}
where $K(u)$ is the boundary matrix.
There is a commutative family of operators~\eqref{Qop}
\begin{align} \nonumber
	Q_n(x) =
	\mathcal{R}_{n\,n-1}(x)\cdots 
	\mathcal{R}_{21}(x)
	\mathcal{K}_1(s,x)\,
	\mathcal{R}_{12}(x)
	\cdots \mathcal{R}_{n-1\,n}(x)\,[\hat{p}_n]^{x-s}
\end{align}
which commute with the element $B(u)$ of monodromy matrix \eqref{Top} and with its antiholomorphic counterpart $\bar{B}(\bar{u})$.
Operator $Q(x)$ obeys all characteristic properties of the Baxter $Q$-operator
\begin{eqnarray*}
	Q_n(x)\,B(u) &=& B(u)\,Q_n(x) \,, \\
	Q_n(x)\,Q_n(y) &=& Q_n(y)\,Q_n(x) \,, \\
	B(u)\,Q_n(u,\bar{u}) &=& \imath^{2(n-1)}\,{\textstyle \left(u-\frac{1}{2}\right)}\,
	Q_n(u+1\,,\bar{u}) \,,
\end{eqnarray*}
the relations between $Q_n(x)$ and $\bar{B}(\bar{u})$ copy their holomorphic analogues with $B(u)$.
We obtained the common eigenfunctions of $B(u)$-, $\bar{B}(\bar{u})$- and $Q$-operators
\begin{eqnarray*}
	B_{n}(u)\,\Psi_{\bm x_n}(\bm z_n) &=&
	\textstyle
	\left(u - \frac{1}{2} \right)\left(u^2-x^2_1\right)\cdots
	\left(u^2-x^2_{n}\right)\,
	\Psi_{\bm x_n}(\bm z_n) \,, \\
	\bar{B}_{n}(\bar{u})\,\Psi_{\bm x_n}(\bm z_n) &=&
	\textstyle
	\left(\bar{u} - \frac{1}{2} \right)\left(\bar{u}^2-\bar{x}^2_1\right)\cdots
	\left(\bar{u}^2-\bar{x}^2_{n}\right)\,
	\Psi_{\bm x_n}(\bm z_n) \,, \\
	Q_n(u)\,\Psi_{\bm x_n}(\bm z_n) &=&
	q(u\,,\bm x_{n}) \, \Psi_{\bm x_n}(\bm z_n) \,,
\end{eqnarray*}
where we use the compact notations for tuples of coordinate and spectral variables $\bm z_n=(z_1,\bar{z}_1, \ldots, z_n, \bar{z}_n)$, $\bm x_n=(x_1,\bar{x}_1, \ldots, x_n, \bar{x}_n)$. The eigenvalue of $Q$-operator is given by expression \eqref{qn}.
The eigenfunctions are constructed inductively 
\begin{equation*}
	\Psi_{\bm x_n}(\bm z_n) = 
	\Lambda_n(x_n) \, \Lambda_{n-1}(x_{n-1}) \cdots \Lambda_{1}(x_1) \cdot 1 \,,
\end{equation*}
with the help of raising operators \eqref{L_k}
\begin{align*}
	& \Lambda_k(x) = \lambda_k(x)\,\mathcal{R}_{k\,k-1}(x)\,
	\mathcal{R}_{k-1\, k-2}(x)
	\ldots \mathcal{R}_{2 1}(x)\,
	\mathcal{K}_1(s,x)\,
	\mathcal{R}_{1 2}(x) \, \mathcal{R}_{2 3}(x) \ldots 
	\mathcal{R}_{k-1\, k}(x) \,, \\
	& \Lambda_1(x)=\lambda_1(x)\,\mathcal{K}_1(s,x)
\end{align*}
where $\lambda_k(x)$ is the normalization constant \eqref{lambda}.
That is, to get $\Psi_{\bm x_n}$ one needs to act by the product of raising operators on the function identically equal to $1$.
The integral operator $\Lambda_k(x_k)$ maps the function $\Psi_{\bm{x}_{k-1}}$ to $\Psi_{\bm{x}_{k}}$
\begin{equation*}
	\Psi_{\bm{x}_{k}}(\bm{z}_k)=[\Lambda_k(x_k)\,\Psi_{\bm{x}_{k-1}}](\bm z_k) = \int \mathrm{d}^{2} \bm w_{k-1} \, 
	\Lambda(\bm z_k,\bm w_{k-1};x_k)\,\Psi_{\bm{x}_{k-1}}(\bm w_{k-1}) \,,
\end{equation*}
where $\mathrm{d}^2\bm{w}_{k-1} = \prod_{m=1}^{k-1} \mathrm{d}^2 w_m$.
One can express its kernel $\Lambda(\bm z_k,\bm w_{k-1};x)$ as a $k$-fold integral of the product of power functions using the explicit formulas for integral operators $\mathcal{R}_{k j}(x)$ and $\mathcal{K}_1(s,x)$.
This way, the described iterative construction gives the representation for an eigenfunction in terms of the multiple integral of the product of double powers.
In graphical language \cite{DKM,DM14,DMV:Gust3} it is the representation in terms of a Feynmann diagram, where the power function plays the role of Feynmann propagator.

We also obtained the Mellin-Barnes integral representation \eqref{PsiMB} for the constructed eigenfunctions of BC-type spin chain.
We decomposed $\Psi_{\bm{x}_n}$ over the complete set of eigenfunctions $\psi_{\bm{y}_n}(\bm{z}_n)$ of the operator $a_n(u)+\gamma\,b_n(u)$, where $a_n(u)$ and $b_n(u)$ are elements of the monodromy matrix \eqref{t} for $SL(2,\mathbb{C})$ spin chain of type~A.
The formula reads
\begin{equation*}
	\Psi_{\bm{x}_n}(\bm{z}_n) =
	\lim\limits_{\varepsilon\to 0_+}
	\int\mathcal{D}\bm{y}_n\,\tilde{\mu}(\bm{y}_n)\,
	P_\varepsilon(\bm{y}_n,\bm{x}_n)\,
	\hat{\psi}_{\bm{y}_n}(\bm{z}_n) \,,
\end{equation*}
where $\hat{\psi}_{\bm{y}_n}(z_1,\ldots,z_n)=\psi_{\bm{y}_n}(z_n,\ldots,z_1)$, the Sklyanin's measure $\tilde{\mu}(\bm{y}_n)$ for A-type spin chain is given by \eqref{mu} and the kernel $P_\varepsilon(\bm{y}_n,\bm{x}_n)$ is expressed in terms of gamma functions of the complex field in \eqref{Peps}.
The integration variables have the form \eqref{x} and the corresponding integral is defined in \eqref{intx}.
The functions $\psi_{\bm{y}_n}$ are given by formula \eqref{reprPsiCl}, they are constructed in a similar manner to $\Psi_{\bm{x}_n}$ by the use of raising integral operators~\eqref{LamCl}.
In contrast to the previous integral representation, the integration is performed with respect to spectral variables $\bm{y}_n$ and the coordinate variables $\bm{z}_n$ are fixed.
The Mellin-Barnes integral representation was also constructed for the eigenfunctions $\hat{\psi}_{\bm{y}_n}$ of A-type spin chain \cite{Val20}, \cite[Section~7]{ADV1}.
Therefore, the function $\Psi_{\bm{x}_n}(\bm{z}_n)$ can be expressed in terms of the multiple gamma function integral of Mellin-Barnes type.

The eigenfunctions of BC-type spin chain are symmetric with respect to the action of the group $\mathbb{Z}_2\ltimes\mathfrak{S}_n$ -- the Weyl group of B and C root systems.
That is, for any permutation $\tau$ and inversion $\sigma_k$ of spectral variables one has
\begin{align*} 
	\Psi_{\tau \bm x_n}(\bm z_n) = 
	\Psi_{\bm x_n}(\bm z_n)\, , \qquad
	\Psi_{\sigma_k \bm x_n}(\bm z_n) = 
	\Psi_{\bm x_n}(\bm z_n) \,,
\end{align*}
where $\tau \bm x_n = (x_{\tau(1)},\bar{x}_{\tau(1)}, \ldots , x_{\tau(n)},\bar{x}_{\tau(n)})$ and $\sigma_k \bm x_n = (x_{1},\bar{x}_1,\ldots , -x_{k},-\bar{x}_{k}, 
\ldots , x_{n},\bar{x}_{n})$. Correspondingly, the eigenvalues of $B$- and $Q$- operators possess the same symmetry.
These properties of eigenfunctions follow from the relations for raising operators
\begin{equation} \nonumber
	\Lambda_k(x)\,\Lambda_{k-1}(y) =  
	\Lambda_k(y)\,\Lambda_{k-1}(x) \,, \qquad
	\Lambda_k(x) = \Lambda_k(-x) \,,
\end{equation}
which are derived in Section~\ref{sect:Symmetry}.
The commutation relation for $\Lambda$-operators is the consequence of the commutativity of $Q$-operators.

We have proven that the constructed system of eigenfunctions is orthogonal and complete. The orthogonality relation has the form
\begin{equation*}
	\langle\Psi_{\bm{y}_n}|\Psi_{\bm{x}_n}\rangle = 
	\mu^{-1}(\bm{x}_n)\, \delta^{(2)}(\bm{x}_n,\bm{y}_n) \,,
\end{equation*}
where we consider the standard scalar product in the space $\mathrm{L}^2(\mathbb{C}^n)$, and the delta function of spectral variables $\delta^{(2)}(\bm{x}_n,\bm{y}_n)$
is given by the formula \eqref{deltaSym}, it possesses the same symmetry under permutations and reflections of these variables as the eigenfunctions.
From the orthogonality formula we obtain the measure $\mu(\bm{x}_n)$ \eqref{mu0} for the completeness relation~\eqref{compl}, which can be written in compact Dirac bra-ket notation as follows
\begin{equation*}
	\int \mathcal{D}\bm{x}_n \, \mu(\bm{x}_n) \,
	|\Psi_{\bm{x}_n}\rangle \langle\Psi_{\bm{x}_n}| = \II\,,
\end{equation*}
where $\II$ is the identity operator in the coordinate space.

There is also a simple transformation rule for eigenfunctions in the case of reflection of model's parameters $(s,g)\to(1-s,1-g)$. Namely,
\begin{multline*}
	\Psi_{\bm{x}_n}(1-s,1-g;\bm{z}_n) \\
	= c(\bm{x}_n)\,[z_{12}]^{2s-1}\,[z_{23}]^{2s-1} 
	\cdots [z_{n-1\,n}]^{2s-1}\,[z_1+\gamma]^{s-g}\,
	[z_1-\gamma]^{s+g-1}\,\Psi_{\bm{x}_n}(s,g;\bm{z}_n) \,.
\end{multline*}
The numerical factor reads
\begin{equation*}
	c(\bm{x}_n) = [-1]^{ng+(n^2+2)s}\,[2\gamma]^{n-2ns}
	\prod\limits_{k=1}^n \mathbf{\Gamma}^{2n-1}(s\pm x_k)\,\mathbf{\Gamma}(1-g\pm x_k) \,,
\end{equation*}
where we use the notation \eqref{pm} for the gamma function of the complex field \eqref{gamma-def}.

In~\cite[Section~5]{IS} the $B$-element of monodromy matrix was diagonalized for a resembling model with boundary interaction -- the B-type Toda lattice.
The boundary matrix considered in that model can be obtained from the present $K$-matrix \eqref{K} by taking $\gamma=0$.
The resulting integral representation for corresponding eigenfunctions is similar to the Mellin-Barnes representation constructed in our case.
The eigenfunctions were expanded over the basis of wavefunctions for the open Toda chain, and the kernel of the integral decomposition was also expressed in terms of gamma functions.

The iterative expression for eigenfunctions in terms of integral raising operators was derived for other integrable models as well, including $SL(2,\mathbb{R})$ \cite{BDM} and $SL(2,\mathbb{C})$ \cite{DKM,DM14} spin chains of type~A and Ruijsenaars hyperbolic system \cite{BDKK1,BDKK2,BDKK3,BDKK4}.
In addition, the dual raising operator was constructed for each of these models \cite{DM:Gust1,Val20,ADV1,BDKK2}.
It is also an integral operator which maps the eigenfunction for $k-1$ particles to the eigenfunction for $k$ particles.
But, in contrast to the ordinary raising operator, the integration variables in the corresponding formula are the spectral variables of the eigenfunction under the integral, and the coordinate variables stay fixed.
The construction of the dual raising operator is one of the obvious open problems for the model in question.
It can be solved by the use of the derived Mellin-Barnes representation for eigenfunctions of BC-type spin chain and the dual raising operator for A-type chain.

It is also worth making an alternative check of the equivalence of Mellin-Barnes integral representation and the iterative formula with raising operators, without using the completeness of eigenfunctions $\psi_{\bm{y}_n}$.
In the case of A-type spin chain the similar representations of eigenfunctions are obtained by using two types of raising operators.
In~\cite[Section~7]{ADV1} their equivalence is proven inductively by the use of $Q$-operator and the complex generalization of A-type Gustafson integral.
Taking into account the structure of Mellin-Barnes formula in our case, the analogous proof will most likely use the $Q$-operator~\eqref{Qcl} and dual raising operator for A-type spin chain along with the BC-type Gustafson integral~\eqref{Gust}.

The $SL(2,\mathbb{C})$ spin chain appears in description of the high-energy behavior of
quantum chromodynamics, see~\cite{L0,L1,L2,FK} and in connection to the two-dimensional version \cite{KO,DKO} of conformal fishnet field theory \cite{KG,BD,BD1,BCF,GGKK,GKK,GKKNS,CK}.
The set of A-type eigenfunctions was used for the calculation of the two-dimensional Basso-Dixon diagramm in \cite{DKO} and has direct analogue in the case of higher dimensions \cite{DO1,DO2,DO3,DFO}. 
After that the multi-leg fishnet integrals in two dimensions were studied 
and its connection to the geometry of Calabi-Yau varieties was established \cite{Duhr1,Duhr2}.
The recent work \cite{LS} shows intriguing similarities of the conformal integrals in different dimensions so that there exists some hope to connect two-dimensional results of this paper, \cite{Duhr1,Duhr2} and \cite{DKO} with higher dimensional analogs \cite{BD,BD1,DO1,DO2,DO3,DFO}. 

A connection is established \cite{DM:Gust1,DMV:Gust2,DMV:Gust3,M} 
between the Gustafson integrals 
\cite{G1,G2,G3} and integrable spin chains and their generalization to the complex field is obtained \cite{DMV:Gust2,DMV:Gust3,DM:Gust4,SS0,SS1,SS2,N,MN}.
We have demonstrated in \cite{ADV1} and in the present paper the Gustafson integrals 
at work and hope that it can help to perform calculations in the program of the  
hexagonalization of Fishnet integrals \cite{O1,O2,AO} or in more complicated 
fishnet models \cite{AFKO,KO1,KFM,Oliv}.

In recent years, significant progress has been made in 
constructing SoV representations for higher rank finite-dimensional models of A-type, see~\cite{Cavaglia:2019pow,GromovRyan20,GromovSizov17,Gromov:2022waj,Levk,MailletNiccoli18,
MR3983970,MailletNiccoli19,Maillet:2020ykb,Ryan:2021duf,RyanVolin19,
Ryan:2020rfk}, but for BC-type models or the case of reflection algebra 
\cite{MNopen,IS} the problem is more complicated.   
It seems that development of an algebraic approach is the first needed step for the generalization of the SoV to the case of BC-type models with 
noncompact symmetry groups of the higher rank.  

The additional source of interest to the BC-type noncompact spin chains with nontrivial boundary interaction are the integrable models of stochastic particle processes 
\cite{Frassek1,Frassek2,Frassek3,Frassek4}.

\section*{Acknowledgements}

We are grateful to N.Belousov, S.Khoroshkin and A.Manashov for fruitful discussions.
The work of P.~Antonenko and S.~Derkachov was supported by the Russian Science Foundation under grant No. \mbox{23-11-00311}.

\appendix

\section{Chain rule and star-triangle relation}

Throughout the text we apply several useful identities. The proofs can be found in \cite[Appendix A]{DKM}. The first one is the \textit{chain relation}.
It can be represented in two equivalent forms -- as an integral identity 
\begin{equation}\label{Chain}
	\int \mathrm{d}^2 w \, \frac{1}{[z-w]^a \, [w-z_0]^{b}}=
	\frac{\pi \, [-1]^c}{\mathbf{\Gamma}(a,b,c)}
	\frac{1}{[z-z_0]^{a+b-1}},
\end{equation}
where $c=2-a-b,\ \bar c=2-\bar a-\bar b$,
and as result of the application of the operator $[\hat{p}]^{a-1}$ to 
the particular function $[z-z_0]^{-b}$
\begin{equation}\label{Chain1}
[\hat{p}]^{a-1}\,[z-z_0]^{-b} =
\frac{[\imath]^{a}\,[-1]^{-a-b}}{\mathbf{\Gamma}(b,2-a-b)}\,
[z-z_0]^{1-a-b},
\end{equation}
where we used \eqref{d} and \eqref{c}.

From the right we have gamma function of the complex field~\cite[Section~1.4]{GGR},~\cite[Section 1.3]{N}
\begin{equation}\label{gamma-def}
	\bm{\Gamma}(a) = \frac{\Gamma(a)}{\Gamma(1 - \bar{a})}
\end{equation}
and we denote its products as
\begin{align*}
	\bm{\Gamma}(a,b) = \bm{\Gamma}(a)  \bm{\Gamma}(b).
\end{align*}
Notice that $\bm{\Gamma}(a)$ depends on two parameters $(a, \bar{a}) \in \mathbb{C}^2$ such that $a - \bar{a} \in \mathbb{Z}$, but for brevity we display only the first one. Moreover, for $\rho \in \mathbb{R}$ we write
\begin{align*}
	\bm{\Gamma}(a + \rho) \equiv \frac{\Gamma(a + \rho)}{\Gamma(1 - \bar{a} - \rho)}.
\end{align*}
From the well-known properties of the ordinary gamma function it is easy to prove the following relations
\begin{align}\label{gamma-diff}
	& \bm{\Gamma}(a + 1) = - a \bar{a} \, \bm{\Gamma}(a), \\[6pt]
	\nonumber 
	& \bm{\Gamma}(a) = [-1]^{a} \, \bm{\Gamma}(\bar{a}), \\[6pt] \label{gamma-refl}
	& \bm{\Gamma}(a) \, \bm{\Gamma}(1 - a) = [-1]^a.
\end{align}
and the complex conjugation rule:
\begin{equation} \label{gamma-conj}
	\mathbf{\Gamma}(\rho+a)^\ast = \bm{\Gamma}(\rho - \bar{a}) = [-1]^a \, \mathbf{\Gamma}(\rho-a),
	\qquad \rho \in \mathbb{R},
\end{equation}
where $a,\bar{a}$ are of the form
\begin{equation*}
	a = \frac{n}{2} + \imath \nu, \qquad
	\bar{a} = -\frac{n}{2} + \imath \nu,
	\qquad n \in \mathbb{Z}, \quad \nu \in \mathbb{R} \,.
\end{equation*}

The second identity is the \textit{star-triangle relation}. 
This identity can be represented in three equivalent forms -- as integral identity
\begin{align*} 
\int \mathrm{d}^2 w \, \frac{1}{[w-z_1]^a[w-z_2]^b [w-z_3]^c}
 = \frac{\pi}{\mathbf{\Gamma}(a,b,c)}
\frac{1}{[z_{12}]^{1-c}[z_{31}]^{1-b}[z_{23}]^{1-a}}\,,
\end{align*}
where again $a+b+c=2\,, \bar a+\bar b+\bar c=2$ and 
as result of application of operator $[\hat{p}_1]^{a-1}$ 
to the particular function $[z_{12}]^{-b} [z_{13}]^{-c}$
\begin{align}\label{Star1}
[\hat{p}_1]^{a-1}\,[z_{12}]^{-b} [z_{13}]^{-c}
= \frac{[\imath]^{-a}}
{\mathbf{\Gamma}(b,c)}\,
[z_{12}]^{c-1}\,[z_{31}]^{b-1}\,[z_{23}]^{a-1}\,,
\end{align}
The last and most compact form is the operator form of the star-triangle relation \cite{Isa,DM09} 
\begin{align}\label{star-tr}
[\hat{p}]^{a}\,[z-z_0]^{a+b}\,[\hat{p}]^{b} = 
[z-z_0]^{b}\,[\hat{p}]^{a+b}\,[z-z_0]^{a}
\end{align}
In the main text we need formula for the application of considered operator to the constant function $\psi(z) = 1$. Using the chain rule 
\eqref{Chain1} we obtain
\begin{align}\label{star-tr1}
[\hat{p}]^{a}\,[z -z_0]^{a+b}\,[\hat{p}]^{b}\cdot 1 = 
[z-z_0]^{b}\,[\hat{p}]^{a+b}\,[z-z_0]^{a}\cdot 1 = \\ 
[z-z_0]^{b}\,\frac{[\imath]^{a+b+1}\,[-1]^{-a-b-1+a}}{\mathbf{\Gamma}(-a,2-a-b-1+a)}\,
[z-z_0]^{1-a-b-1+a} = 
\frac{[\imath]^{a+b+1}\,[-1]^{-b}}{\mathbf{\Gamma}(-a,1-b)}\,
\end{align}
	
Finally, we give two representations for the $\delta$ function.
The first one is
\begin{equation}\label{delta1}
	\delta^2(z)=\lim_{\varepsilon\to 0}\frac{\varepsilon}{\pi} \frac{1}{[z]^{1-\varepsilon}}
\end{equation}
and the second formula
\begin{equation}\label{delta2}
	\int \mathrm{d}^2 w\frac{1}{[z_1-w]^{2-\alpha}[w-z_2]^\alpha} =
	\frac{\pi^2}{\mathbf{\Gamma}(\alpha,2-\alpha)}\,\delta^2(z_1-z_2)\,
\end{equation}
results from the chain relation~(\ref{Chain}) and (\ref{delta1}). In operator form the identity \eqref{delta2} is represented as the action of operator $[\hat{p}]^{1-\alpha}$ on the function $[z-z_0]^{-\alpha}$
\begin{equation} \label{delta3}
	[\hat{p}]^{1-\alpha}\,[z-z_0]^{-\alpha} = \frac{\pi\,[\imath]^{-\alpha}}{\mathbf{\Gamma}(\alpha)}\,\delta^2(z-z_0) \,.
\end{equation}

\section{Reflection operator $\mathcal{K}(s,x)$}
\label{Reflection}

The explicit expression for the reflection operator has the following form
\begin{align} \nonumber
\mathcal{K}(s,g;x) =
[z+\gamma]^{g-s} \,
[z-\gamma]^{1-s-g} \,
[\hat{p}]^{x-s} \,
[z+\gamma]^{x-g} \,
[z-\gamma]^{x+g-1} = \\  
\label{K1}
[z_+]^{g-s} \,
[z_-]^{1-s-g} \,
[\hat{p}]^{x-s} \,
[z_+]^{x-g} \,
[z_-]^{x+g-1}.
\end{align}
where in the second line we used the 
compact notation $z_{\pm} = z\pm\gamma$ for simplicity. 
Note that we will omit the explicit dependence on 
parameters $s$ and $g$ when it is evident from context. 

We are going to prove the commutativity relation for 
the $Q$-operators for one site 
\begin{align*}
\mathcal{K}(x)\,[\hat{p}]^{x-s}\,
\mathcal{K}(y)\,[\hat{p}]^{y-s} = 
\mathcal{K}(y)\,[\hat{p}]^{y-s}\,
\mathcal{K}(x)\,[\hat{p}]^{x-s}
\end{align*}
using two equivalent representations for $Q$-operator 
\begin{align*}
Q(x) = \mathcal{K}(s,x)\,[\hat{p}]^{x-s} = 
[z_+]^{g-s}\,
[z_-]^{1-s-g}\,
[\hat{p}]^{x-s}\,
[z_+]^{x-g}\,
[z_-]^{x+g-1}\,[\hat{p}]^{x-s} = \\ 
[\hat{p}]^{x+s-1}\,
[z_+]^{g+x-1}\,
[z_-]^{x-g}\,[\hat{p}]^{x-s}\,
[z_+]^{1-s-g}\,
[z_-]^{g-s}\,[\hat{p}]^{1-2s}
\end{align*}
To prove equivalence of these representations one 
uses the star-triangle relation in operator form \eqref{star-tr} (for simplicity we use compact notation $z_{\pm} = z\pm\gamma$)
\begin{multline*} 
[z_-]^{1-s-g}\,
{\blue [z_+]^{g-s}\,[\hat{p}]^{x-s}\,
[z_+]^{x-g}}\,
[z_-]^{x+g-1}\,[\hat{p}]^{x-s} = \\ 
[z_-]^{1-s-g}\,[\hat{p}]^{x-g}\,[z_+]^{x-s}
\,{\blue [\hat{p}]^{g-s}\,[z_-]^{x+g-1}\,[\hat{p}]^{x+s-1}}
\,[\hat{p}]^{1-2s} = \\
{\blue [z_-]^{1-s-g}\,[\hat{p}]^{x-g}\,[z_-]^{x+s-1}}\,
[z_+]^{x-s}\,[\hat{p}]^{x+g-1}\,[z_-]^{g-s}
\,[\hat{p}]^{1-2s} = \\ 
[\hat{p}]^{x+s-1}\,[z_-]^{x-g}\,
{\blue [\hat{p}]^{1-s-g}\,[z_+]^{x-s}\,[\hat{p}]^{x+g-1}}\,[z_-]^{g-s}
\,[\hat{p}]^{1-2s} = \\ 
[\hat{p}]^{x+s-1}\,[z_-]^{x-g}\,
[z_+]^{x+g-1}\,[\hat{p}]^{x-s}\,[z_+]^{1-s-g}
\,[z_-]^{g-s}\,[\hat{p}]^{1-2s}  
\end{multline*}
Note, that this transformation can be rewritten 
in an equivalent form 
\begin{align}\label{pK}
[\hat{p}]^{1-x-s}\,\mathcal{K}(s;x) = 
\mathcal{K}(1-x;1-s)\,[\hat{p}]^{1-x-s}
\end{align}

The expression for the product of two $Q$-operators has the form 
\begin{multline*}
Q(x)\,Q(y) = [z_-]^{1-s-g}\,[z_+]^{g-s}\,
[\hat{p}]^{x-s}\,
[z_+]^{x-g}\,[z_-]^{x+g-1}\,[\hat{p}]^{x-s}\,\\
[\hat{p}]^{y+s-1}\,[z_-]^{y-g}\,
[z_+]^{y+g-1}\,[\hat{p}]^{y-s}\,[z_+]^{1-s-g}
\,[z_-]^{g-s}\,[\hat{p}]^{1-2s}   
\end{multline*}
where we used the first representation 
for $Q(x)$ and the second representation for $Q(y)$.
We have to prove that the whole expression 
in the middle which depends on $x$ and $y$ can be transformed 
to the same form with exchange $x \rightleftarrows y$. 
This transformation is based on the star-triangle relation 
\eqref{star-tr} again 
\begin{multline*}
[\hat{p}]^{x-s}\,
[z_+]^{x-g}\,{\blue [z_-]^{x+g-1}\,
[\hat{p}]^{x+y-1}\,[z_-]^{y-g}}\,
[z_+]^{y+g-1}\,[\hat{p}]^{y-s} = \\    
[\hat{p}]^{x-s}\,
[z_+]^{x-g}\,[\hat{p}]^{y-g}\,[z_-]^{x+y-1}\,
{\blue [\hat{p}]^{x+g-1}\,
[z_+]^{y+g-1}\,[\hat{p}]^{y-x}}\,[\hat{p}]^{x-s} = \\
[\hat{p}]^{x-s}\,
{\blue [z_+]^{x-g}\,[\hat{p}]^{y-g}\,[z_+]^{y-x}}\,[z_-]^{x+y-1}\,
[\hat{p}]^{y+g-1}\,[z_+]^{x+g-1}\,[\hat{p}]^{x-s} = \\
[\hat{p}]^{y-s}\,
[z_+]^{y-g}\,{\blue [\hat{p}]^{x-g}\,[z_-]^{x+y-1}\,
[\hat{p}]^{y+g-1}}\,[z_+]^{x+g-1}\,[\hat{p}]^{x-s} = \\ 
[\hat{p}]^{y-s}\,[z_+]^{y-g}\,
[z_-]^{y+g-1}\,[\hat{p}]^{x+y-1}\,[z_-]^{x-g}\,
[z_+]^{x+g-1}\,[\hat{p}]^{x-s}
\end{multline*}
Now we are going to prove the general reflection relation for K-operators
\begin{multline*}
[z_{12}]^{1-s-y}\,
[\hat{p}_1]^{s^{\prime}-s}\,[\hat{p}_2]^{x-y}\,
[z_{12}]^{x+s^{\prime}-1}\,
\mathcal{K}_1(s,x)\,[z_{12}]^{1-s-s^{\prime}}\,
[\hat{p}_1]^{x-s^{\prime}}\,[\hat{p}_2]^{y-s}\,
[z_{12}]^{x+y-1}\, \mathcal{K}_1(s^{\prime},y) = \\ 
\mathcal{K}_1(s^{\prime},y)\,[z_{12}]^{1-s-s^{\prime}}\,
[\hat{p}_1]^{y-s}\,[\hat{p}_2]^{x-s^{\prime}}\,
[z_{12}]^{x+y-1}\, \mathcal{K}_1(s,x)\,
[z_{12}]^{1-s-y}\,
[\hat{p}_1]^{x-y}\,[\hat{p}_2]^{s^{\prime}-s}\,
[z_{12}]^{x+s^{\prime}-1} 
\end{multline*}
The first step is based on the following main relation
\begin{multline}\label{main1}
[z_{12}]^{x+s^{\prime}-1}\,
\mathcal{K}_1(s,x)\,[z_{12}]^{1-s-s^{\prime}}\,
[\hat{p}_1]^{x-s^{\prime}}\,[\hat{p}_2]^{y-s}\,
[z_{12}]^{x+y-1}\, \mathcal{K}_1(s^{\prime},y) = \\ 
\mathcal{K}_1(s,1-s^{\prime})\,[z_{12}]^{x-s}\,
[\hat{p}_1]^{x-s^{\prime}}\,[\hat{p}_2]^{y-s}\,
[z_{12}]^{y-s^{\prime}}\, \mathcal{K}_1(1-x,y)\,
[z_{12}]^{x+s^{\prime}-1} 
\end{multline}
The movement of the operator $[z_{12}]^{x+s^{\prime}-1}$ from the left to the right leads to the change of parameters: 
$x \to 1- s^{\prime}$ and $s^{\prime} \to 1-x$.

We use this transformation in the left hand side of our identity 
and then cancel operator $[z_{12}]^{x+s^{\prime}-1}$ from the right
\begin{multline*}
[z_{12}]^{1-s-y}\,
[\hat{p}_1]^{s^{\prime}-s}\,[\hat{p}_2]^{x-y}\,
\mathcal{K}_1(s,1-s^{\prime})\,[z_{12}]^{x-s}\,
[\hat{p}_1]^{x-s^{\prime}}\,[\hat{p}_2]^{y-s}\,
[z_{12}]^{y-s^{\prime}}\, \mathcal{K}_1(1-x,y) = \\ 
\mathcal{K}_1(s^{\prime},y)\,[z_{12}]^{1-s-s^{\prime}}\,
[\hat{p}_1]^{y-s}\,[\hat{p}_2]^{x-s^{\prime}}\,
[z_{12}]^{x+y-1}\, \mathcal{K}_1(s,x)\,
[z_{12}]^{1-s-y}\,
[\hat{p}_1]^{x-y}\,[\hat{p}_2]^{s^{\prime}-s}\,
\end{multline*}
The second step is very similar and is based on the second main relation
\begin{multline}\label{main2}
\mathcal{K}_1(s^{\prime},y)\,[z_{12}]^{1-s-s^{\prime}}\,
[\hat{p}_1]^{y-s}\,[\hat{p}_2]^{x-s^{\prime}}\,
[z_{12}]^{x+y-1}\, \mathcal{K}_1(s,x)\,
[z_{12}]^{1-s-y} = \\
[z_{12}]^{1-s-y}\,
\mathcal{K}_1(s^{\prime},1-s)\,[z_{12}]^{y-s^{\prime}}\,
[\hat{p}_1]^{y-s}\,[\hat{p}_2]^{x-s^{\prime}}\,
[z_{12}]^{x-s}\, \mathcal{K}_1(1-y,x) 
\end{multline}
The movement of the operator $[z_{21}]^{1-s-y}$ from the right to the 
left leads to the change of parameters: 
$y \to 1- s$ and $s \to 1-y$.
We use this transformation in the right hand side of our identity 
and then cancel operator $[z_{21}]^{1-s-y}$ from the left
\begin{multline*}
[\hat{p}_1]^{s^{\prime}-s}\,[\hat{p}_2]^{x-y}\,
\mathcal{K}_1(s,1-s^{\prime})\,[z_{12}]^{x-s}\,
[\hat{p}_1]^{x-s^{\prime}}\,[\hat{p}_2]^{y-s}\,
[z_{12}]^{y-s^{\prime}}\, \mathcal{K}_1(1-x,y) = \\ 
\mathcal{K}_1(s^{\prime},1-s)\,[z_{12}]^{y-s^{\prime}}\,
[\hat{p}_1]^{y-s}\,[\hat{p}_2]^{x-s^{\prime}}\,
[z_{12}]^{x-s}\, \mathcal{K}_1(1-y,x)
[\hat{p}_1]^{x-y}\,[\hat{p}_2]^{s^{\prime}-s}\,
\end{multline*}
Next step we use the identity \eqref{pK} in the following forms 
\begin{align*}
[\hat{p}_1]^{s^{\prime}-s}\,\mathcal{K}_1(s,1-s^{\prime}) = \mathcal{K}_1(s^{\prime},1-s)\,[\hat{p}_1]^{s^{\prime}-s}\ ;\  \\ 
\mathcal{K}_1(1-y,x)\,
[\hat{p}_1]^{x-y} = [\hat{p}_1]^{x-y}\,\mathcal{K}_1(1-x,y)
\end{align*}
and obtain
\begin{multline*}
[\hat{p}_2]^{x-y}\,
\mathcal{K}_1(s^{\prime},1-s)\,[\hat{p}_1]^{s^{\prime}-s}\,[z_{12}]^{x-s}\,
[\hat{p}_1]^{x-s^{\prime}}\,[\hat{p}_2]^{y-s}\,
[z_{12}]^{y-s^{\prime}}\, \mathcal{K}_1(1-x,y) = \\ 
\mathcal{K}_1(s^{\prime},1-s)\,[z_{12}]^{y-s^{\prime}}\,
[\hat{p}_1]^{y-s}\,[\hat{p}_2]^{x-s^{\prime}}\,
[z_{12}]^{x-s}\,[\hat{p}_1]^{x-y}\, \mathcal{K}_1(1-x,y)
\,[\hat{p}_2]^{s^{\prime}-s}\,
\end{multline*}
Operator $\hat{p}_2$ commutes with operator $\mathcal{K}_1$ 
so that it is possible to cancel $\mathcal{K}_1(s^{\prime},1-s)$ from the left and 
$\mathcal{K}_1(1-x,y)$ from the right
\begin{multline*}
[\hat{p}_2]^{x-y}\,[\hat{p}_1]^{s^{\prime}-s}\,[z_{12}]^{x-s}\,
[\hat{p}_1]^{x-s^{\prime}}\,[\hat{p}_2]^{y-s}\,
[z_{12}]^{y-s^{\prime}} = \\ 
[z_{12}]^{y-s^{\prime}}\,
[\hat{p}_1]^{y-s}\,[\hat{p}_2]^{x-s^{\prime}}\,
[z_{12}]^{x-s}\,[\hat{p}_1]^{x-y}\,[\hat{p}_2]^{s^{\prime}-s}\,
\end{multline*}
The remaining identity is proved with the use of star-triangle relation 
\begin{multline*}
[\hat{p}_2]^{x-y}\,{\blue [\hat{p}_1]^{s^{\prime}-s}\,[z_{12}]^{x-s}\,
[\hat{p}_1]^{x-s^{\prime}}}\,[\hat{p}_2]^{y-s}\,
[z_{12}]^{y-s^{\prime}} = \\ 
[\hat{p}_2]^{x-y}\,
[z_{12}]^{x-s^{\prime}}\,[\hat{p}_1]^{x-s}\,
{\blue [z_{12}]^{s^{\prime}-s}\,[\hat{p}_2]^{y-s}\,
[z_{12}]^{y-s^{\prime}}} = \\ 
{\blue [\hat{p}_2]^{x-y}\,
[z_{12}]^{x-s^{\prime}}\,[\hat{p}_2]^{y-s^{\prime}}}\,
[\hat{p}_1]^{x-s}\,[z_{12}]^{y-s}[\hat{p}_2]^{s^{\prime}-s} = \\ 
[z_{12}]^{y-s^{\prime}}\,[\hat{p}_2]^{x-s^{\prime}}\,
{\blue [z_{12}]^{x-y}\,[\hat{p}_1]^{x-s}\,
[z_{12}]^{y-s}}\,[\hat{p}_2]^{s^{\prime}-s} = \\ 
[z_{12}]^{y-s^{\prime}}\,[\hat{p}_2]^{x-s^{\prime}}\,
[\hat{p}_1]^{y-s}\,[z_{12}]^{x-s}\,[\hat{p}_1]^{x-y}\,[\hat{p}_2]^{s^{\prime}-s}
\end{multline*}
Let us return to the relation \eqref{main1}
\begin{multline*}
[z_{12}]^{x+s^{\prime}-1}\,
\mathcal{K}_1(s,x)\,[z_{12}]^{1-s-s^{\prime}}\,
[\hat{p}_1]^{x-s^{\prime}}\,[\hat{p}_2]^{y-s}\,
[z_{12}]^{x+y-1}\, \mathcal{K}_1(s^{\prime},y) = \\ 
\mathcal{K}_1(s,1-s^{\prime})\,[z_{12}]^{x-s}\,
[\hat{p}_1]^{x-s^{\prime}}\,[\hat{p}_2]^{y-s}\,
[z_{12}]^{y-s^{\prime}}\, \mathcal{K}_1(1-x,y)\,
[z_{12}]^{x+s^{\prime}-1} 
\end{multline*}
which can be rewritten in an equivalent form
\begin{multline*}
[z_{12}]^{x+s^{\prime}-1}\,[\hat{p}_1]^{x-s}\,
[z_{12}]^{1-s-s^{\prime}}\,
\mathcal{K}_1(1-x,1-g; 1-s^{\prime})\,
[\hat{p}_2]^{y-s}\,
[z_{12}]^{x+y-1}\,[\hat{p}_1]^{y-s^{\prime}} = \\ 
[\hat{p}_1]^{1-s-s^{\prime}}\,
[z_{12}]^{x-s}\,
\mathcal{K}_1(s^{\prime},1-g; x)\,
[\hat{p}_2]^{y-s}\,[z_{12}]^{y-s^{\prime}}\,
[\hat{p}_1]^{x+y-1}\,
[z_{12}]^{x+s^{\prime}-1} 
\end{multline*}
In order to derive the second relation we substitute 
in initial relation explicit expression \eqref{K1}
for all reflection operators and cancel the factor 
$[z_{1+}]^{g-s}\,[z_{1-}]^{1-g-s}$ from the left 
and factor $[z_{1+}]^{y-g}\,[z_{1-}]^{y+g-1}$ from the right.
The remaining factors are naturally combined 
into reflection operators with transformed parameter $g \to 1-g$.  

The proof of obtained relation is reduced to the use of 
the star-triangle relation and formula \eqref{pK} in the form 
\begin{align*}
[\hat{p}_1]^{x+s^{\prime}-1}\,
\mathcal{K}_1(1-x,1-g; 1-s^{\prime}) = 
\mathcal{K}_1(s^{\prime},1-g; x)\,
[\hat{p}_1]^{x+s^{\prime}-1}
\end{align*}
We have
\begin{multline*}
{\blue [z_{12}]^{x+s^{\prime}-1}\,[\hat{p}_1]^{x-s}\,
[z_{12}]^{1-s-s^{\prime}}}\,
\mathcal{K}_1(1-x,1-g; 1-s^{\prime})\,
[\hat{p}_2]^{y-s}\,
[z_{12}]^{x+y-1}\,[\hat{p}_1]^{y-s^{\prime}} = \\ 
[\hat{p}_1]^{1-s-s^{\prime}}\,
[z_{12}]^{x-s}\,{\blue [\hat{p}_1]^{x+s^{\prime}-1}\,
\mathcal{K}_1(1-x,1-g; 1-s^{\prime})}\,
[\hat{p}_2]^{y-s}\,
[z_{12}]^{x+y-1}\,[\hat{p}_1]^{y-s^{\prime}} = \\ 
[\hat{p}_1]^{1-s-s^{\prime}}\,
[z_{12}]^{x-s}\,\mathcal{K}_1(s^{\prime},1-g; x)
\,[\hat{p}_2]^{y-s}\,
{\blue [\hat{p}_1]^{x+s^{\prime}-1}\,
[z_{12}]^{x+y-1}\,[\hat{p}_1]^{y-s^{\prime}}} = \\ 
[\hat{p}_1]^{1-s-s^{\prime}}\,
[z_{12}]^{x-s}\,
\mathcal{K}_1(s^{\prime},1-g; x)\,
[\hat{p}_2]^{y-s}\,[z_{12}]^{y-s^{\prime}}\,
[\hat{p}_1]^{x+y-1}\,
[z_{12}]^{x+s^{\prime}-1}
\end{multline*}
The proof of the second relation \eqref{main2} is very similar.

In the main text we also use the formula for action of reflection operator on delta function
\begin{equation} \label{Kdelta}
	\mathcal{K}(x)\,\delta^2(z-\gamma) = \frac{[2\imath\gamma]^{x-s}\,\mathbf{\Gamma}(2-g-s)}{\mathbf{\Gamma}(2-g-x)}\,\delta^2(z-\gamma) \,.
\end{equation}
It is derived by the use of explicit expression for $\mathcal{K}$. We have
\begin{equation} \nonumber
	\mathcal{K}(x) = [z_+]^{g-s}[z_-]^{1-g-s}[\hat{p}]^{x-s}[z_-]^{x+g-1}[z_+]^{x-g} = [z_+]^{g-s}[\hat{p}]^{x+g-1}[z_-]^{x-s}[\hat{p}]^{1-g-s}[z_+]^{x-g} \,.
\end{equation}
Thus,
\begin{equation} \nonumber
	\mathcal{K}(x)\,\delta^2(z-\gamma) = [z_+]^{g-s}[\hat{p}]^{x+g-1}[z_-]^{x-s}[\hat{p}]^{1-g-s}[z_+]^{x-g}\,\delta^2(z-\gamma) \,.
\end{equation}
First, we calculate the action of $[\hat{p}]^{1-g-s}$
\begin{align}
	& \nonumber
	[\hat{p}]^{1-g-s}[z_+]^{x-g}\,\delta^2(z-\gamma) =
	\pi^{-1}\,[\imath]^{1-g-s}\,\mathbf{\Gamma}(2-g-s)
	\int \mathrm{d}^2w\,\frac{[w+\gamma]^{x-g}\,\delta^2(w-\gamma)}{[z-w]^{2-g-s}} \\
	& \nonumber
	= \pi^{-1}\,[\imath]^{1-g-s}\,\mathbf{\Gamma}(2-g-s)\,[2\gamma]^{x-g}\,[z_-]^{g+s-2}
\end{align}
and arrive at
\begin{equation} \nonumber
	\mathcal{K}(x)\,\delta^2(z-\gamma) = \pi^{-1}\,[\imath]^{1-g-s}\,\mathbf{\Gamma}(2-g-s)\,[2\gamma]^{x-g}\,[z_+]^{g-s}[\hat{p}]^{x+g-1}[z_-]^{x+g-2}\,.
\end{equation}
Second, we use \eqref{delta3}
\begin{equation} \nonumber
	[\hat{p}]^{x+g-1}[z_-]^{x+g-2} = [\hat{p}]^{x+g-1}[z-\gamma]^{x+g-2} =
	\frac{\pi\,[\imath]^{x+g-2}}{\mathbf{\Gamma}(2-g-x)}\,\delta^2(z-\gamma) \,.
\end{equation}
To obtain \eqref{Kdelta} it remains to rewrite
\begin{equation} \nonumber
	[z_+]^{g-s}\,\delta^2(z-\gamma) = [z+\gamma]^{g-s}\,\delta^2(z-\gamma) = [2\gamma]^{g-s}\,\delta^2(z-\gamma) \,.
\end{equation}

\section{$\mathcal{R}$-operator identities}

Recall the explicit expressions for $\mathcal{R}$-operator
\begin{equation} \label{Rexpl}
	\mathcal{R}_{12}(x) = [z_{12}]^{1-2s}\,[\hat{p}_1]^{x-s}\,[z_{12}]^{x+s-1}
	= [\hat{p}_1]^{x+s-1}\,[z_{12}]^{x-s}\,[\hat{p}_1]^{1-2s} \,.
\end{equation}
In the proof of $Q$-operators' commutativity we used the following relation
\begin{equation} \label{I}
	\mathcal{R}_{23}(x)\,[\hat{p}_3]^{x-s}\,
	\mathcal{R}_{32}(y)\,\mathcal{R}_{21}(y) = 
	\mathcal{R}_{32}(y)\,[\hat{p}_2]^{x-s}\,
	\mathcal{R}_{21}(y)\,[\hat{p}_2]^{s-x}\,
	\mathcal{R}_{23}(x)\,[\hat{p}_3]^{x-s} \,,
\end{equation}
To derive it we apply \eqref{Rexpl} to $\mathcal{R}_{23}$ and $\mathcal{R}_{32}$ in the left hand side
\begin{multline*}
	\mathcal{R}_{23}(x)\,[\hat{p}_3]^{x-s}\, \mathcal{R}_{32}(y)\,\mathcal{R}_{21}(y) \\
	= [-1]^{y-s}[z_{23}]^{1-2s}[\hat{p}_2]^{x-s}[z_{23}]^{x+s-1}[\hat{p}_3]^{x+y-1}[z_{23}]^{y-s}[\hat{p}_3]^{1-2s}\,\mathcal{R}_{21}(y) \,,
\end{multline*}
then use the star-triangle relation in the operator form \eqref{star-tr}
\begin{equation} \nonumber
	[z_{23}]^{x+s-1}[\hat{p}_3]^{x+y-1}[z_{23}]^{y-s} =
	[\hat{p}_3]^{y-s}[z_{23}]^{x+y-1}[\hat{p}_3]^{x+s-1}
\end{equation}
and rewrite $[-1]^{y-s}[z_{23}]^{1-2s}[\hat{p}_3]^{y-s} = \mathcal{R}_{32}(y)\,[\hat{p}_3]^{1-s-y}$.
Thus,
\begin{equation} \nonumber
	\mathcal{R}_{23}(x)\,[\hat{p}_3]^{x-s}\, \mathcal{R}_{32}(y)\,\mathcal{R}_{21}(y) =
	\mathcal{R}_{32}(y)\,[z_{23}]^{1-s-y}\,[\hat{p}_2]^{x-s}\,[z_{23}]^{x+y-1}\,\mathcal{R}_{21}(y)\,[\hat{p}_3]^{x-s} \,.
\end{equation}
Applying \eqref{star-tr} to the product between $\mathcal{R}$-operators and rewriting $\mathcal{R}_{21}$ in the explicit form one gets the expression
which is finally reduced to the right hand side of \eqref{I}
\begin{multline} \nonumber
	\mathcal{R}_{32}(y)\,[\hat{p}_2]^{x+y-1}\,[z_{23}]^{x-s}\,[z_{21}]^{y-s}\,[\hat{p}_2]^{1-2s}\,[\hat{p}_3]^{x-s} \\
	= \mathcal{R}_{32}(y)\,[\hat{p}_2]^{x-s}\,[\hat{p}_2]^{y+s-1}\,[z_{21}]^{y-s}[\hat{p}_2]^{1-2s}[\hat{p}_2]^{s-x}[\hat{p}_2]^{x+s-1}\,[z_{23}]^{x-s}\,[\hat{p}_2]^{1-2s}\,[\hat{p}_3]^{x-s} \\
	= \mathcal{R}_{32}(y)\,[\hat{p}_2]^{x-s}\,
	\mathcal{R}_{21}(y)\,[\hat{p}_2]^{s-x}\,
	\mathcal{R}_{23}(x)\,[\hat{p}_3]^{x-s} \,.
\end{multline}

Now we are going to prove the following commutation relation which is widely used throughout the text
\begin{align}\label{comm}
	&\mathcal{R}_{12}(x)\cdots\mathcal{R}_{k-1\,k}(x)\,[\hat{p}_k]^{x-s}\,
	\mathcal{R}_{12}(y)\cdots\mathcal{R}_{k-1\,k}(y)\,[\hat{p}_k]^{y-s} = 
	\\
	\nonumber
	&\mathcal{R}_{12}(y)\cdots\mathcal{R}_{k-1\,k}(y)\,[\hat{p}_k]^{y-s}\,
	\mathcal{R}_{12}(x)\cdots\mathcal{R}_{k-1\,k}(x)\,[\hat{p}_k]^{x-s} \,.
\end{align}
To derive the simplest identity for $k=1$
\begin{equation} \label{comm3}
	\mathcal{R}_{12}(x)\,[\hat{p}_2]^{x-s}\,\mathcal{R}_{12}(y)\,[\hat{p}_2]^{y-s} =
	\mathcal{R}_{12}(y)\,[\hat{p}_2]^{y-s}\,\mathcal{R}_{12}(x)\,[\hat{p}_2]^{x-s}
\end{equation}
one needs to use the explicit formula \eqref{Rexpl} for $\mathcal{R}$-operator and apply the star-triangle relation \eqref{star-tr} several times.
In the general case the proof
is based on the following relation for $\mathcal{R}$-operators
\begin{align}\label{YB}
	\mathcal{R}_{12}(x,y)\,\mathcal{R}_{23}(x)\,\mathcal{R}_{12}(y) = 
	\mathcal{R}_{23}(y)\,\mathcal{R}_{12}(x)\,\mathcal{R}_{23}(x,y) \,,
\end{align}
where 
\begin{align*}
	\mathcal{R}_{12}(x,y) = [z_{12}]^{1-s-y}\,[\hat{p}_1]^{x-y}\,[z_{12}]^{s+x-1} = \mathcal{R}_{12}(x)\,\mathcal{R}^{-1}_{12}(y) = 
	\mathcal{R}^{-1}_{12}(y)\,\mathcal{R}_{12}(x) \,.
\end{align*}
To show how it works we will use representative example $k=3$
\begin{align} \label{comm4}
	\mathcal{R}_{12}(x)\,\mathcal{R}_{23}(x)\,[\hat{p}_3]^{x-s}\,
	\mathcal{R}_{12}(y)\,\mathcal{R}_{23}(y)\,[\hat{p}_3]^{y-s} = 
	\mathcal{R}_{12}(y)\,\mathcal{R}_{23}(y)\,[\hat{p}_3]^{y-s}\,
	\mathcal{R}_{12}(x)\,\mathcal{R}_{23}(x)\,[\hat{p}_3]^{x-s}
\end{align}
and hope that generalization will be evident. 
We have 
\begin{align*}
	\mathcal{R}_{12}(x)\,\mathcal{R}_{23}(x)\,[\hat{p}_3]^{x-s}\,
	\mathcal{R}_{12}(y)\,\mathcal{R}_{23}(y)\,[\hat{p}_3]^{y-s} = \\
	\mathcal{R}_{12}(y)\,
	\mathcal{R}_{12}(x,y)\,\mathcal{R}_{23}(x)\,
	\mathcal{R}_{12}(y)\,[\hat{p}_3]^{x-s}\,\mathcal{R}_{23}(y)\,
	[\hat{p}_3]^{y-s} = \\ 
	\mathcal{R}_{12}(y)\,\mathcal{R}_{23}(y)\,
	\mathcal{R}_{12}(x)\,\mathcal{R}_{23}(x,y)\,[\hat{p}_3]^{x-s}\,\mathcal{R}_{23}(y)\,
	[\hat{p}_3]^{y-s} = \\
	\mathcal{R}_{12}(y)\,\mathcal{R}_{23}(y)\,
	\mathcal{R}_{12}(x)\,\mathcal{R}^{-1}_{23}(y)\,
	\mathcal{R}_{23}(x)\,[\hat{p}_3]^{x-s}\,\mathcal{R}_{23}(y)\,
	[\hat{p}_3]^{y-s} \,.
\end{align*}
Now we use the $n=1$ commutation \eqref{comm3} to permute the combinations $\mathcal{R}_{23}(x)\,[\hat{p}_3]^{x-s}$ and $\mathcal{R}_{23}(y)\,[\hat{p}_3]^{y-s}$, so the last expression takes the form
\begin{align*}
	\mathcal{R}_{12}(y)\,\mathcal{R}_{23}(y)\,
	\mathcal{R}_{12}(x)\,\mathcal{R}^{-1}_{23}(y)\,
	\mathcal{R}_{23}(y)\,[\hat{p}_3]^{y-s}\,\mathcal{R}_{23}(x)\,
	[\hat{p}_3]^{x-s} = \\ 
	\mathcal{R}_{12}(y)\,\mathcal{R}_{23}(y)\,[\hat{p}_3]^{y-s}\,
	\mathcal{R}_{12}(x)\,\mathcal{R}_{23}(x)\,[\hat{p}_3]^{x-s} \,.
\end{align*}
Thus the relation \eqref{comm4} is proven.

In section~\ref{sect:Decomp} we used the formula
\begin{multline} \label{RRp}
	\mathcal{R}_{12}(y)\ldots\mathcal{R}_{n-1\,n}(y)\,[\hat{p}_n]^{y-s} =
	[-1]^{(n-1)(y+s)}\,[\hat{p}_1]^{y+s-1}\,\mathcal{R}_{21}^{-1}(1-y)\ldots\mathcal{R}_{n\,n-1}^{-1}(1-y) \\
	\times\mathcal{R}_{12}(1-s)\ldots\mathcal{R}_{n-1\,n}(1-s)\,[\hat{p}_n]^{1-2s} \,,
\end{multline}
It is proven by induction. The basic case $n=2$
\begin{equation} \label{RRp1}
	\mathcal{R}_{12}(y)\,[\hat{p}_2]^{y-s} =
	[-1]^{y+s}\,[\hat{p}_1]^{y+s-1}\,\mathcal{R}_{21}^{-1}(1-y)\,\mathcal{R}_{12}(1-s)\,[\hat{p}_2]^{1-2s}
\end{equation}
is verified by direct calculation with the help of \eqref{Rexpl}
\begin{align}
	& \nonumber
	\mathcal{R}_{12}(y)\,[\hat{p}_2]^{y-s} = [\hat{p}_1]^{y+s-1}[z_{12}]^{y-s}[\hat{p}_1]^{1-2s}[\hat{p}_2]^{y-s} \\
	& \nonumber
	= [\hat{p}_1]^{y+s-1}[z_{12}]^{y-s}[\hat{p}_2]^{y+s-1}[\hat{p}_1]^{1-2s}[\hat{p}_2]^{1-2s} \\
	& \nonumber
	= [\hat{p}_1]^{y+s-1}{\blue [z_{12}]^{y-s}[\hat{p}_2]^{y+s-1}[z_{12}]^{2s-1}}\;
	{\red [z_{12}]^{1-2s}[\hat{p}_1]^{1-2s}}[\hat{p}_2]^{1-2s} \\
	& \nonumber
	= [-1]^{y+s}\,[\hat{p}_1]^{y+s-1}\mathcal{R}_{21}^{-1}(1-y)
	\mathcal{R}_{12}(1-s)[\hat{p}_2]^{1-2s} \,,
\end{align}
where the coloured combinations are equal to $\mathcal{R}_{21}^{-1}(1-y)$ and $\mathcal{R}_{12}(1-s)$.

The induction step $n-1\to n$ is proven as follows. We begin with inserting mutually inverse operators $[\hat{p}_{n-1}]^{y-s}$ and $[\hat{p}_{n-1}]^{s-y}$
\begin{equation} \nonumber
	\mathcal{R}_{12}(y)\ldots\mathcal{R}_{n-1\,n}(y)\,[\hat{p}_n]^{y-s} = 
	\mathcal{R}_{12}(y)\ldots\mathcal{R}_{n-2\,n-1}(y)\,[\hat{p}_{n-1}]^{y-s}[\hat{p}_{n-1}]^{s-y}\mathcal{R}_{n-1\,n}(y)\,[\hat{p}_n]^{y-s} \,.
\end{equation}
Using the induction proposition we rewrite
\begin{align}
	\nonumber
	\mathcal{R}_{12}(y)\ldots\mathcal{R}_{n-2\,n-1}(y)\,[\hat{p}_{n-1}]^{y-s} & =
	[-1]^{(n-2)(y+s)}\,[\hat{p}_1]^{y+s-1}\,\mathcal{R}_{21}^{-1}(1-y)\ldots\mathcal{R}_{n-1\,n-2}^{-1}(1-y) \\
	& \nonumber
	\times\mathcal{R}_{12}(1-s)\ldots\mathcal{R}_{n-2\,n-1}(1-s)\,[\hat{p}_{n-1}]^{1-2s} \,,
\end{align}
and transform the product $[\hat{p}_{n-1}]^{s-y}\mathcal{R}_{n-1\,n}(y)\,[\hat{p}_n]^{y-s}$ employing the basic formula \eqref{RRp1} with $z_1, z_2$ replaced by $z_{n-1}, z_n$
\begin{equation} \nonumber
	[\hat{p}_{n-1}]^{s-y}\mathcal{R}_{n-1\,n}(y)\,[\hat{p}_n]^{y-s} =
	[-1]^{y+s}\,[\hat{p}_{n-1}]^{2s-1}\,\mathcal{R}_{n\,n-1}^{-1}(1-y)\,\mathcal{R}_{n-1\,n}(1-s)\,[\hat{p}_n]^{1-2s} \,.
\end{equation}
Hence, we obtain
\begin{align}
	& \nonumber
	\mathcal{R}_{12}(y)\ldots\mathcal{R}_{n-1\,n}(y)\,[\hat{p}_n]^{y-s} =
	[-1]^{(n-1)(y+s)}\,[\hat{p}_1]^{y+s-1}\,\mathcal{R}_{21}^{-1}(1-y)\ldots\mathcal{R}_{n-1\,n-2}^{-1}(1-y) \\
	& \nonumber
	\times\mathcal{R}_{12}(1-s)\ldots\mathcal{R}_{n-2\,n-1}(1-s)\,[\hat{p}_{n-1}]^{1-2s}[\hat{p}_{n-1}]^{2s-1}\,\mathcal{R}_{n\,n-1}^{-1}(1-y)\,\mathcal{R}_{n-1\,n}(1-s)\,[\hat{p}_n]^{1-2s} \\
	& \nonumber
	= [-1]^{(n-1)(y+s)}\,[\hat{p}_1]^{y+s-1}\,\mathcal{R}_{21}^{-1}(1-y)\ldots\mathcal{R}_{n-1\,n-2}^{-1}(1-y)\,\mathcal{R}_{n\,n-1}^{-1}(1-y) \\
	& \nonumber
	\times\mathcal{R}_{12}(1-s)\ldots\mathcal{R}_{n-2\,n-1}(1-s)\,\mathcal{R}_{n-1\,n}(1-s)\,[\hat{p}_n]^{1-2s} \,,
\end{align}
which was to be proved. At the last step the operators $[\hat{p}_{n-1}]^{1-2s}$ and $[\hat{p}_{n-1}]^{2s-1}$ canceled out, and we commuted the operator $\mathcal{R}_{n\,n-1}^{-1}(1-y)$ with $\mathcal{R}_{12}(1-s)\ldots\mathcal{R}_{n-2\,n-1}(1-s)$.

\end{document}